\documentclass{emulateapj}
\usepackage{apjfonts}
\usepackage{times} 
\usepackage{graphicx}
\usepackage{rotating}
\usepackage[mathscr]{euscript}
\usepackage{array}
\usepackage[usenames]{color}
\usepackage{epsfig}
\usepackage{amsmath}
\usepackage{natbib}
\def\mbh{\ifmmode{{\mathrm M}_{BH}\,}\else{M$_{BH}$\,}\fi}
\def\msigma{\ifmmode{{\mathrm M}_{BH}-{\sigma}\,}\else{M$_{BH}- \sigma$\,}\fi}
\def\msun{\ifmmode{{\mathrm M}_{\odot}}\else{M$_{\odot}$\,}\fi} 
\def\kms{\ifmmode{{\mathrm{km \, s^{-1}}}}\else{${\mathrm{km \, s^{-1}}}$}\fi}

\def\lesssim{\mathrel{\hbox{\rlap{\hbox{\lower4pt\hbox{$\sim$}}}\hbox{$<$}}}}
\def\gtrsim{\mathrel{\hbox{\rlap{\hbox{\lower4pt\h$N$-bodybox{$\sim$}}}\hbox{$>$}}}}

\def\beq{\begin{equation}}
\def\eeq{\end{equation}}

\shorttitle{Unifying $N$-body bars and triaxial ellipsoids}
\shortauthors{Valluri et al. }

\begin{document}

\title{A unified framework for the orbital structure of bars and triaxial ellipsoids}

\author{Monica Valluri \altaffilmark{1}}
\affil{Department of Astronomy, University of Michigan, Ann Arbor, MI 48109, USA}
\author{Juntai Shen\altaffilmark{2}}
\affil{Key Laboratory for Research in Galaxies and Cosmology, Shanghai Astronomical Observatory, Chinese Academy of Sciences, 80 Nandan Road, Shanghai 200030, China}
\author{Caleb Abbott \altaffilmark{3}}
\affil{Department of Astronomy, University of Michigan, Ann Arbor, MI 48109, USA}
\and
 \author{Victor P. Debattista \altaffilmark{4}}
 \affil{Jeremiah Horrocks Institute, University of Central Lancashire,  Preston, PR1 2HE, UK}
\altaffiltext{1}{{E-mail: mvalluri@umich.edu}}
\altaffiltext{2}{{E-mail: jshen@shao.ac.cn}}
\altaffiltext{3}{{E-mail: calebga@umich.edu}}
\altaffiltext{4}{{E-mail: vpdebattista@uclan.ac.uk}}




\begin{abstract}
 We examine a large random sample of orbits in self-consistent simulations of $N$-body bars. Orbits in the bars are classified both visually and with {  a new automated orbit classification method based on frequency analysis. The well known prograde x1 orbit family originates from the same parent orbit as the box orbits in stationary and rotating triaxial ellipsoids. However only a small fraction of bar orbits $\sim$4\% have predominately prograde motion like their periodic parent orbit.} Most bar orbits arising from the x1 orbit have little net angular momentum in the bar frame making them equivalent to box orbits in rotating triaxial potentials.  A small fraction of bar orbits ($\sim$7\%)  are long axis tubes that behave exactly like  those in triaxial ellipsoids:they are tipped about the intermediate-axis due to the Coriolis force, with the sense of tipping determined by the sign of their angular momentum about the long axis. No orbits parented by prograde periodic x2 orbits are found in the pure bar model, but a tiny population ($\sim$2\%) of short axis tube orbits parented by retrograde x4 orbits are found.  When a central point mass representing a supermassive black hole (SMBH) is grown adiabatically at the center of the bar, those orbits that lie in the immediate vicinity of the SMBH are transformed into precessing Keplerian orbits (PKOs) which belong the same major families (short axis tubes, long axis tubes and boxes) occupying the bar at larger radii. During the growth of a SMBH  the inflow of mass and outward transport of angular momentum transforms some x1 and long axis tube orbits into prograde short axis tubes. This study has important implications for future attempts to constrain the masses of SMBHs in barred galaxies using orbit based methods like the Schwarzschild orbit superposition scheme and for understanding the observed features in barred galaxies.
\end{abstract}


\section{Introduction}
\label{sec:intro}

Historically the study of orbits in potentials has focused on periodic orbits.   In systems like disk galaxies small perturbations to closed periodic orbits (e.g. the epicyclic and vertical perturbations of circular orbits) provided a good analytic description of most orbits. Self-consistent distribution functions are thought to be  ``parented''  by  stable periodic orbits \citep{arnold_78}. Early works \citep[e.g.][]{contopoulos_papayannopoulos_80}  identified and characterized the stability properties of the periodic orbit families in  rapidly rotating bars. The most important periodic families in two dimensional bars were identified as the  {  prograde} x1 family which is elongated along the major axis of the bar, the  prograde stable  x2  family and  unstable x3 families which are elongated perpendicular to the bar (primarily found at small radii).  The retrograde x4 (stable) orbit family is also elongated perpendicular to the bar at small radii, but becomes rounder as it extends to larger radii \citep[for detailed description of orbit families and how they are identified see][]{contopoulos_grosbol_89, sellwood_wilkinson_93,BT08,sellwood_14RMP}. In the frame of reference rotating with the bar, all of these families are characterized by a 1:2 resonance between the tangential oscillation frequency ($\Omega_\phi$) and the radial or epicyclic frequency ($\Omega_R$).  Indeed studies of orbits in 2D $N$-body bars  largely confirmed the picture arising from the study of periodic orbits and showed that many regular orbits elongated along the bar were {  parented by x1 orbits, a small fraction were parented by retrograde x4 orbits \citep{sparke_sellwood_87} and none were parented by prograde x2 orbits.}  The realization that bars can also undergo buckling instabilities \citep{com_san_81,Raha91} which makes them develop substantial vertical thickness and peanut shaped morphologies, led to the study of periodic orbits in three dimensional bars \citep{pfenniger_84a,martinet_dezeeuw_88,pfenniger_friedli_91,skokos1,skokos2}. It was shown that  appearance of specific morphological features in images of bars, such as the X-shape and peanut features seen in edge-on bars and the boxy/rectangular isophotes and  ``ansae'' of face-on bars could be explained by orbits trapped around specific periodic orbit families \citep{patsis1, patsis2, patsis_10}.   The introduction of a third dimension did not drastically change the picture of the nature of periodic orbits and it was found that 3D bars are comprised primarily of vertical bifurcations (resonances) of the x1 family  and a few additional families \citep[e.g.][]{pfenniger_friedli_91,skokos1,skokos2}. Most studies of periodic orbits in analytic potentials consider prograde x2 orbits (but not retrograde x4 orbits) to be another fundamental building block of bars \citep{skokos1, BT08}. In our study we do not find any orbits parented by the periodic prograde x2  orbit in our initial bar model but we do find orbits that are parented by the periodic retrograde x4 -- a result that is consistent with previous studies \citep{sparke_sellwood_87,pfenniger_friedli_91,voglis_etal_07}. 

In contrast with the study of bars which has largely focused on planar periodic orbits and their vertical bifurcations, the orbits in triaxial ellipsoids are considered to belong to four regular families which occupy three dimensions: (a) short axis tubes\footnote{Hereafter we will use a coordinate system in which the long axis of the potential is aligned with the $x$-axis, the short axis of the potential is aligned with the $z$-axis and the intermediate axis is aligned with the $y$-axis.}; (b) inner long axis tubes; (c) outer long axis tubes and (d) box orbits and resonant boxlets \citep{dezeeuw_85b,statler_87,miralda-escude_schwarzschild_89}. These families are primarily distinguished their net angular momentum:  short ($z$) axis tubes have non-zero angular momentum about the short axis ($\langle{J_z}\rangle \ne 0$);  both families of  long axis tubes have a net angular momentum about the long ($x$) axis ($\langle{J_x}\rangle \ne 0$);  boxes and boxlets have no net angular momentum about any axis. In addition, in a realistic triaxial potential, a significant fraction of orbits may be chaotic \citep[e.g.][]{schwarzschild_93,merritt_fridman_96}. 

The earliest studies of orbits in a triaxial system also examined orbital structure under the premise that the major families arise from perturbations of stable periodic orbits.  For example when discussing the numerical distributions derived by \citet{schwarzschild_79}, \citet{dezeeuw_85a} states ``... phase-space is well ordered and [...] most  orbits belong  to one of  a few major families, each connected with a simple periodic orbit...''.  \citet{dezeeuw_85b} showed that equations of motion in a triaxial ellipsoid were separable in ellipsoidal coordinates and that the major orbit families (the boxes and three families of tubes) had well defined shapes and where characterized by well defined relationships between their integrals of motion. 

In this framework box orbits arise from perturbations (in two perpendicular directions $y$ and $z$) to the $x$-axial orbit (i.e. they are Lissajous curves in 3 dimensions). Since box orbits are comprised of orthogonal perturbations of the $x$-axial orbit they have no net angular momentum about any axis of symmetry. In contrast short axis tubes arise from perturbations (in the radial and $z$ direction) to a closed (periodic) elliptical orbit that circulates about the $z$-axis in the $x-y$ plane. Consequently short axis tubes have a net angular momentum about the $z$-axis. Likewise long axis tubes arise from perturbations to a closed (periodic) elliptical orbit in the $y-z$ plane and have a net angular momentum about the $x$-axis. Intermediate axis tube orbits do not exist since elliptical orbits that circulate in the $x-z$ plane perpendicular to the intermediate axis are unstable \citep{heiligman_schwarzschild_79,adams_etal_07}.  

When a triaxial potential is subjected to rotation of the figure about the $z$-axis, the $x$-axial orbit no longer oscillates back and forth along the long axis of the figure. Instead, in the frame that is co-rotating with the potential, such a particle experiences a Coriolis force whose sign changes each time the orbit reverses direction. Consequently the particle follows an elliptical trajectory \citep{schwarzschild_82,dezeeuw_merritt_83, martinet_dezeeuw_88} in a prograde sense about the $z$-axis. 

\citet{heisler_etal_82} showed that when subjected to figure rotation (about the short axis), the clockwise and anti-clockwise 1:1 periodic loop orbits circulating about the long axis are tipped about the  intermediate axis by the Coriolis force, giving them  additional retrograde angular momentum about the short axis.  These orbits are referred to as ``anomalous'' and stable and unstable versions exist. The direction in which these orbits  are tipped about the $y$-axis depends on the sign of $J_x$. {   At large energies these orbits become complex unstable \citep{martinet_dezeeuw_88, patsis_zachilas_90}.  The anomalous orbits are also connected with the x4 family \citep{pfenniger_friedli_91}. Since the loop orbits circulating around the long axis are the parents of the long axis tube family in a stationary triaxial potential it follows that the stable anomalous orbits parent the long axis tube families in a rotating triaxial potential.} Consequently long axis tube orbits are also tipped about the intermediate axis in a rotating system \citep{deibel_etal_11}. 

Finally, while short axis tubes remain stable under figure rotation the phase space occupied by prograde short axis tubes decreases  dramatically with increasing pattern speed and are increasingly replaced by retrograde short axis tubes \citep{martinet_dezeeuw_88,deibel_etal_11}. 

{  Since bars form via secular instabilities from rapidly rotating disks while stationary or slowly tumbling triaxial ellipsoids are thought to form primarily from collisionless mergers, 
in the current literature these systems are usually considered to be fundamentally different from each other \citep[e.g.][]{BT08}.  However, early studies of rotating triaxial figures \citep[e.g.]{schwarzschild_82,heisler_etal_82, dezeeuw_merritt_83,martinet_dezeeuw_88}  predicted that the behavior of orbits in rotating triaxial systems and bars were fundamentally similar.}

{  In this paper we examine  high resolution  $N$-body simulations of bars with and without a point mass representing  a supermassive black hole (SMBH) and show that all the families of orbits in triaxial ellipsoids are present in self-consistent distribution functions of $N$-body bars as previously predicted. Unlike studies which analyze the periodic orbits in an analytic potential our goal is to examine the orbits that comprise the actual distribution functions of $N$-body bars with and without SMBH. Our main goals are (a) to develop an automated orbital classification method for bar orbits, and (b) to use it to understand how the orbit populations are modified by the growth of a SMBH.}

In Section~\ref{sec:methods} we describe the $N$-body simulations of self-consistent bars and the prolate triaxial Dehnen model used to illustrate the similarity with orbits in  $N$-body bars. In Section~\ref{sec:bar_orbits} we characterize the main orbit families present in our two $N$-body bars and compare them with orbits integrated in stationary and rotating triaxial Dehnen models. In Section~\ref{sec:population_DF} we compare the orbit populations in the pure bar with the populations in the bar after the growth of a central point mass representing a SMBH. We also characterize the phase space distribution of the bar models using both surfaces-of-section and frequency maps and show how the orbit populations vary with orbital apocenter radius. In Section~\ref{sec:conclude} we discuss the implications of this work and summarize our results.  In Appendix~\ref{sec:frequencies} we briefly describe the orbital frequency analysis method and in Appendix~\ref{sec:autoclassification} we describe our new automatic classification scheme for orbits in $N$-body bars.

\section{Methods}
\label{sec:methods}

\subsection{Simulations}
\label{sec:simulations}

The $N$-body disk models studied in this paper are almost identical to those presented in \citet[][ hereafter SS04]{shen_sellwood_04} and were previously  analyzed by \citet{brown_etal_13} who examined the effects of the growth of SMBHs on  the measurement of the observable velocity dispersion within the effective radius ($\sigma_e$) and the resultant consequences for the scatter and slope of the well known scaling relationship between SMBH mass ($\mbh$) and stellar velocity dispersion ($\sigma$): the ``$\mbh-\sigma$ relationship''.  Below we give a brief description of the simulations and refer the reader to the above papers for a more detailed discussion of the set up and simulation process.

In this paper we restrict our orbital analysis to a single set of initial conditions comprised of a Kuzmin disk embedded in a static logarithmic halo \citep[for details see][]{shen_sellwood_04}. Following standard practice the units used in the simulations are $G=M_d=R_d=1$ where $G$ is Newton's gravitational constant, $M_d$ is the mass of the disk, and $R_d$ is the initial disk scale-radius. By dimensional analysis the unit of time is $t_{\mathrm{dyn}} =(R_d^3/GM_d)^{1/2}$. In this paper all figures are presented in these units. Physically relevant scalings can be obtained by choosing observationally motivated values for $M_d$ and $R_d$. For example $M_d = 5 \times 10^{10} M_{\odot}$ and $R_d = 1$~kpc, would yield a unit of time $t_{\mathrm{dyn}} \sim$ 2.1~Myr.  In these units the semi major axis length of the bar is about  3 -- 4~kpc in length and the disk has a total radius of about 25~kpc, making it similar to the Milky Way.

The initial  disk distribution function consisted of  2.8 $\times 10^6$  disk particles set up with a Toomre $Q$  parameter $\simeq$ 1.5, a condition that ensures that it is unstable to bar formation \citep{Athanassoula86}.  The initial conditions were  evolved using a three-dimensional, cylindrical, polar grid--based $N$-body code described in \citet{sellwood_valluri_97} and \citet{sellwood14}.

The disk formed a bar which subsequently experienced vertical thickening via the buckling instability  \citep{Toomre66, Raha91, sellwood_wilkinson_93} and developed the peanut-shaped bulge that is characteristic of this instability.  
After the disk reached a  nearly steady state (simulation time $t_1=700$) a central point mass\footnote{with softening parameter $\epsilon_{\rm cmc}= 0.001R_d$} representing a SMBH  of mass  0.2\%$M_d$ was grown adiabatically \citep[for details see][]{brown_etal_13}.  At $t_2=1200$ the transients due to the changing central point mass (SMBH) potential had dissipated and the simulation was frozen and orbits were examined and compared with orbits from the frozen potential  at $t_1$. 

Hereafter we refer to the frozen snapshots at $t_1$ and $t_2$ as Model A and Model B respectively. The bar pattern speed at $t_1$ and $t_2$ was computed from several successive time steps before and after each snapshot. {  The galaxy potential for each snapshot is derived from the full $N$-body distribution and is not described by an analytic form. Since our objective is to understand the actual orbits that populate the $N$-body distribution function we  randomly select 10,000 particles (from the total of $\sim$2.8 million) uniformly sampling the entire distribution function. The instantaneous positions and velocities of the selected particles (transformed to the rotating frame of the bar) were used to advance each of the 10,000 particles  individually while keeping the rest of the particles fixed, by using a modification of the $N$-body code used to run the simulations. Orbit integration was carried out as described in footnote 10.} The orbits were integrated in Cartesian coordinates\footnote{The Cartesian coordinate system was oriented with the major axis of the bar aligned with the $x$-axis, and with the angular momentum of the disk aligned with the $z$-axis, and the intermediate axis of the bar along the $y$-axis.} in the rotating frame using the accelerations derived from the frozen potential of that snapshot, and adding the appropriate Coriolis and centrifugal pseudo-forces determined by the bar pattern speed (see equations 2-4).  Each orbit was integrated for 1000 time units (corresponding to $\sim 2.1\times 10^9$years for the units adopted above) and sampled at 20,000 equally spaced time intervals. Since orbits at different radii have different orbital periods this corresponds to  hundreds of orbital periods for the innermost particles but to just tens of orbital periods for orbits in the outskirts of the disk. 

Since the 10,000 particles for which orbits were integrated were selected at random from the entire distribution function, disk particles were included. The bar length in $N$-body simulations is generally estimated to be the radius at which the strength of the bar mode $A_2$ drops to some fraction (e.g. 1/2) of its maximum  value \citep{ath_mis_02}.  By this criterion the bar length in this simulation is estimated to be about 3 (in program length units). However visual classification of all 10000 orbits in Model A showed that a significant fraction of bar-like orbits continue to exist all the way out to $\sim 4$ units in length (see Appendix).  

\subsection{The Triaxial Dehnen Potential}
\label{sec:triax}

We compare orbits from the bar snapshots with orbits in stationary and rotating triaxial potentials which are generalizations of the spherical ``Dehnen models''    \citep{dehnen_93,tremaine_etal_94}. These models have density profiles which provide good fits to luminosity profiles of elliptical galaxies and the bulges of spiral galaxies. The density distribution
 in these models is stratified on concentric ellipsoids with principle axes aligned with Cartesian coordinates $x, y, z$ and with the semi-major, semi-intermediate and semi-minor axes lengths $a$, $b$, $c$ respectively.  The parameter $\gamma$ determines the logarithmic
slope of the central density cusp, and ranges observationally from $\gamma = 0.1-1$ in bright galaxies with shallow cusps to $\gamma=2$ in fainter galaxies with steep cusps \citep{gebhardt_etal_96,lauer_etal_07}. We  restrict our selves to models with $\gamma = 0.1$ since more cuspy models produce many resonant boxlet orbit families  \citep{miralda-escude_schwarzschild_89,merritt_valluri_99} that are not found in our $N$-body bars. Equations describing the potential and the accelerations for this model are given in \citet{merritt_fridman_96} and \citet{deibel_etal_11}. 

In a rotating frame,  the Jacobi integral ($E_J$) is a conserved quantity (equivalent to energy in a stationary frame):
\begin{equation}
E_J = {1\over 2}|\dot{\mathbf{x}}|^2+\Phi-{1\over 2}|\Omega_p \times {\mathbf{x}}|^2,
\end{equation}
where $\mathbf{x}$ and $\mathbf{\dot{x}}$ are 3 dimensional spatial and velocity vectors respectively. 
When the rotation is about the short axis ($z$)  the equations of motion in Cartesian coordinates are  given by:
\begin{align}
\ddot{x}  =&  - {{\partial\Phi\over{\partial x}}} + 2\Omega_p \dot{y}+ \Omega_p^2 x,\\
\ddot{y}  = & - {{\partial\Phi\over{\partial y}}} - 2\Omega_p \dot{x}+ \Omega_p^2 y,\\
\ddot{z}  =  &- {{\partial\Phi\over{\partial z}}}, 
\end{align}
where $2\Omega_p \dot{y}$ and $-2\Omega_p \dot{x}$ are  components of the Coriolis force and $\Omega_p^2x$ and $\Omega_p^2y$ are components of the centrifugal force. Following standard practice we adopt the right-handed rule where motion anticlockwise  about the $z$ axis (as viewed from positive $z$) has positive angular momentum, while clockwise motion about the $z$ axis has negative angular momentum. In this terminology the direction of the pattern rotation of the simulated bars and the direction in which the triaxial Dehnen model is rotating when viewed from an inertial frame are both positive (anti-clockwise). Similarly when we discuss  short (long) axis tubes, anti-clockwise motion about the $z$ ($x$) axis corresponds to positive angular momentum $J_z$ ($J_x$).

\citet{deibel_etal_11} studied orbits in a  triaxial Dehnen models with a variety of shapes subjected to a range of pattern speeds. Here we restrict our comparison to orbits in  a prolate-triaxial potential with axis ratios $c/a=0.4$, $b/a=0.48$ and hence triaxiality parameter $T= (a^2 - b^2)/(a^2 - c^2) =0.916$  and a weak cusp ($\gamma=0.1$). This shape is quite close to that of the bars in our $N$-body simulations. As in the case of the $N$-body simulation we adopt a set of units where the total mass of the model $M$, the semi-major axis scale-length $a$, and the
gravitational constant $G$ are set to unity. 

Following standard practice, the patten speed used to describe the tumbling of the triaxial figure is defined in terms of the ``co-rotation radius'', hereafter $R_{CR}$. In a nearly axisymmetric potential the co-rotation radius is the radius at which the azimuthal frequency  of a closed (almost circular) orbit in the equatorial ($x-y$) plane of the potential is the same as the pattern frequency (generally called ``pattern speed'') $\Omega_p$. The pattern speeds of bars have been measured by applying the Tremaine-Weinberg method \citep{tre_wei_84,tremaine_weinberg_meth_08} to stellar kinematical velocity fields (as well as  H$\alpha$ and H\,{\small I}). These measurements find that ratio of the co-rotation radius to the length of the bar $R_{CR}/R_{\rm bar} = [1, 1.4]$ for  a bar of length $R_{\rm bar}$ \citep[e.g.][]{mer_kui_95,debattista_etal_02,deb_wil_04,aguerri_etal_03, corsini_10,aguerri_etal_15}.   For the bar in the simulations described in the previous section, $R_{CR}/R_{\rm bar} \sim 1$.  For the orbits in the triaxial Dehnen model the pattern speed was chosen so that the co-rotation radius  was at roughly the same radial distance from the center as it is in the case of the bar simulations (i.e. at $\sim 3$ units or one bar length). Note that the triaxial ellipsoidal surface with major-axis length $\sim 3$ in the Dehnen model encloses only  1/2 of the total mass of the model. To ensure a fair comparison the orbits from the stationary and rotating triaxial Dehnen model were selected to have similar radial extents as the bar orbits that they are compared with.

\subsection{Orbit classification}
\label{sec:visual}

\begin{figure*}
\centering
\includegraphics[trim=0.pt 0.pt 163.pt 0.pt ,clip, angle=0.,width=0.6\textwidth]{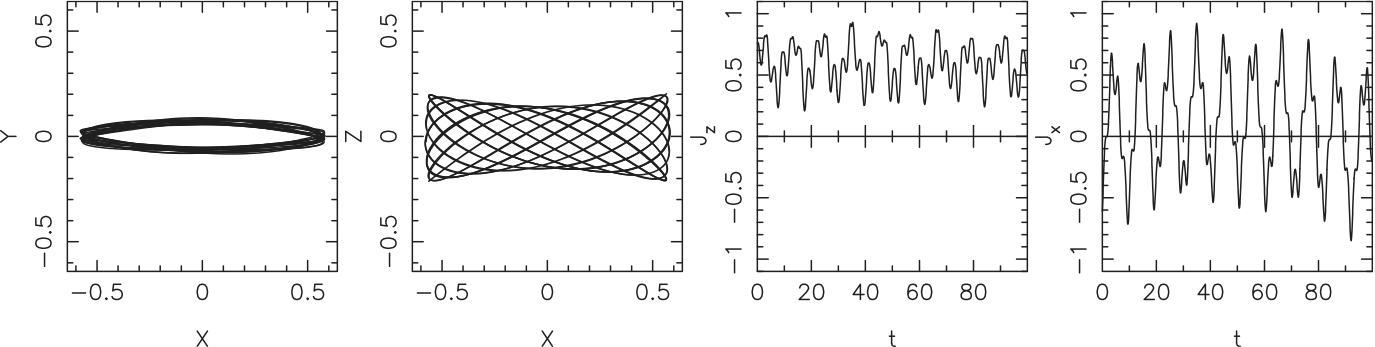}
\includegraphics[trim=0.pt 0.pt 163.pt 0.pt ,clip, angle=0.,width=0.6\textwidth]{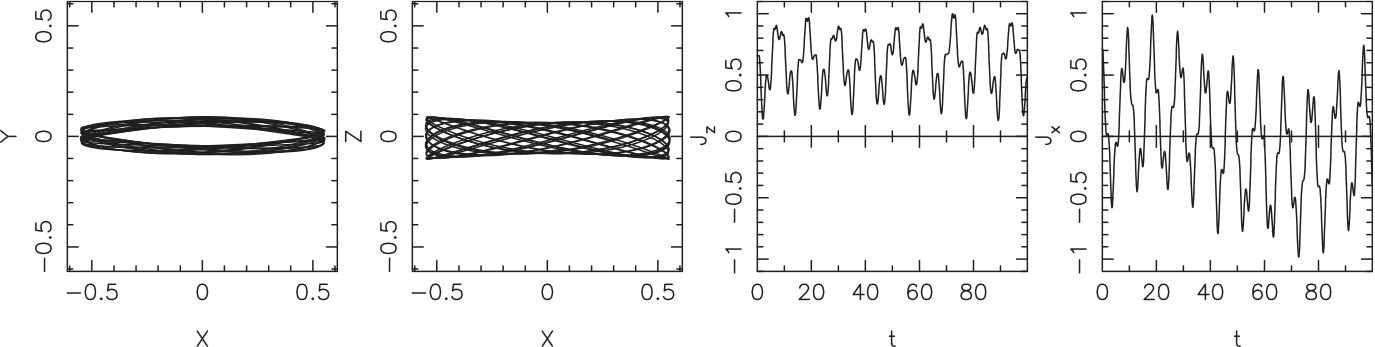}
\includegraphics[trim=0.pt 0.pt 163.pt 0.pt ,clip, angle=0.,width=0.6\textwidth]{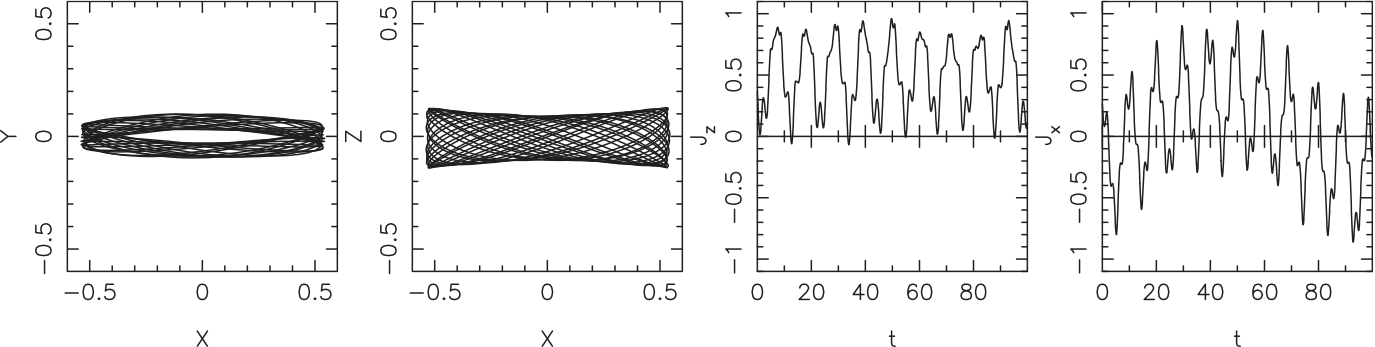}
\includegraphics[trim=0.pt 0.pt 163.pt 0.pt ,clip, angle=0.,width=0.6\textwidth]{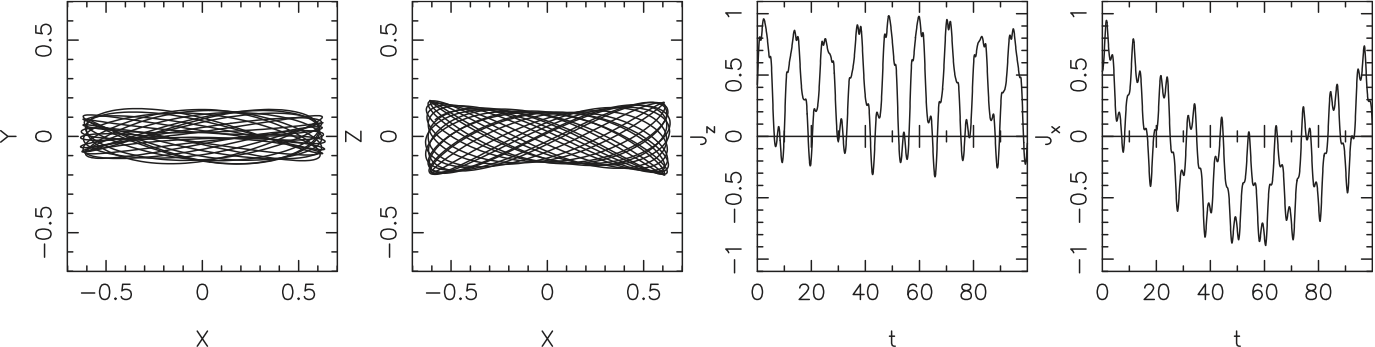}
\includegraphics[trim=0.pt 0.pt 163.pt 0.pt ,clip, angle=0.,width=0.6\textwidth]{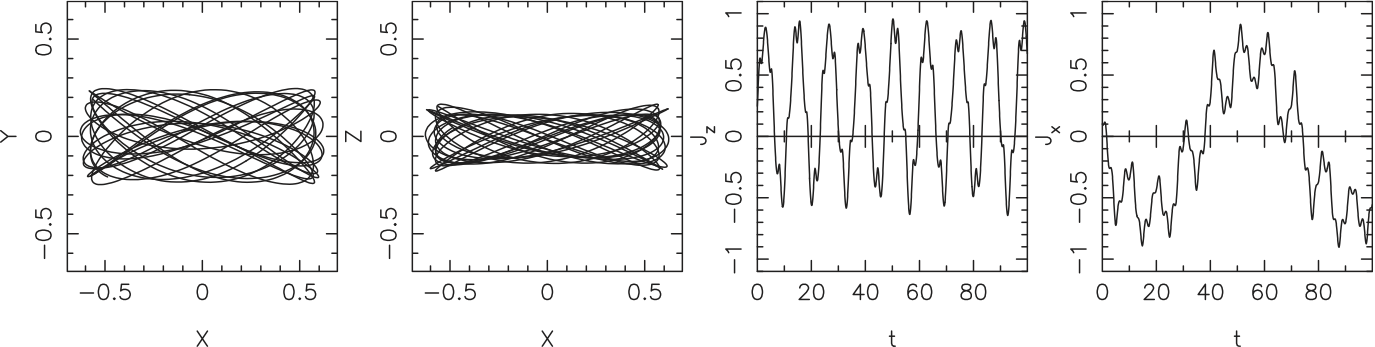}
\includegraphics[trim=0.pt 0.pt 163.pt 0.pt ,clip, angle=0.,width=0.6\textwidth]{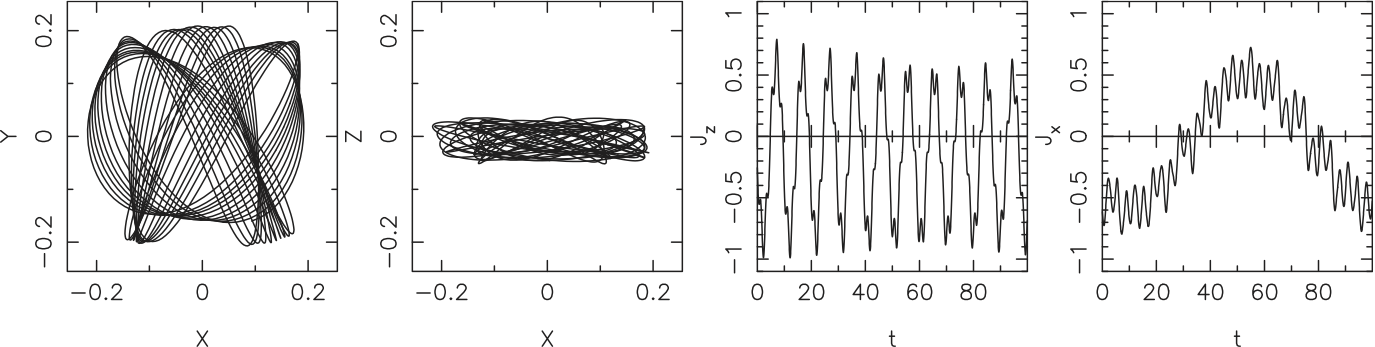}
\caption{Left to right: $x-y$ and $x-z$ projections and normalized angular momentum  $J_z(t)$ for six orbits from Model A that satisfy the condition  $|\Omega_\phi/\Omega_R - 1/2|< 10^{-3}$ (standard definition for x1 orbits). The orbit in the top most row is closest to the closed periodic parent x1 orbit while orbits in successive rows get further and further from the parent orbit.  Orbits have a narrow range of Jacobi integral $-0.89<E_J <-0.83$.}
\label{fig:x1-Box-bar}
\end{figure*}

Orbits from the $N$-body bars were classified both visually and using our new automated orbit classification algorithm.  All 10,000 orbits in Model A and B were visually classified by CA. In addition the innermost $\sim 4000$ particles of Model A were visually classified by MV. Visual classification was based on  $x-y$, $x-z$, $y-z$ projections and plots of angular momenta as a function of time ($J_x(t), J_z(t)$)  (examples of such plots are given in Figures~\ref{fig:x1-Box-bar}, \ref{fig:boxlets}, \ref{fig:xtubes}, \ref{fig:x4}). 
We adopted classifications based on the full orbit (integration time $t=1000$) which is $\sim 200$ orbital periods for the innermost bar orbits but only $\sim 30$ orbital periods for the outermost disk particles. 
{  Although all orbits in an $N$-body system are mildly chaotic \citep[e.g.][]{miller_64,hemsendorf_merritt_02} for the purpose of testing our automated classification scheme we also attempt to visually classify chaotic orbits, although this is not possible to do in a robust manner.}  An orbit was visually classified as chaotic only if it could not be easily identified with a major orbit family (box, short axis tube, long axis tube), or if it showed signs of changing from one family to another during the integration time (indicating that it is sticky chaotic). For the $\sim 4000$ orbits that were visually classified by two of us (CA \& MV) we found that fewer than 2\% of orbits were classified differently. Where classifications disagreed, the orbits were generally transitional orbits - probably lying close to the separatrix between two families. 

{  Since the visual classification of chaotic orbits was based on conservative criteria it resulted in fewer chaotic orbits than the automatic classification method while relies on orbital frequency drift (see Appendix A). This is expected since sticky chaotic orbits at the edge of a resonant island can appear regular for long times. }

In Section~\ref{sec:phasespace} we show frequency maps  with several  minor resonant families. In frequency maps resonant orbits appear clustered along thin lines that satisfy a resonance condition $l\Omega_x+m\Omega_y+n\Omega_z=0$ (where $l, m, n$ are small integers). The signs of the integers indicate  the phase relationship between the frequency components. The automatic classification method easily identifies  resonant orbits, but these are visual identified only if they are quite close to the resonant parent orbit. Weakly resonant orbits are visually classified as boxes. 

 \citet{voglis_etal_07} used the  orbital fundamental frequencies in cylindrical polar coordinates ($\Omega_R, \Omega_\phi, \Omega_z$) to classify orbits in an $N$-body bar. In Appendix~B we show that 
a much clearer separation of orbit families is obtained with fundamental frequencies computed in {\it Cartesian coordinates}.    Our new automatic bar orbit classification scheme is described in detail in Appendix~\ref{sec:autoclassification} and a detailed comparison between the automatic and visual classification schemes is deferred to Section~\ref{sec:vis_auto}.
 
\section{Orbital building blocks of bars}
\label{sec:bar_orbits}

\subsection{x1 orbits and box orbits}
\label{sec:box}

The family of orbits parented by the periodic x1 orbit is considered to be the main bar-supporting orbit family \citep[c.f.][, hereafter BT08]{BT08}. As mentioned previously, a particle on such an orbit makes two excursions in radius during each complete circuit in the azimuthal angle $\phi$ \citep[e.g.][]{contopoulos_papayannopoulos_80} and therefore the radial oscillation frequency and tangential frequency are resonant $\Omega_R: \Omega_\phi = 2:1$.  Since its apocenter radius increases with increasing Jacobi integral $E_J$,  at large $E_J$,  x1 orbits develop loops at their extremities. A particle on an x1 orbit travels in a prograde sense about the center of the galaxy (i.e. in the same sense that the bar pattern is rotating) except  in the loops where the motion is retrograde relative to the figure.  

    In triaxial ellipsoids the main orbit family responsible for providing the high density along the long axis is the 3 dimensional box orbit family (BT08). In a stationary potential boxes have no net angular momentum about any axis. However in a rotating frame Coriolis forces result in ``envelope doubling'' which imparts a small net angular momentum to the parent  $x$-axis orbit as well as box orbits \citep{schwarzschild_82, dezeeuw_merritt_83}.  {  The ``x1 orbit'' is therefore also the parent of the box orbit family in a rotating triaxial potential \citep{martinet_dezeeuw_88}. In this section we illustrate this with orbits selected from the self-consistent $N$-body bar simulations and show that the vast majority of orbits in these simulations are boxes with little net rotation in the bar frame unlike their prograde parent x1 orbit. }

\begin{figure}
\centering
\includegraphics[trim=20.pt 0.pt 20.pt 0pt,width=0.35\textwidth]{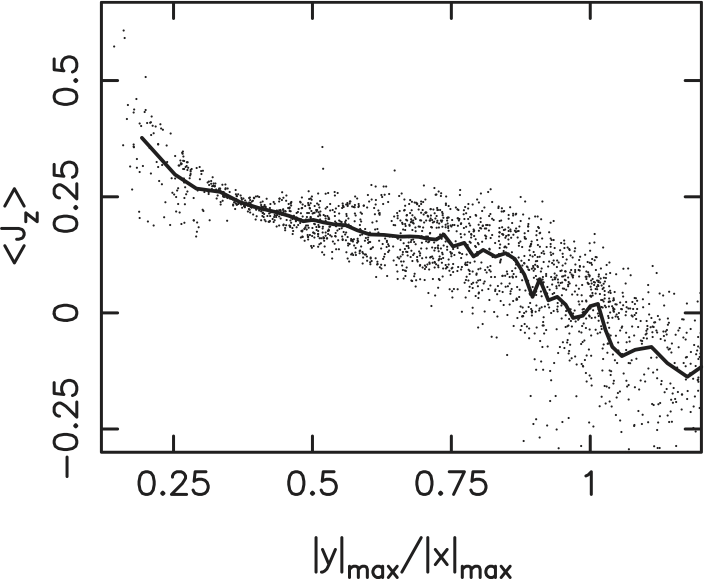}
\caption{Each dot represents the time average of the normalized angular momentum $\langle {J_z}\rangle$ for a single orbit as a function of $|y|_{\rm max}/|x|_{\rm max}$ (see text for details). Dots are plotted for all orbits in Model A that were visually classified as x1 or box (but resonant 3:-2:0 boxlets were excluded). The solid line is the mean value of the distribution represented by dots.}
\label{fig:Jzvsybyx}
\end{figure}

\begin{figure*}
\centering
\includegraphics[trim=0.pt 0.pt 163.pt 0.pt,angle=0, clip,width=0.45\textwidth]{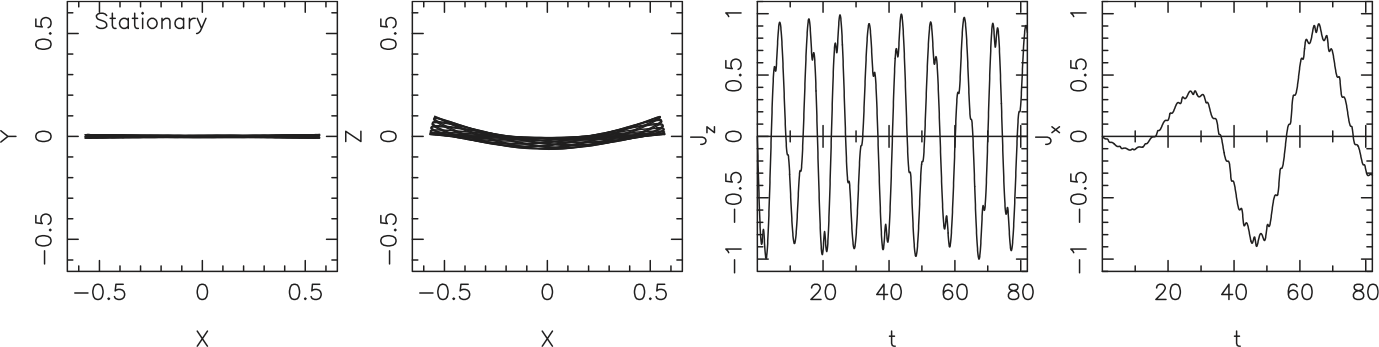}\hspace{12pt}
\includegraphics[trim=0.pt 0.pt 163.pt 0.pt,angle=0, clip,width=0.45\textwidth]{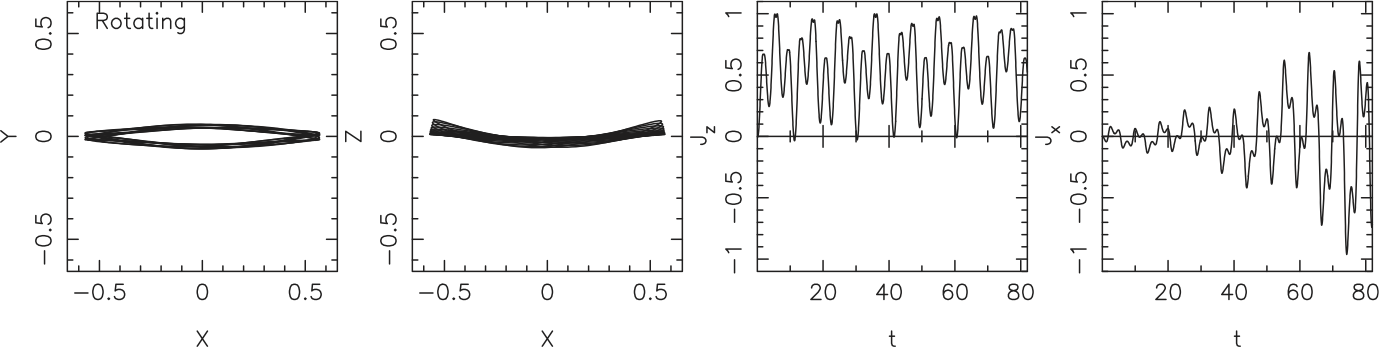} 
\includegraphics[trim=0.pt 0.pt 163.pt 0.pt,angle=0, clip,width=0.45\textwidth]{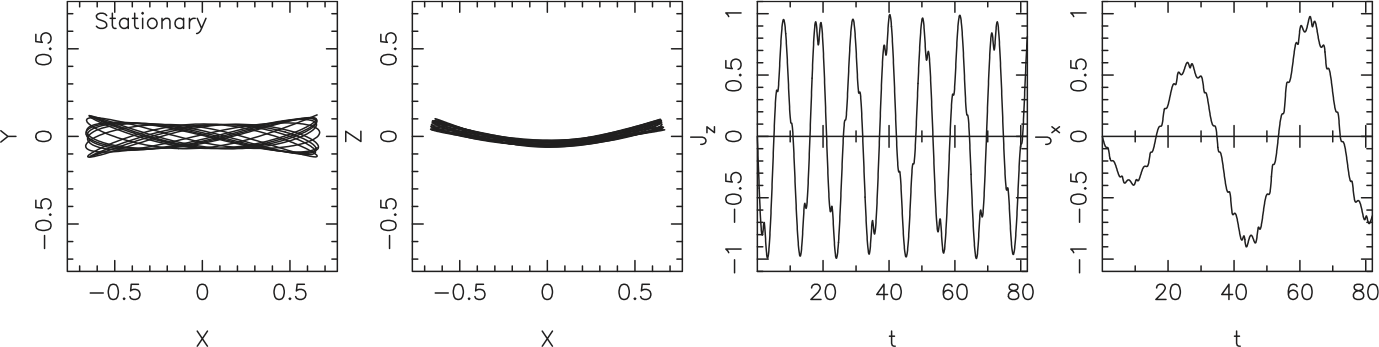}\hspace{12pt}
\includegraphics[trim=0.pt 0.pt 163.pt 0.pt,angle=0, clip,width=0.45\textwidth]{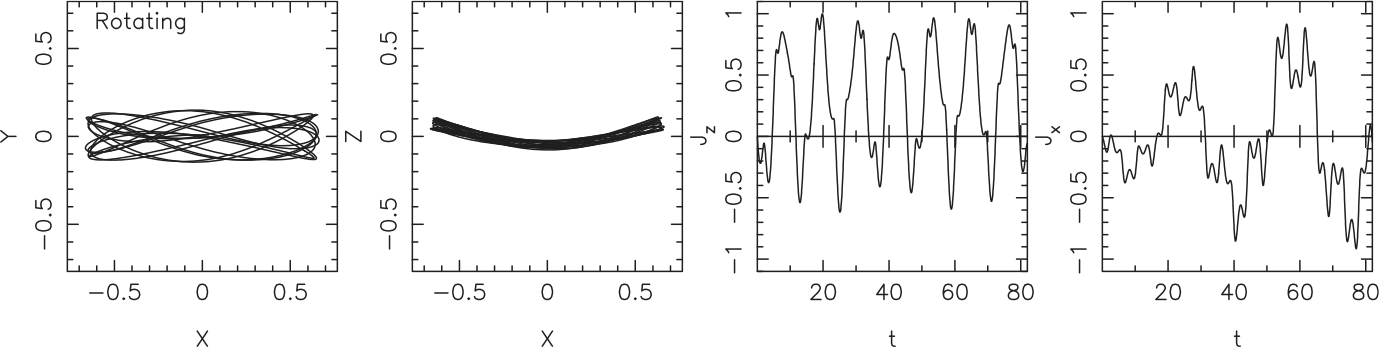} 
\includegraphics[trim=0.pt 0.pt 163.pt 0.pt,angle=0, clip,width=0.45\textwidth]{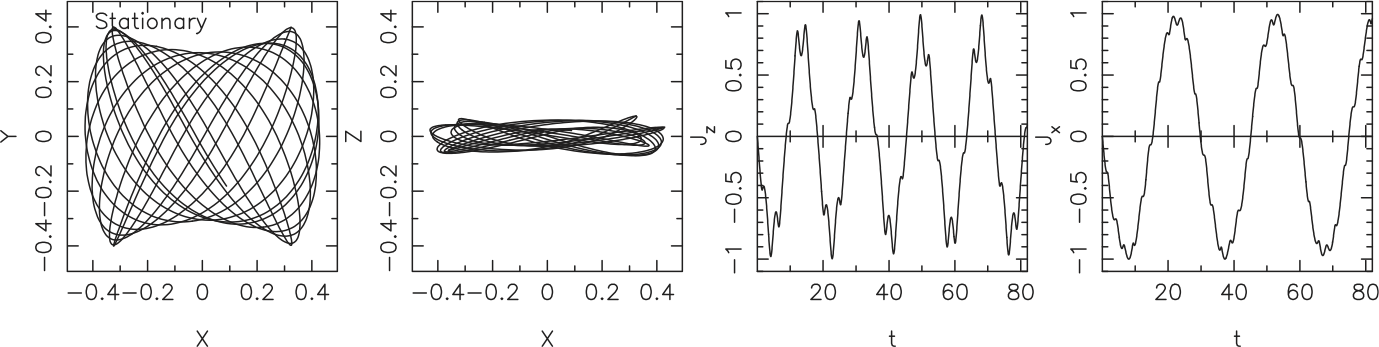}\hspace{12pt}
\includegraphics[trim=0.pt 0.pt 163.pt 0.pt,angle=0, clip,width=0.45\textwidth]{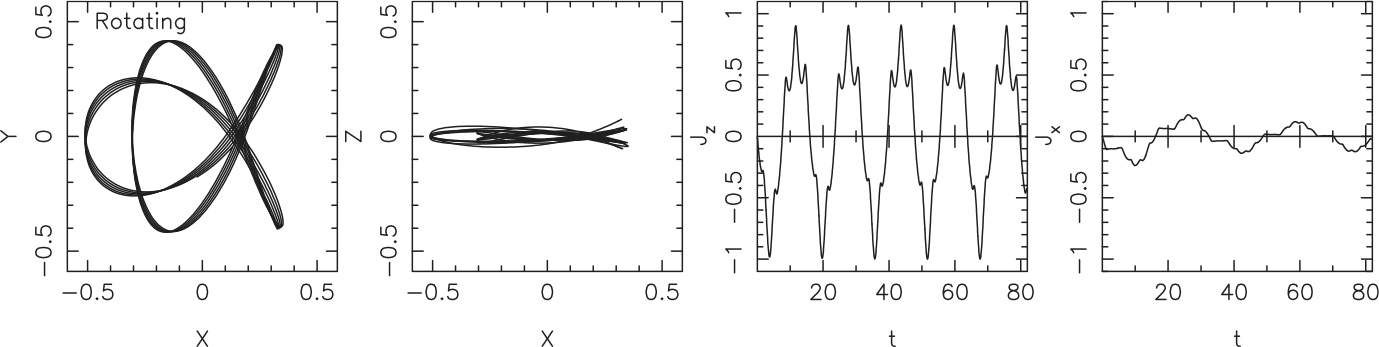}
\includegraphics[trim=0.pt 0.pt 163.pt 0.pt,angle=0, clip,width=0.45\textwidth]{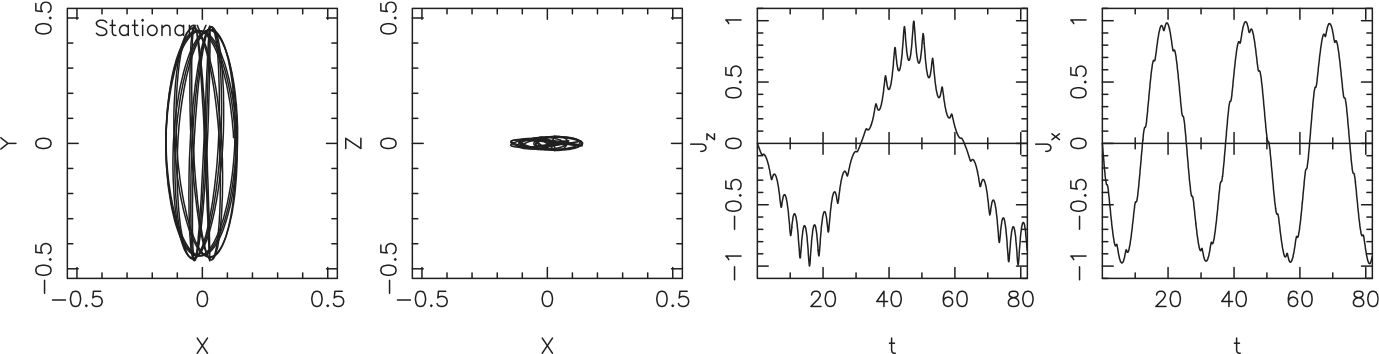}\hspace{12pt}
\includegraphics[trim=0.pt 0.pt 163.pt 0.pt,angle=0, clip,width=0.45\textwidth]{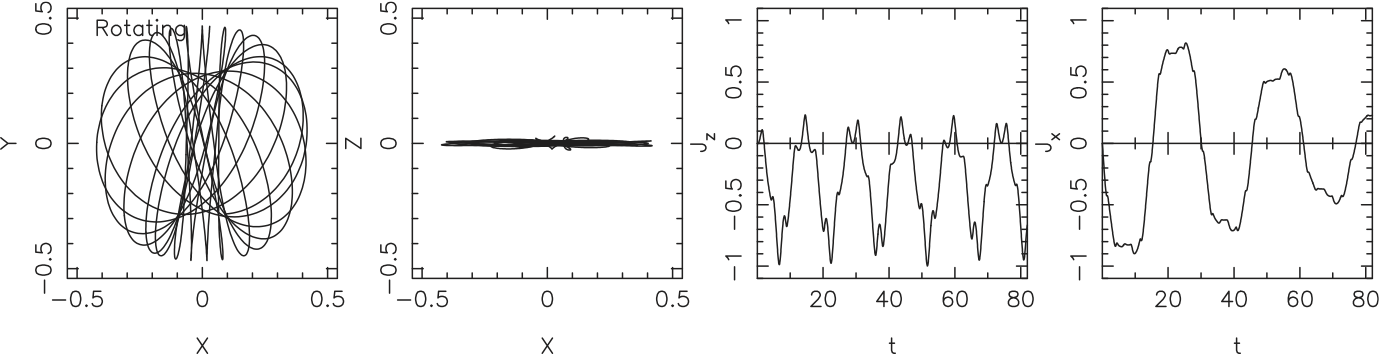}
\caption{Each row shows two orbits launched from identical initial conditions in a prolate triaxial Dehnen model with $c/a=0.4, T=0.916,\gamma=0.1$. The three left columns show $x-y$ and $x-z$ projections and normalized angular momenta $J_z(t)$  for orbits in the stationary model, while the three right columns show these quantities in a rotating Dehnen model with the same co-rotation radius as the bar. The top row shows orbits launched along the $x$ axis. The 2nd row shows standard box orbits launched above the $x-y$ plane. The third row shows a box orbit in the stationary model (left 3 columns) that is transformed to a resonant 3:-2:0 orbit in the rotating frame. The fourth row shows an orbit launched as close to the $y$-axis as the orbit in the first row is from the $x$-axis (the $y$-axis orbit itself is unstable).}
\label{fig:Dehnenboxes}
\end{figure*}

Figure~\ref{fig:x1-Box-bar} shows five different orbits from the pure $N$-body bar (Model A) computed in the frame of reference co-rotating with the bar. On each row we show different projections/ quantities for a single orbit. From left to right: projection on the $x-y$ plane, projection on the $x-z$ plane, the angular momentum as a function of time $J_z(t)$ and $J_x(t)$  normalized relative to their maximum absolute values. The top two rows show orbits near the closed periodic x1 parent {   (it is difficult to find strictly periodic x1 orbits in an $N$-body simulation)}. For these two orbits we see that the angular momentum $J_z$ oscillates between two positive values throughout its orbit. The next three rows show orbits that travel increasingly further from the parent x1 orbit.  For these orbits  $J_z(t)$  does not remain positive  but becomes negative for those portions of the orbit when the motion is retrograde.  The examples in Figure~\ref{fig:x1-Box-bar} demonstrate that as orbits deviate further from the periodic x1 parent (i.e. as their extent in the $y$ direction increases) they spend more and more time moving retrograde.  


One way to quantify the deviation from the parent x1 orbit is to measure the extent of an orbit in the $y$ direction relative to its extent in the $x$ direction. We do this by computing the ratio of the  maximum absolute $y$ value attained over the orbit, $|y|_{\rm max}$, relative to the maximum absolute $x$ value attained over the orbit, $|x|_{\rm max}$. Based on the visual classification of orbits we find that  classical x1 orbits have $|y|_{\rm max}/|x|_{\rm max} < 0.35$ (they are significantly more elongated along the $x$-axis than along the $y$-axis).  

Figure~\ref{fig:Jzvsybyx} shows {   $\langle {J_z}\rangle$ (the time-average of the angular momentum $J_z$ normalized to its maximum value)} as a function of $|y|_{\rm max}/|x|_{\rm max}$. Each dot represents an orbit in Model A that was classified as x1 or box (we exclude the resonant boxlet orbits that are discussed in the next section). The solid line is the mean value of the distribution represented by dots.  Orbits  with the smallest values of $|y|_{\rm max}/|x|_{\rm max}$ have the highest net angular momentum and are classical x1 orbits.   As orbits get thicker  in $y$ this average angular momentum decreases. For reference an orbit with constant angular momentum would have $\langle{J_z}\rangle =\pm 1$;  the orbits in the top two rows of Figure~\ref{fig:x1-Box-bar} have $\langle{J_z}\rangle = 0.6$, while the orbit in the bottom row of Figure~\ref{fig:x1-Box-bar} has $\langle{J_z}\rangle =-0.09$. Only a small fraction of orbits have both the high $\langle{J_z}\rangle$ and the small $|y|_{\rm max}/|x|_{\rm max}$ that is characteristic of classical x1 orbits. In fact from Figure~\ref{fig:Jzvsybyx} we see that most orbits have $\langle{J_z}\rangle < 0.25$ and a significant fraction have negative $\langle{J_z}\rangle$ implying they spend more time traveling retrograde than prograde (for example the orbit in the last row of Fig.~\ref{fig:x1-Box-bar}). 

\begin{figure*}
\centering
\includegraphics[trim=0.pt 0.pt 163.pt 0.pt, angle=0, clip, width=0.6\textwidth]{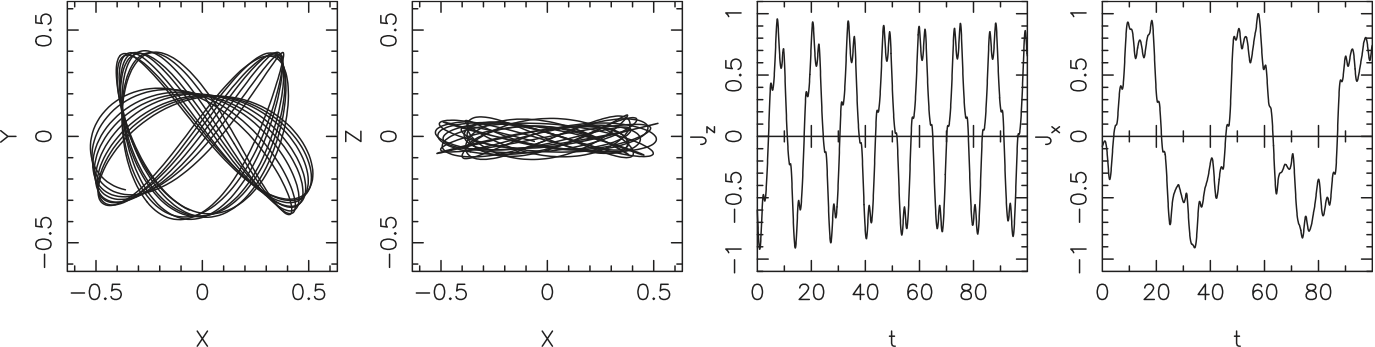}
\includegraphics[trim=0.pt 0.pt 163.pt 0.pt , angle=0, clip, width=0.6\textwidth]{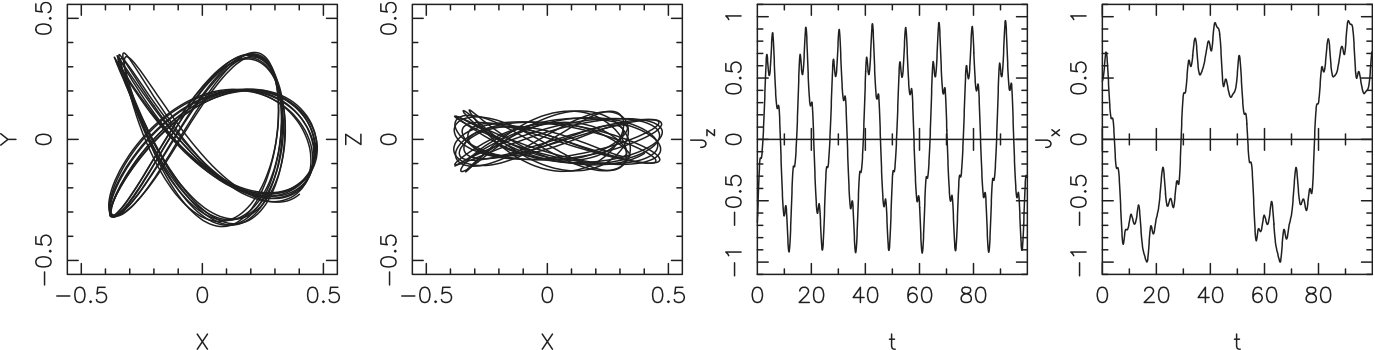}
\includegraphics[trim=0.pt 0.pt 163.pt 0.pt , angle=0, clip, width=0.6\textwidth]{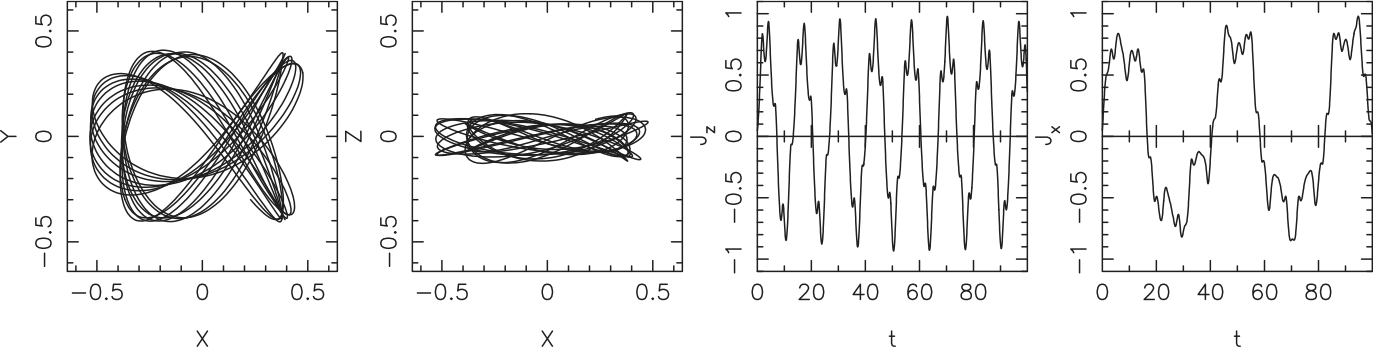} 
\includegraphics[trim=0.pt 0.pt 163.pt 0.pt , angle=0, clip, width=0.6\textwidth]{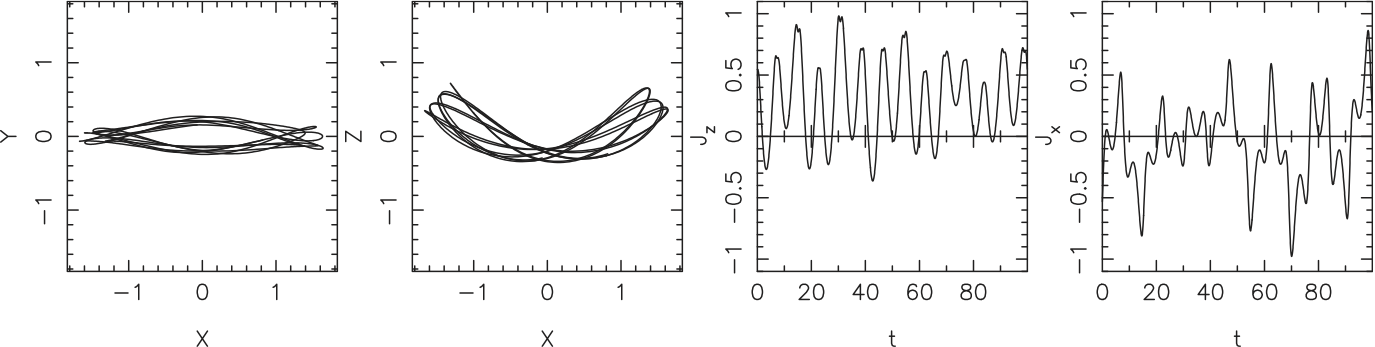} 
\includegraphics[trim=0.pt 0.pt 163.pt 0.pt , angle=0, clip, width=0.6\textwidth]{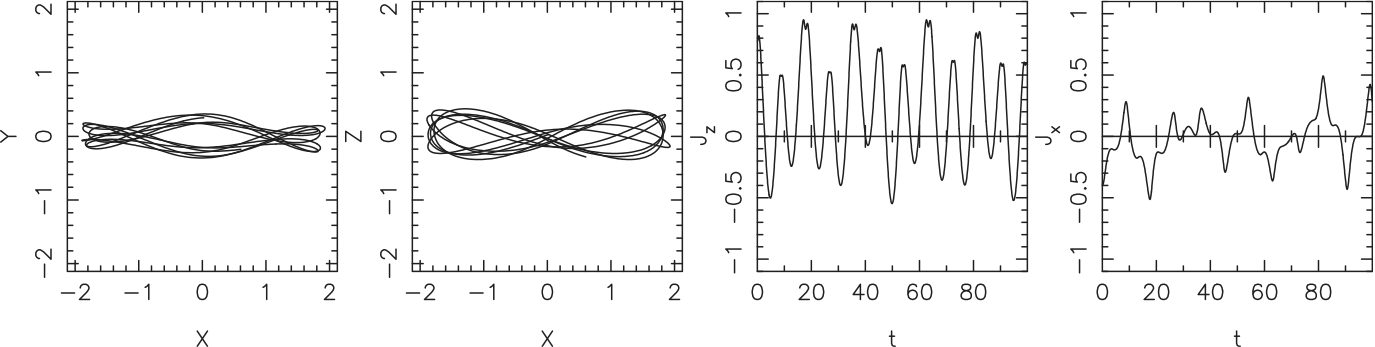}
\includegraphics[trim=0.pt 0.pt 163.pt 0.pt , angle=0, clip, width=0.6\textwidth]{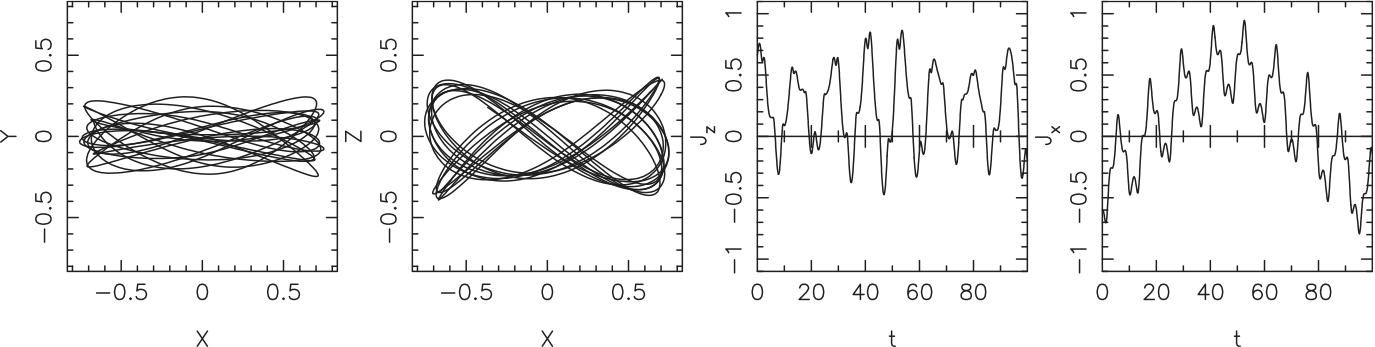}
\caption{Same as in Fig.~\ref{fig:x1-Box-bar}  for near-resonant boxlet orbits selected from $N$-body bar.  The orbits in the first three rows are  all associated with a 3:-2:0 resonance (although they bear superficial resemblance to the  4:-3:0 ``pretzel''  and 3:0:-2 ``fish'' orbits found in triaxial potentials they belong to a different family). The orbits in the 4th and 5th rows are  2:0:-1  ``x1-banana'' and ``x1- anti-banana'' resonant orbits which are also found in triaxial ellipsoids \citep{merritt_valluri_99}. The orbit in the 6th row is classified by our automatic classifier as a resonant boxlet with $\Omega_x:\Omega_y:\Omega_z = 3:0:-5$. }
\label{fig:boxlets}
\end{figure*}

We illustrate this point further by launching orbits from box orbit initial conditions in a prolate-triaxial ellipsoid with the Dehnen density profile (Section~\ref{sec:triax}), with shape parameters close to that of the $N$-body bar in Model A ($c/a=0.4, b/a=0.48, T=0.916, \gamma=0.1$).  In Figure~\ref{fig:Dehnenboxes} each row shows two orbits launched from the same initial conditions in a stationary Dehnen model (three left columns) and in the Dehnen model with a fast pattern speed  ($R_{CR}/r_{1/2} = 1.1$, where $r_ {1/2}$ is the half-mass radius of the model) (three right columns). For each orbit we show projections of the orbit ($x-y$, $x-z$) and its angular momentum with time ($J_z(t)$). The orbits in the top row were launched very close to the long ($x$) axis of the model. On the left is a long axis orbit  (parent of the box family) with no net angular momentum  $J_z$ or $J_x$. The three right-hand columns show what happens to this orbit in a rotating frame: the Coriolis force now causes the orbit to loop around the center in an anticlockwise sense, and hence $J_z(t)$ becomes  strictly positive. 
  In the 2nd row the three left columns show a standard box orbit in a stationary potential while the three right columns show the same orbit in the rotating frame. This orbit behaves in a manner identical to the ``thick x1'' orbits from Model A (e.g. in rows 4 \& 5 of Fig.~\ref{fig:x1-Box-bar}). (The  box orbits in the 3rd \& 4th rows of Figure~\ref{fig:Dehnenboxes} will be discussed later.)
 
The differences seen in the shapes and angular momentum distributions of the orbits in the stationary potential (3 left-hand columns) and in the rapidly rotating potential (3 right hand columns) are {\em entirely} a consequence of the pseudo forces in the rotating frame.  A comparison of orbits in the first two rows of this figure with the x1 family orbits drawn from the $N$-body bar in Figure~\ref{fig:x1-Box-bar} shows that both box orbits and x1 orbits are parented by the linear long axis orbit. Figure~10 of SS04 gives very clear examples of quasi-periodic orbits parented by an  x1 orbits as well as orbits that they refer to as ``fat x1'' (but that we call boxes).  Despite their visual difference (classical x1  versus boxlike orbits), SS04 show that these orbits form a continuous sequence  in a surface-of-section (SoS) (e.g. upper panel of Fig 9 in SS04). We discuss this further in Section~\ref{sec:sos}.  

{  To summarize: it is well known that the x1 orbit in bars is the same as parent of the box family in rotating triaxial potentials \citep{schwarzschild_82, martinet_dezeeuw_88}. What is less well appreciated is that in bars, as in triaxial ellipsoids, the dominant bar supporting family are boxlike with little or no  angular momentum unlike their x1 parent which is prograde}.

\subsubsection{Boxlets: fish, pretzels and banana orbits}
\label{sec:boxlets}

\citet{miralda-escude_schwarzschild_89} \& \citet{merritt_valluri_99} showed that when an integrable triaxial potential is perturbed, e.g. by the introduction of a cusp or a central point mass, almost all orbits that originate from ``boxlike'' initial conditions (i.e. launched with zero initial velocity from an equipotential surface) become either resonant or chaotic. Resonant orbits are easily identified by frequency mapping. In stationary triaxial potentials they include the well known  3:0:-2  ``fish'' resonance and the 4: -3: 0 ``pretzel'' resonance  - both of which are absent in our rapidly rotating Dehnen model and $N$-body bars. Commonly found resonant orbits in the rotating Dehnen model are shown in the 3rd row, three right columns of  Fig.~\ref{fig:Dehnenboxes} ($\Omega_x:\Omega_y: \Omega_z = 3:-2:0$, fish/pretzel) and in the 1st and 2nd rows of three right columns of Fig.~\ref{fig:Dehnenboxes} (the ``banana'' resonance). {  A few  other resonances are seen in the frequency maps in Figure~\ref{fig:freqmap_dehnen}}.

\begin{figure*}
\centering
\includegraphics[trim=0.pt 0.pt 0.pt 0.pt, angle=0, clip, width=0.8\textwidth]{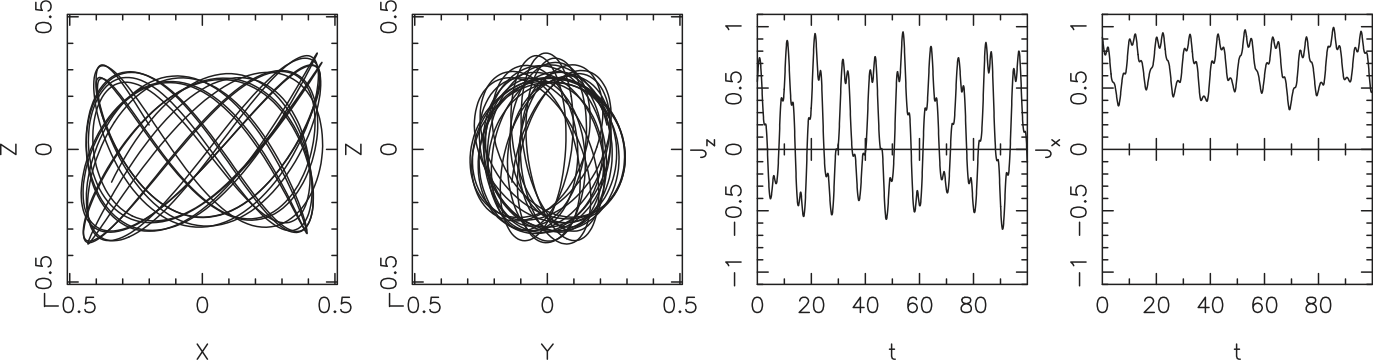}
\includegraphics[trim=0.pt 0.pt 0.pt 0.pt, angle=0, clip, width=0.8\textwidth]{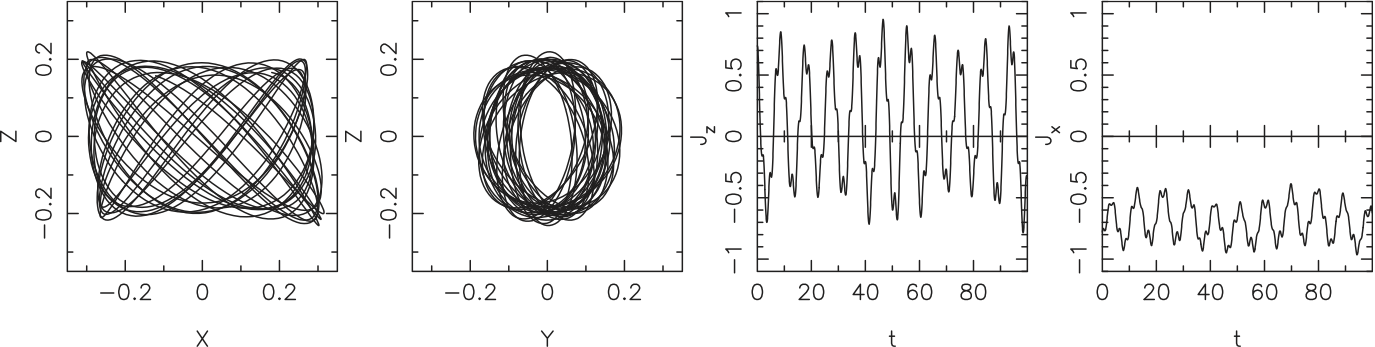}
\includegraphics[trim=0.pt 0.pt 0.pt 0.pt, angle=0, clip, width=0.8\textwidth]{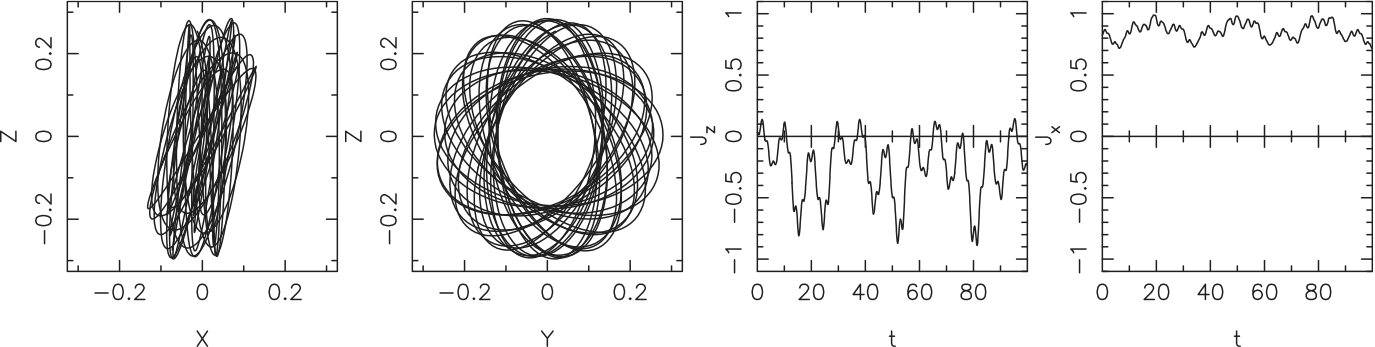}
\includegraphics[trim=0.pt 0.pt 0.pt 0.pt, angle=0, clip, width=0.8\textwidth]{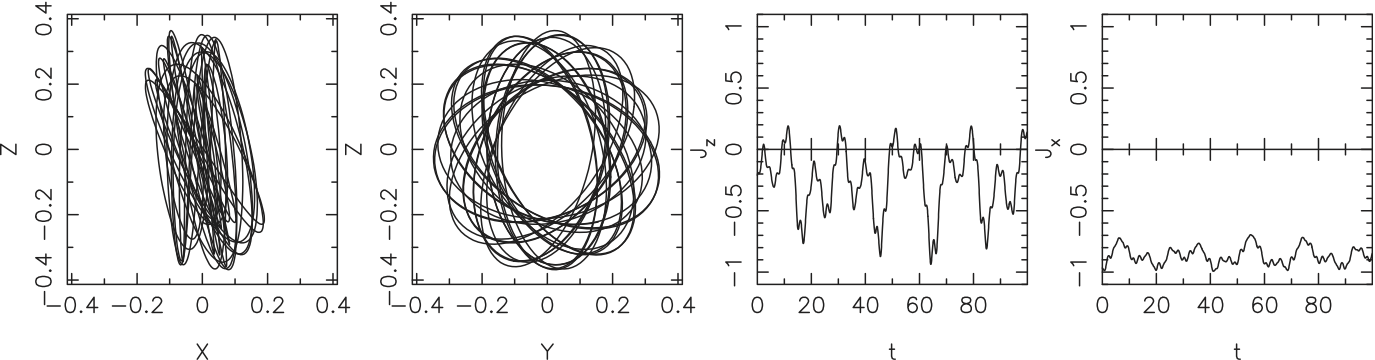} 
\caption{Left to right: $x-z$, $y-z$ projections and angular momenta as a function of time ($J_z(t), J_x(t)$) for long axis tube orbits selected from Model A. Orbits in the top two rows are inner long axis tubes, while the orbits in the lower two rows are outer-long axis tubes. Due to the Coriolis forces, the orbit is tipped clockwise (anticlockwise) about the $y$-axis when $J_x(t)$ is positive (negative). Notice that the two outer-long axis tubes are tipped so far about the $y$-axis that they acquire significant net negative $J_z$, as predicted by  \citet{heisler_etal_82}}.
\label{fig:xtubes}
\end{figure*}

\begin{figure*}
\centering
\includegraphics[trim=0.pt 0.pt 0.pt 0.pt , angle=0, clip, width=0.6\textwidth]{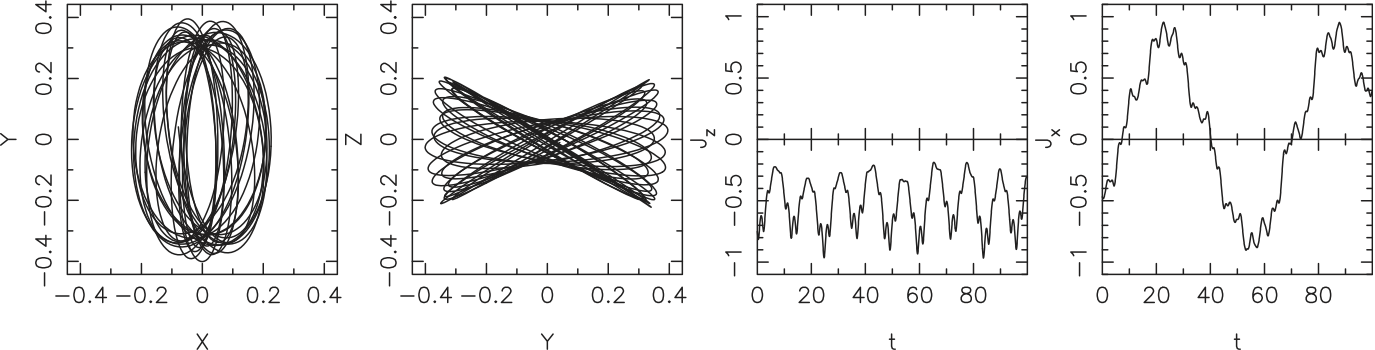}
\includegraphics[trim=0.pt 0.pt 0.pt 0.pt , angle=0, clip, width=0.6\textwidth]{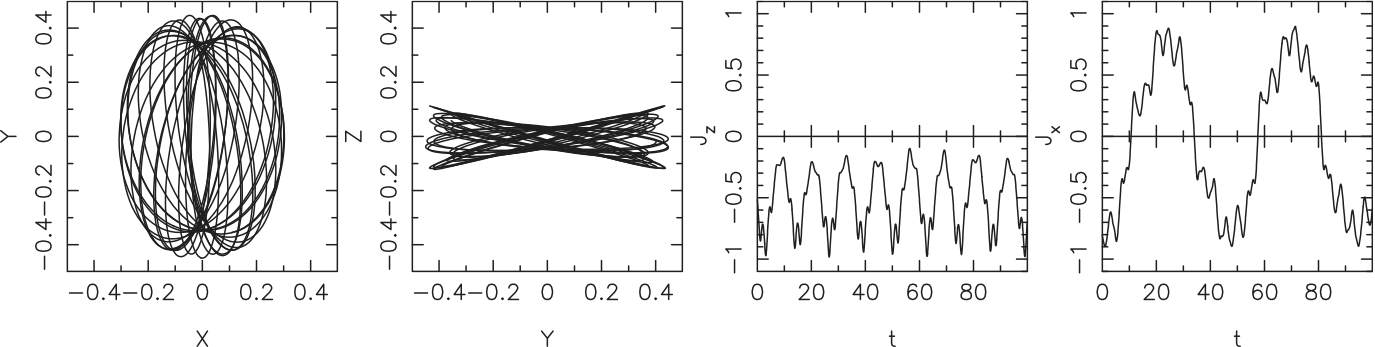}
\includegraphics[trim=0.pt 0.pt 0.pt 0.pt , angle=0, clip, width=0.6\textwidth]{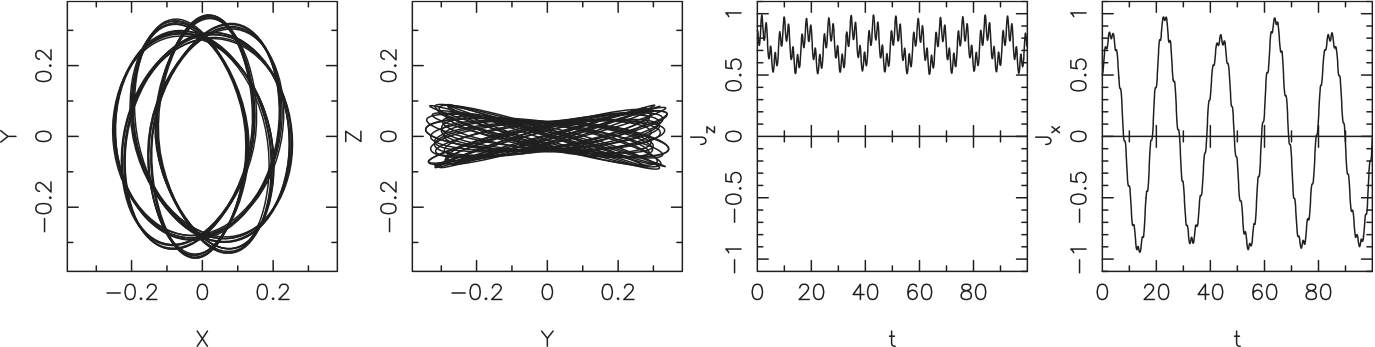}
\includegraphics[trim=0.pt 0.pt 0.pt 0.pt , angle=0, clip, width=0.6\textwidth]{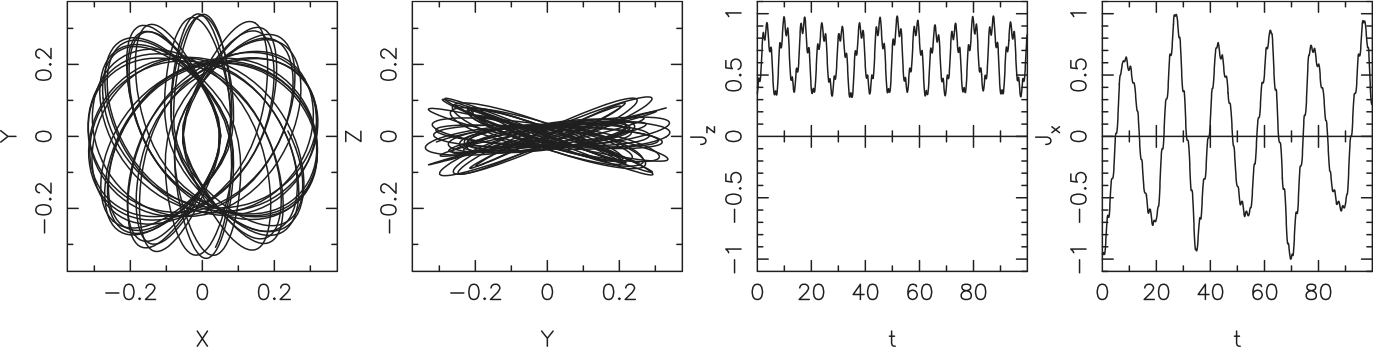}
\includegraphics[trim=0.pt 0.pt 0.pt 0.pt , angle=0, clip, width=0.6\textwidth]{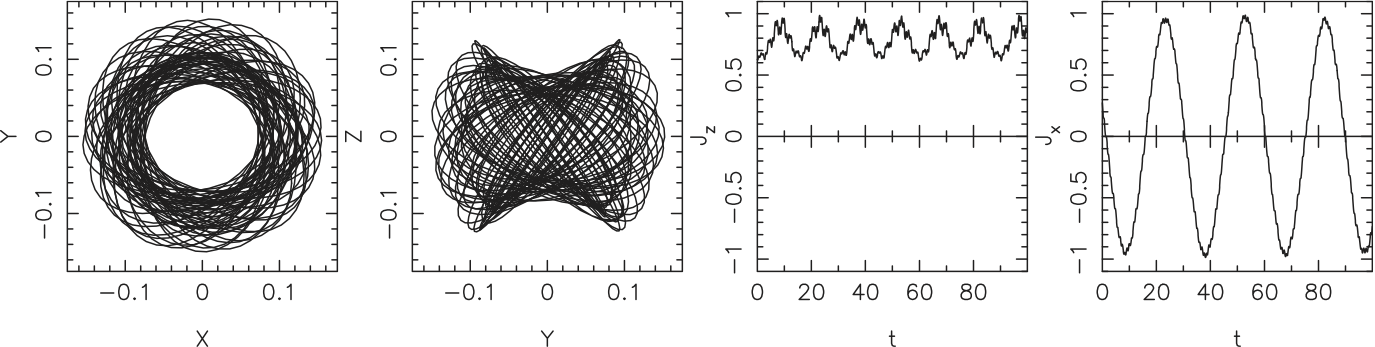}
\caption{$x-y$ and $y-z$ projections and normalized angular momentum $J_z(t)$  for orbits with short axis tube like characteristics. Top two rows show retrograde $z$-tube orbits selected from Model A  (this model has no prograde $z$-tubes).  The  last three rows show prograde $z$-tube orbits all from Model B.}
\label{fig:x4}
\end{figure*}

Our examination of orbits from the $N$-body bar showed that they can also be associated with resonances.  Figure~\ref{fig:boxlets} shows examples of resonant orbits found in the $N$-body bar. Orbits in the first three rows are all characterized by
$\Omega_x:\Omega_y: \Omega_z = 3:-2:0$ and look like fish/pretzels (and correspond to the same resonance as in the 3rd row of Fig.~\ref{fig:Dehnenboxes}). The 4th row shows an orbits associated with ``x1-banana'' resonance ($\Omega_x:\Omega_y:\Omega_z = 2:-2:-1$), \footnote{Note that in some papers on bars, a family of orbits  that lie close to the 4th and 5th Lagrange points of the bar are also referred to as ``banana'' \citep{athanassoula_92} but is not a member of this resonant family.} and the 5th row is an ``x1-anti-banana'' \citep{miralda-escude_schwarzschild_89,pfenniger_friedli_91} which satisfies the same  resonance conditions a banana orbit, but passes through $x=0, z=0$ ({  the automatic classification code is unable to distinguish between banana and anti-banana which are treated as one group}). As pointed out previously by \citet{pfenniger_friedli_91} the banana and anti-banana orbits are vertical bifurcations of the x1-family (as can be seen from their $x-y$ projections in the left hand column). These orbits are referred to as the x1v1 family by \citet{patsis1}. Orbits like those in the 2nd row of Figure~\ref{fig:Dehnenboxes} are boxes with a banana shape in $x-z$ are associated with 2:0:-1 resonance). The 3rd and 4th columns of Figure~\ref{fig:boxlets} show that like box orbits in triaxial potentials, the sign of  $J_z$ changes at each extremum implying that the orbits  have little or no net angular momentum about the $z$ axis, despite being parented by x1 orbits. The frequency maps in Figure~\ref{fig:freqmap} show several other resonances. The presence of resonant boxlet orbits was noted by SS04 (see their  Fig 13), especially at high $E_J$ values.

{  Vertical bifurcations of the x1 orbit associated with the 2:-2:-1 resonance may be responsible for the ``X-shaped'' structures seen in buckled edge-on bars \citep{patsis1,Athanassoula05}. \citet{Patsis_Katsanikas_14a, Patsis_Katsanikas_14b} also present a dynamical mechanism for building X-shaped peanuts with families of periodic orbits that are not  bifurcations of x1 orbits. It was suggested by \citet{portail_etal_15} that a resonant family of $\Omega_x:\Omega_y:\Omega_z = 3:0:-5$  ``brezel'' orbits (bottom row of Figure~\ref{fig:boxlets}) are the backbone of the ``X-shaped'' structures in their made-to-measure $N$-body bar models, but we found only 88 ``brezels''   (1.5\% of the bar orbits) in both Model A and B. A more detailed analysis of the orbits contributing to the X-shape will be discussed in a future work (Abbott et al. in preparation).} 

\subsection{Long axis tube orbits}
\label{sec:LATs}

  \citet{dezeeuw_85b} showed that for integrable triaxial potentials two types of long axis tubes exist, the inner long axis tubes and the outer long axis tubes. This family  is ``parented'' by closed 1:1 periodic orbits that lie in the $y-z$ plane with angular momentum along the $x$-axis. \citet{heisler_etal_82} studied the stability of these periodic orbits and found that they are stable to figure rotation but the Coriolis force tips them about the $y$-axis in a  direction that depends on the sign of $J_x$. Two such stable periodic orbits exist rotating  clockwise  and anti-clockwise about the $x$-axis, the orbits with positive  $J_x$  are tipped clockwise about the $y$-axis while orbits with negative $J_x$ are tipped anti-clockwise about the $y$-axis. These orbits were termed ``anomalous'' by \citet{vanalbada_etal_82} since they both acquire retrograde motion the $z$ axis in the rotating frame as a result of being tipped. 
Since the long axis tube family  is ``parented'' by  anomalous orbits they too are  expected to  be stable and ``tipped'' about the $y$-axis in a rotating frame, and may acquire some retrograde angular momentum about the $z$ axis.  

Figure~\ref{fig:xtubes} shows examples of inner long axis tubes (top two rows) and outer long axis tubes (3rd and 4th rows) selected from bar Model A. The second column  ($y-z$ projection) shows that these orbits circulate about the long ($x$) axis. The right most column shows the angular momentum  $J_x$ about the $x$ axis which, as expected, is strictly positive or strictly negative. An inspection of the first column shows  the orbits are tipped about the $y$ axis exactly as predicted by \citet{heisler_etal_82}, further supporting our claim that bars and rotating triaxial ellipsoids have  fundamentally similar orbital building blocks.

\subsection{Short Axis Tubes, x2 and x4 orbits}
\label{sec:SATs}
  
Short axis ($z$) tubes constitute an important family in oblate-triaxial ellipsoids but are somewhat less important in more prolate systems. Short axis tubes are parented by closed periodic 1:1 orbits that lie in the $x-y$ plane and circulate about the $z$-axis. These orbits are known to be stable to figure rotation \citep{dezeeuw_merritt_83}.
However, \citet{deibel_etal_11} found that as the pattern speed of a triaxial figure was increased, prograde short axis tubes were replaced by retrograde ones. {  This was also found by \citet{martinet_dezeeuw_88} for loop orbits in the $x-y$ plane of a rotating triaxial ellipsoid and occurs because  tube/loop orbits launched prograde ``fall behind'' the figure, becoming retrograde as the pattern speed increases. }

Periodic x2, and x4 orbits (like the classical x1 orbit) satisfy the condition $\Omega_R: \Omega_\phi= 2:1$. In the characteristic diagrams \citep[e.g.][]{sellwood_wilkinson_93} the prograde x2 family is found at low $E_J$ and is elongated along the $y$ axis of the bar while the x4 family is retrograde and is found over a wide entire range of $E_J$ values.  Both the x2 and x4 orbits are elongated along the $y$-axis at small $E_J$. The x4 family is elongated along the $y$-axis at small $E_J$ but becomes rounder at large $E_J$. {  Since x2 and x4 orbits are planar periodic orbits they are not expected to be found in significant numbers in an $N$-body model but they do parent families of prograde and retrograde short axis tubes which can be easily identified if they exist. Henceforth we refer to all non-planar orbits with a fixed sign of angular momentum about the short axis as $z$-tubes.}

\begin{figure*}
\centering
\includegraphics[trim=0.pt 0.pt 0.pt 0.pt,angle=0, clip,width=0.49\textwidth]{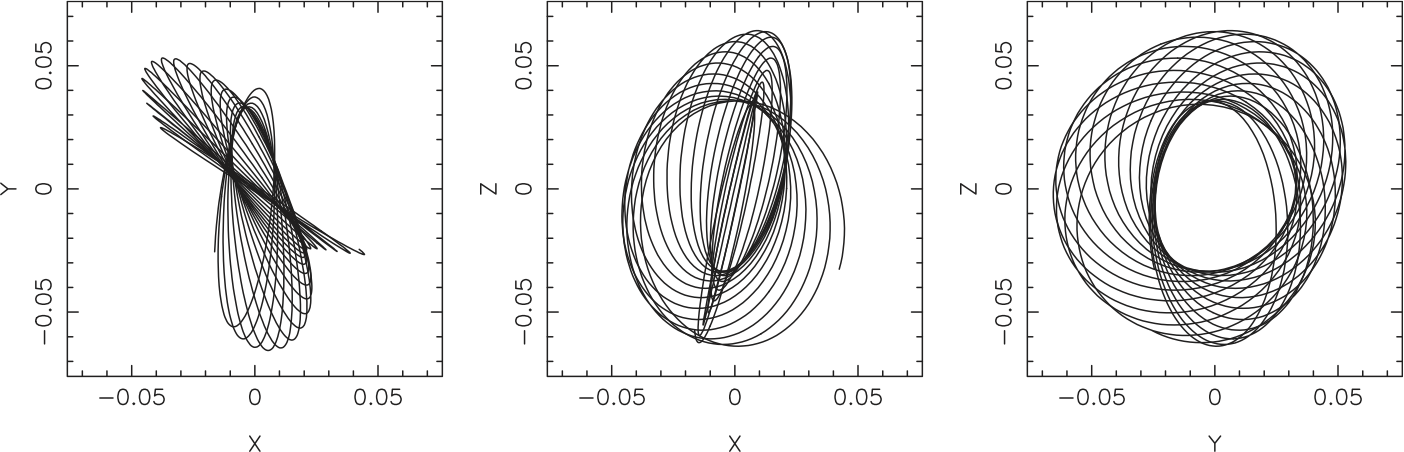}
\includegraphics[trim=0.pt 0.pt 0.pt 0.pt,angle=0, clip,width=0.49\textwidth]{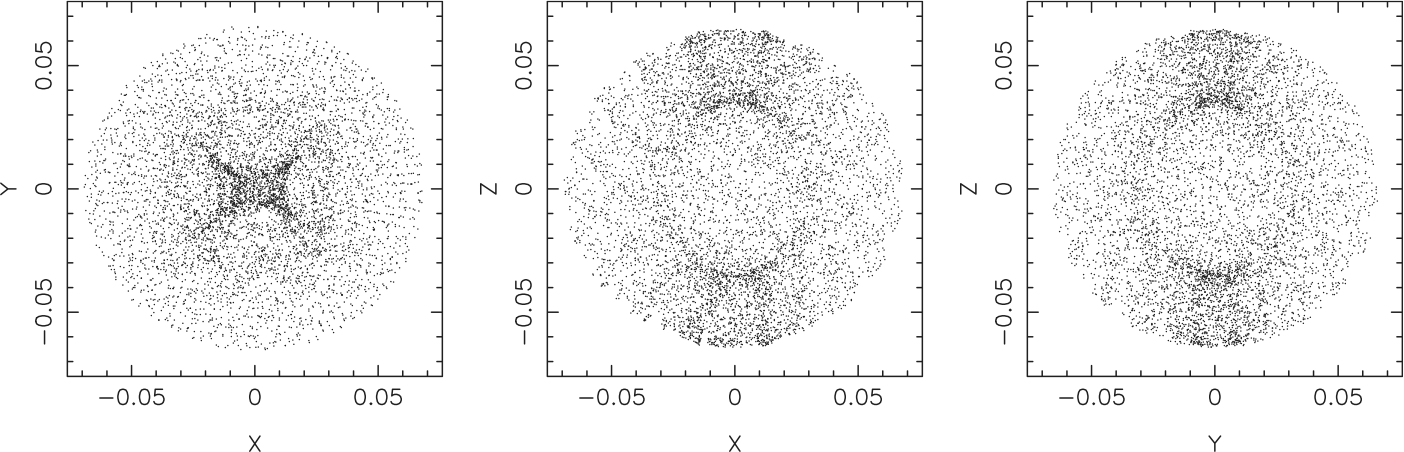}
\includegraphics[trim=0.pt 0.pt 0.pt 0.pt,angle=0, clip,width=0.49\textwidth]{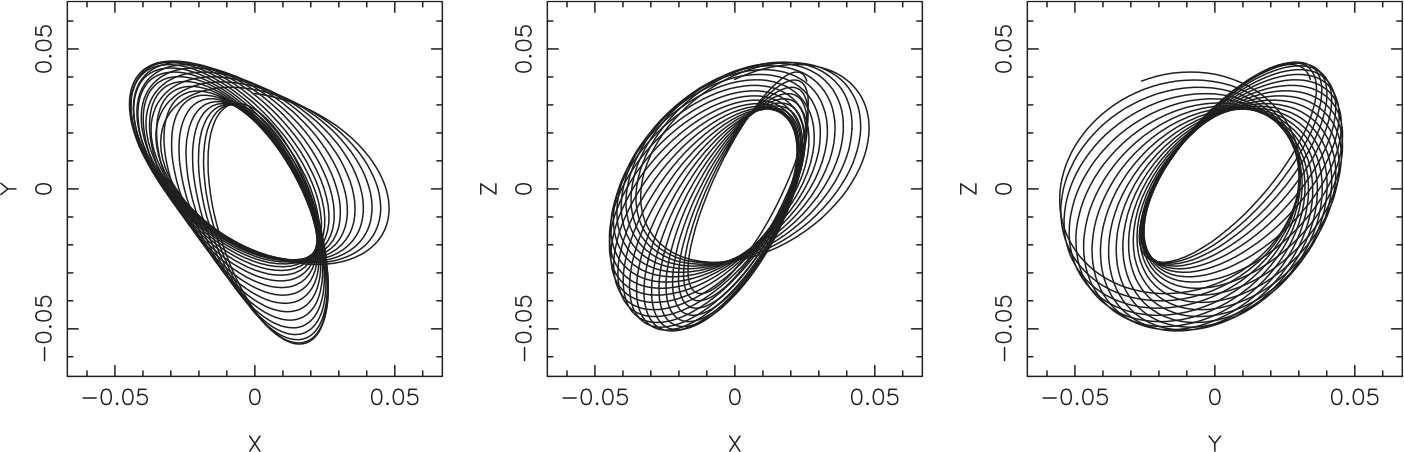}
\includegraphics[trim=0.pt 0.pt 0.pt 0.pt,angle=0, clip,width=0.49\textwidth]{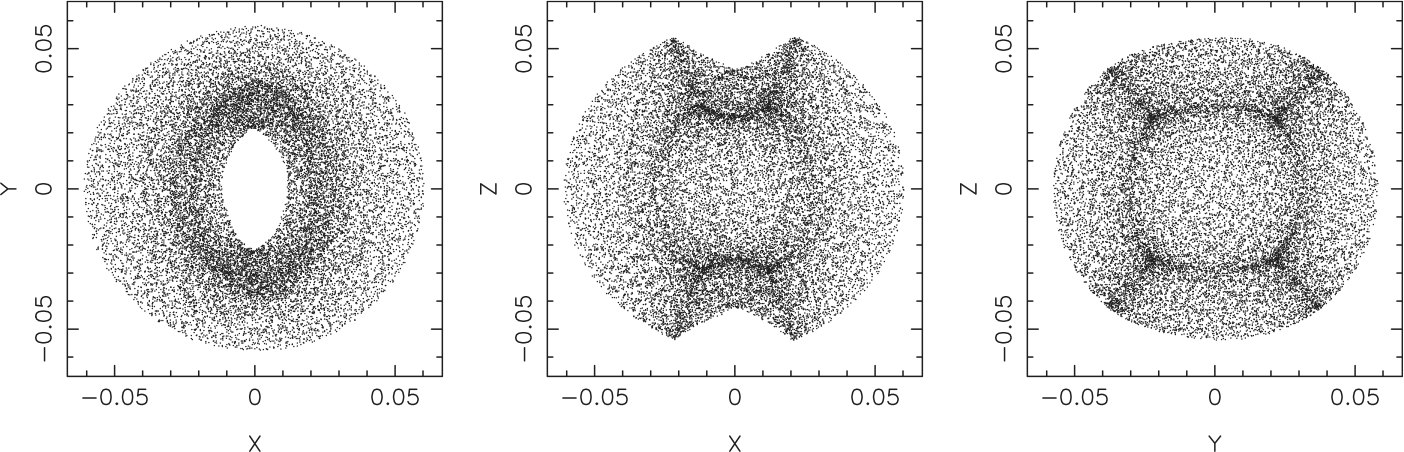}
\includegraphics[trim=0.pt 0.pt 0.pt 0.pt,angle=0, clip,width=0.49\textwidth]{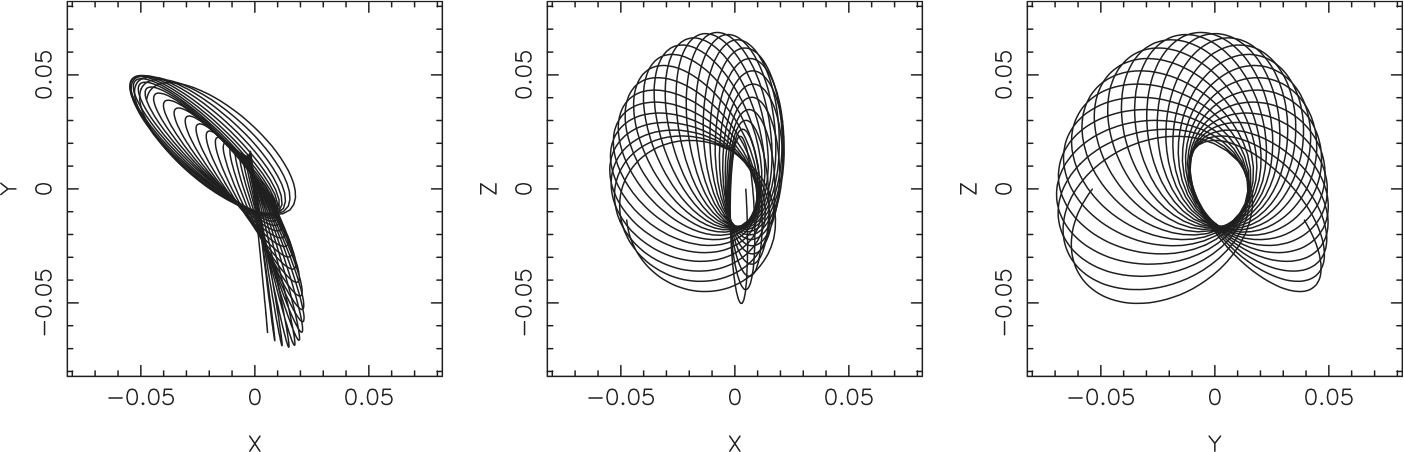}
\includegraphics[trim=0.pt 0.pt 0.pt 0.pt,angle=0, clip,width=0.49\textwidth]{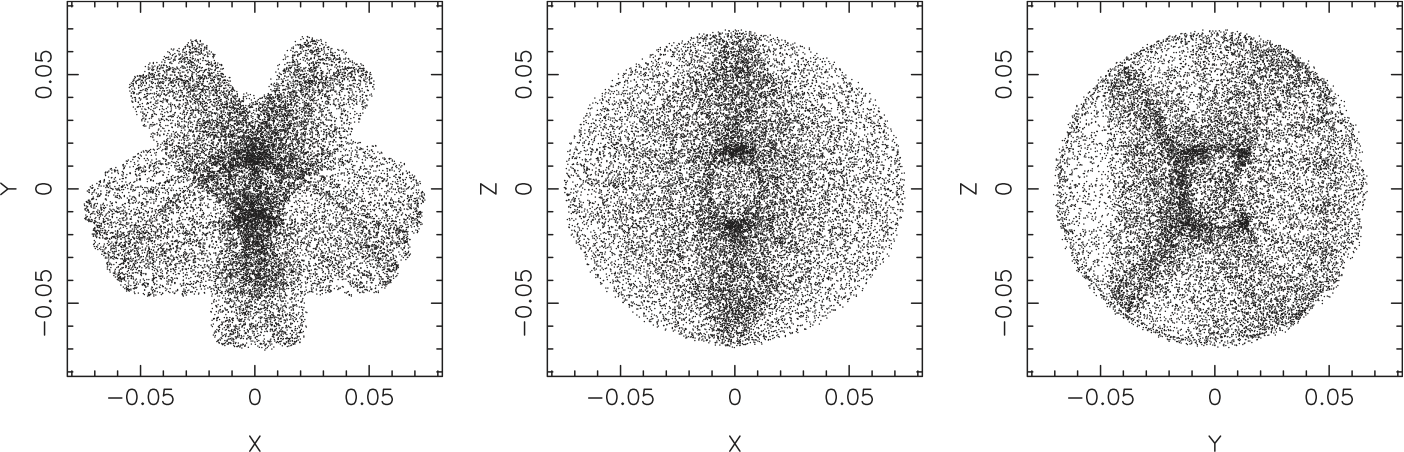}
\includegraphics[trim=0.pt 0.pt 0.pt 0.pt,angle=0, clip,width=0.49\textwidth]{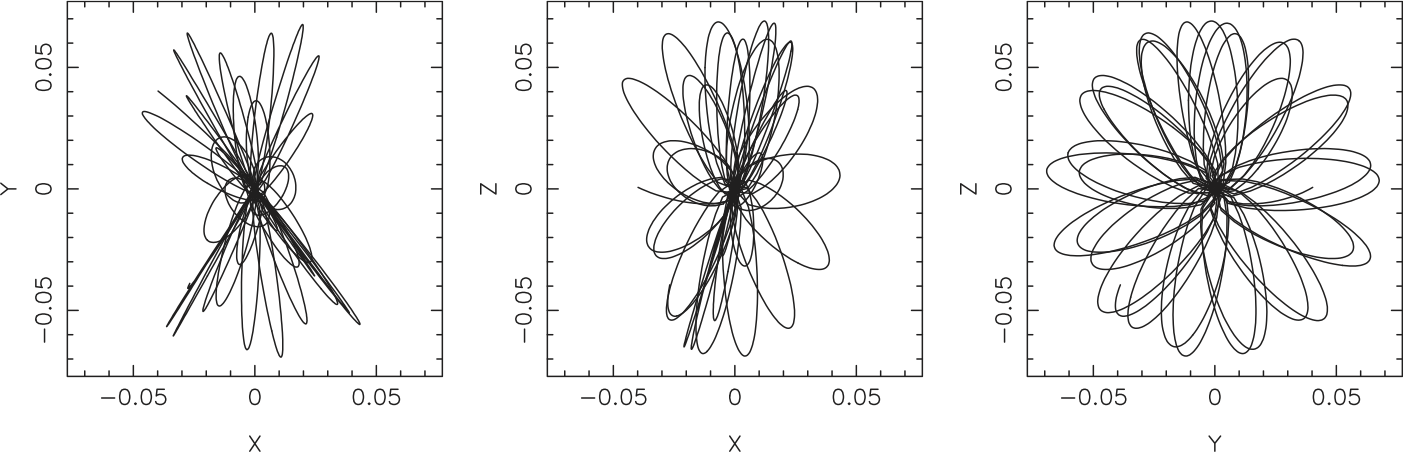}
\includegraphics[trim=0.pt 0.pt 0.pt 0.pt,angle=0, clip,width=0.49\textwidth]{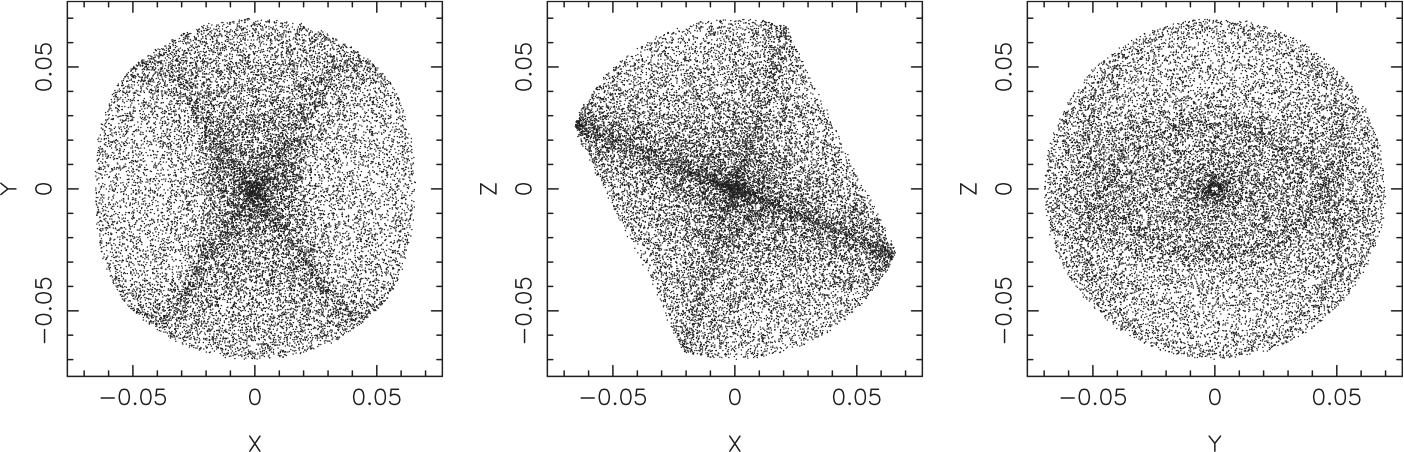}
\caption{Each row shows one orbit close to the SMBH in Model B. The three panels on the left show projections of the orbit plotted for $\sim 20$ orbital periods. On these short integration times the orbits appear like slowly precessing Keplerian orbits. The right three panels show the projected surface density of the orbit integrated over a long time $t=1000$ which in all cases corresponds to over 1000 orbital periods. These right hand panels show that on long integration times the orbit are a box (top row); short axis ($z$) tube (2nd row); resonant $z$-tube (3rd row); $x$-tube (4th row).}
\label{fig:BHorbits}
\end{figure*}

{  The visual classification of 20,000 orbits in Model A and Model B identified $\sim 1.5$\% of bar orbits in Model A  as retrograde $z$-tubes but {\it none} of the orbits in this model were found to be prograde $z$-tubes. The top two rows of Figure~\ref{fig:x4} shows examples of retrograde $z$-tubes from Model A.} The fact that we were unable to find a single example of a prograde $z$-tube orbit is consistent with the findings of \citet{sparke_sellwood_87} and  \citet{voglis_etal_07}, who only found retrograde x4 orbits in their $N$-body bar models.  {  x2 orbits (and orbits parented by them) are expected from the locations of the inner Lindblad resonance and the co-rotation resonance in Model A, however this family appears to be severely underpopulated.} A significant decline in prograde $z$-tubes at high pattern speeds is however predicted by \citet{martinet_dezeeuw_88} and \citet{deibel_etal_11}. 

We examined whether it was possible to generate x2 and x4 orbits from perturbations to a linear $y$-axis orbit in the same way as it was possible to generate the x1 orbit from the linear $x$-axis orbit.  The $y$-axis linear orbit is known to be unstable \citep{adams_etal_07}, hence launching orbits from the $y$-axis of the Dehnen model yielded unstable chaotic orbits.  An example of an orbit launched from the ``stationary start space'' (i.e with no initial velocity) near (but not on) the $y$-axis  is shown in the fourth row of Figure~\ref{fig:Dehnenboxes}. In the stationary frame (3 left columns) the orbit is a stable box elongated along the $y$-axis. In the rotating frame (3 right columns) this orbit becomes an even thicker box but now with $\langle{J_z}\rangle <0$ similar to the orbit in the last row of Figure~\ref{fig:x1-Box-bar}.  If $\left|J_z\right|$ was smaller than a minimum value, boxlike orbits elongated along the $y$-axis always resulted. It was only possible to generate retrograde x4 orbits by launching an orbit from near the $y$-axis with a substantial $\left|J_z\right|$.  All the retrograde $z$-tubes in Model A are found at small values of $E_J$ and are more elongated along the $y$-axis than along the $x$-axis.


In Model B, however we {\it do} find prograde and retrograde $z$-tube orbits. Some are more elongated along the $y$-axis while others are more elongated along the $x$-axis than along the $y$-axis. We defer a discussion of this to Section~\ref{sec:stats} where we argue, based on the work of \citet{brown_etal_13}, that this new population of prograde $z$-tubes is a result of the growth of the SMBH in the bar which induces angular momentum redistribution. 

Figure~\ref{fig:x4} shows three prograde $z$-tubes from Model B. In the 3rd row the orbit is elongated along the $y$ axis (i.e. it is parented by the periodic x2 orbit); in the 5th row it is elongated along the $x$-axis, and the 4th row shows a transitional orbit. We assert that all these orbits belong to the same family since they must have significant initial net angular momentum to avoid becoming boxes, in contrast with x1 orbits which are related to boxes and derive their prograde motion from the Coriolis forces.

\subsection{Precessing Keplerian Orbits around the SMBH}
\label{sec:BHorbits}

Previous studies of orbits in triaxial and axisymmetric potentials have shown that as one approaches the massive central point mass in the region of the potential where the mass of stars is comparable to the mass of the SMBH, orbits begin to resemble  Keplerian ellipses which are perturbed by the large scale triaxial or axisymmetric stellar potential and therefore precess slowly \citep{sridhar_touma_99, poon_merritt_01,merritt_vasiliev_10,Li_Bockelmann_Khan_15}.  Orbits associated with black holes include ``saucers'',  ``pyramids'' and a variety of resonant families  \citep{merritt_valluri_99,merritt_vasiliev_10}. In this work we will refer to these collectively as ``precessing Keplerian orbits'' (PKOs). 

The spatial resolution of our simulations was high enough\footnote{black hole softening length $\epsilon=0.001$ units (1pc in physical units); see  \citet{brown_etal_13} \& SS04 for more details.} that random sampling of the distribution function of Model B yielded a small number of PKOs in the vicinity of the black hole.  Figure~\ref{fig:BHorbits} shows examples of   PKOs that were found in Model B.  Each row shows a single orbit: the 3 panels on the left show the orbit integrated for about 20 orbital periods\footnote{Orbits were integrated with NAG Mathematical Libraries subroutine D02CJF, a variable-order, variable-step implementation of Adams method, which is very accurate.} while the three panels on the right show the projected density distribution of the orbit integrated over hundreds of orbital periods (for $t=1000$ units). Notice that in the short integrations the orbits in the  top three rows look very similar to each other:  they are clearly slowly precessing Keplerian ellipses. However the  projected density distributions of the same orbits  show that they are morphologically quite different from each other. The orbit in the top row is a box (although $J_z(t)$ and $J_x(t)$ are not shown it also has no net angular momentum about either axis), the 2nd row shows a short axis ($z$) tube for which $J_z(t) >0$,  and the 3rd row shows a resonant short axis tube. The bottom row shows a long axis tube ($J_x(t)<0$) which is tipped anticlockwise about the $y$-axis. (The orbits in the last two rows appear to pass through the origin but zoomed in plots show that they do not.) In total 113 orbits (1.8\% of bar orbits) in Model B were found to be PKOs.  At the very small radii at which these orbits are found, the effects of the centrifugal forces from the rotating pattern should be quite small, but the Coriolis forces are large enough (because the velocities are high) to produce effects similar to those seen on orbits at larger radii. A detailed study of the effects of figure rotation on these nearly Keplerian orbits is beyond the scope of this paper and will be investigated in a future paper.

\section{Orbit Population Statistics and Phase Space Distribution}
\label{sec:population_DF}

\subsection{Radial distribution of orbit families}
\label{sec:stats}
 
In this section we discuss how the orbit populations in  Model A and  B change with $R_{apo}$, the apocenter radius of orbits in the $x-y$ plane and with $E_J$.  We use the automated classification for this analysis. As we show in Section~\ref{sec:vis_auto} the automated and visual classifications give similar results. 

\begin{figure}
\centering
\includegraphics[trim=0.pt 43.pt 0.pt 0.pt,angle=0., clip,width=0.45\textwidth]{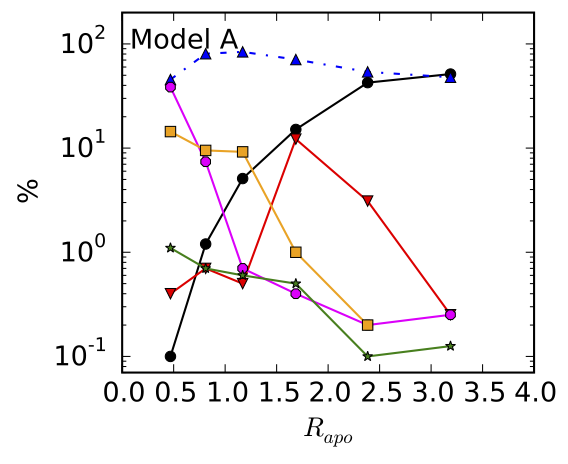} 
\includegraphics[trim=0.pt 0.pt 0.pt 0.pt,angle=0., clip,width=0.45\textwidth]{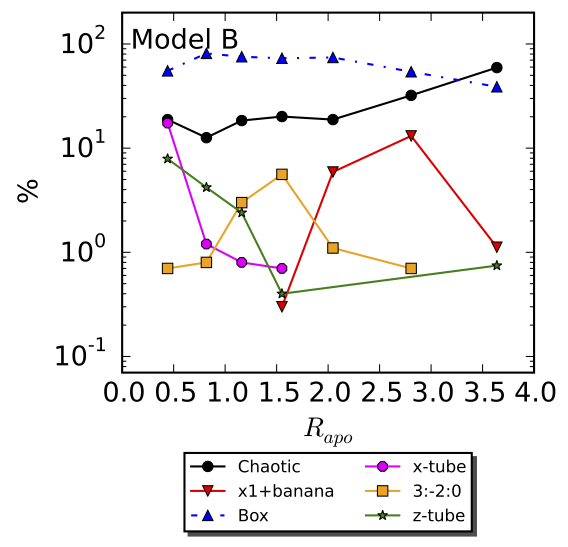}
\caption{Percentage of orbits  in various orbit families as a function of $R_{apo}$ as determined via automated orbit classification. Bins were selected to have equal number of orbits per bin. Only particles with $R_{apo}<4$~kpc are plotted. The percentages within a bin sum to 100\%. If the percentage of orbits in a given family falls to zero no symbol is plotted.}
\label{fig:orbfracs}
\end{figure}

To study change in the membership in the main orbit families within the bar, we bin  orbits with $R_{apo}<4$~kpc  into 6 bins. Figure~\ref{fig:orbfracs} shows the percentage of orbits  in each bin (on a logarithmic scale) as a function of $R_{apo}$ in each of the main families (as indicated by the legends).   

Figure~\ref{fig:orbfracs} shows that the populations in Model A and B  resemble each other, but with important differences. Box orbits (blue triangles, connected by dot-dashed lines) are the most important population in both models.  There are significantly more chaotic orbits (black circles connected by black lines) in the inner regions of Model B than in Model A. The fraction of x1+banana orbits (red triangles connected with solid red lines) is higher in Model A than in Model B especially in the inner most bins. The resonant 3:-2:0 resonant orbits (yellow squares connected  by solid yellow lines) are also  more prominent in Model A than in Model B.  $x$-tube orbits  (pink circles connected by pink lines) decline with radius in both models, and there are slightly more of them in Model A. The short axis tube family shown by green stars connected by solid lines are slightly more important in Model B than in Model A. All of these differences are a consequence of the growth of the central point mass which modifies the mass distribution making it more oblate (SS04 and \citet{merritt_valluri_99}). Overall it is apparent that box-like orbits and chaotic orbits make the most significant contribution to the overall population within the  bar in both models.
 
\subsection{Phase Space Structure of Bars}
\label{sec:phasespace}

\subsubsection{Poinc\'are Surfaces-of-Section}
\label{sec:sos}

\begin{figure*}
\centering
\includegraphics[trim=0.pt 0.pt 0.pt 0.pt,angle=0., clip,width=0.45\textwidth]{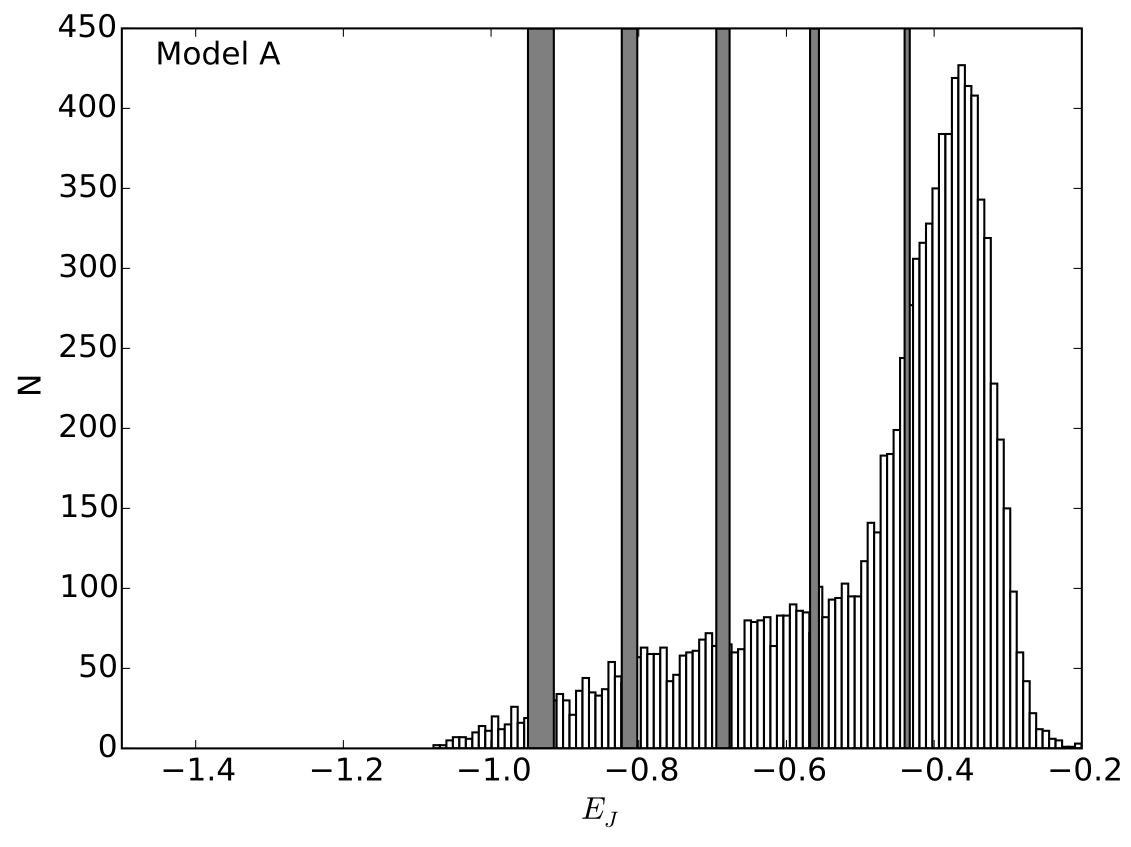} 
\includegraphics[trim=0.pt 0.pt 0.pt 0.pt,angle=0., clip,width=0.45\textwidth]{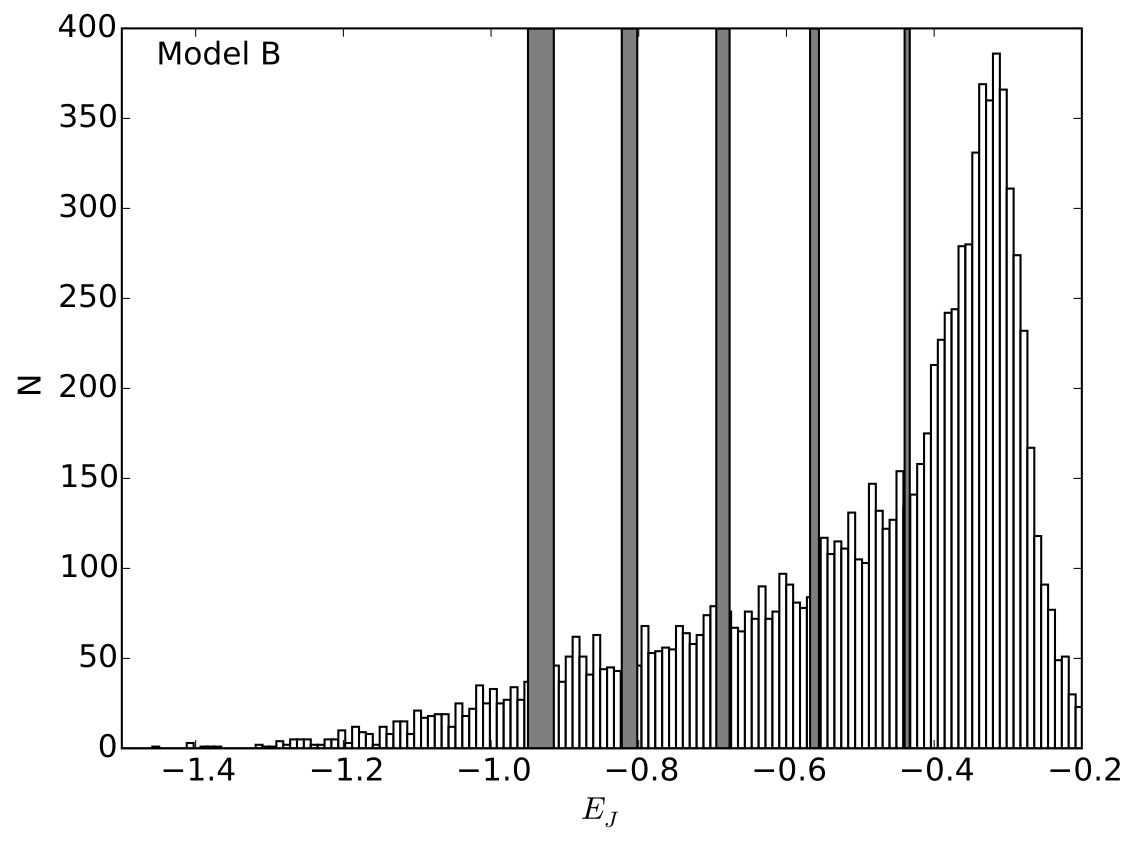}
\caption{Histograms of the distribution of Jacobi Integral ($E_J$) for the 10,000 orbits in Models A and B. The five grey bands mark the range of $E_J$ values that include the 100 orbits plotted in the surfaces-of-section in Fig.~\ref{fig:sosA} and Fig~\ref{fig:sosB}.}
\label{fig:Ejhist}
\end{figure*}

The distribution of $E_J$ values for all the orbits in each of the two models is shown in Figure~\ref{fig:Ejhist}. Since the particles were selected at random from the simulations, each distribution is representative of the distributions of $E_J$ values for the entire model.  Five bins containing 100 orbits each were defined to lie roughly at equal intervals in $E_J$ between the minimum value of $E_J$ and the value at the  co-rotation radius of the bar. The bin limits are shown as grey bands overlying the histograms. These bins were used to select orbits to plot the surfaces-of-section (SoSs) that follow.

\begin{figure*}
\centering
\includegraphics[trim=10.pt 0.pt 0.pt 0.pt,angle=0., clip,width=0.46\textwidth]{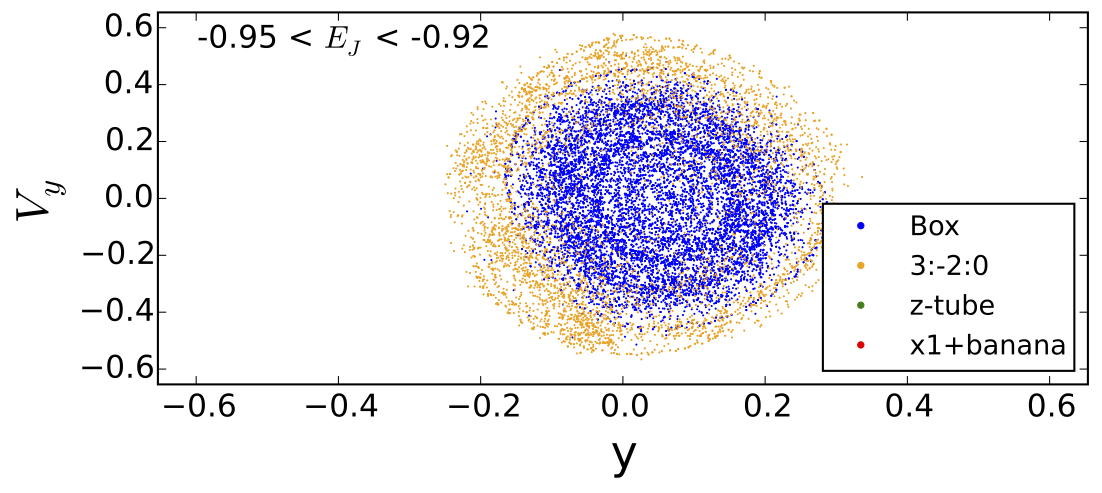} 
\includegraphics[trim=10.pt 0.pt 0.pt 0.pt,angle=0., clip,width=0.46\textwidth]{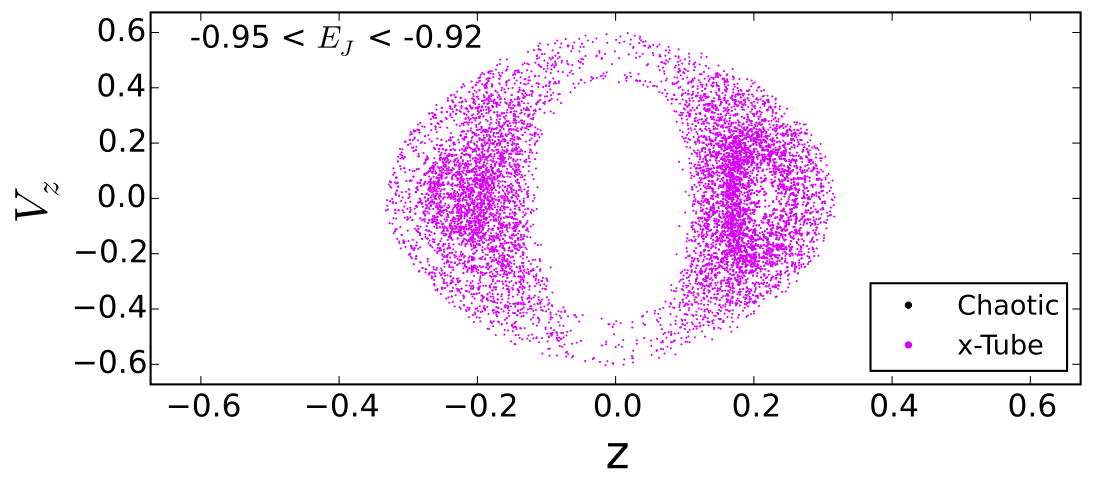} 
\includegraphics[trim=10.pt 0.pt 0.pt 0.pt,angle=0., clip,width=0.46\textwidth]{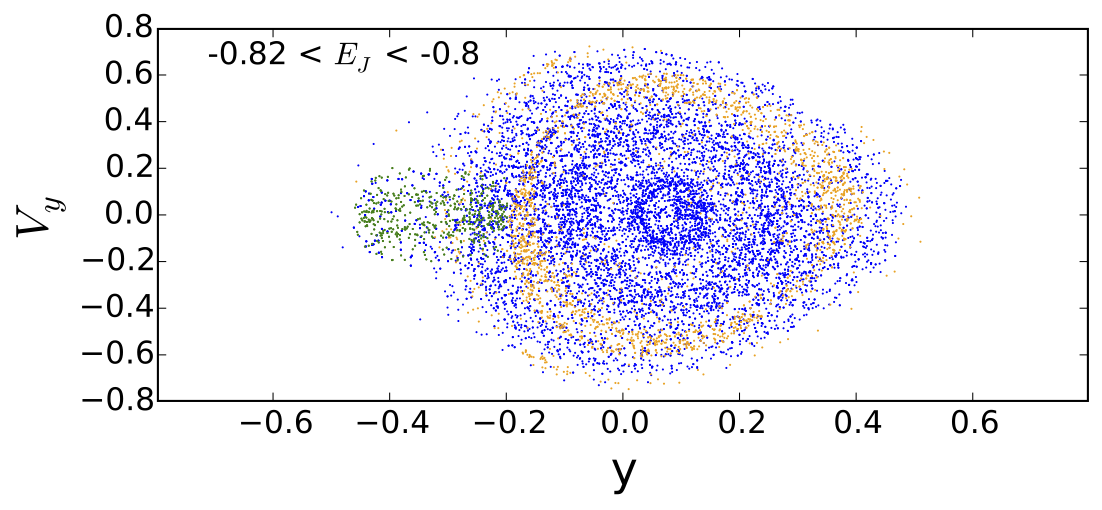} 
\includegraphics[trim=10.pt 0.pt 0.pt 0.pt,angle=0., clip,width=0.46\textwidth]{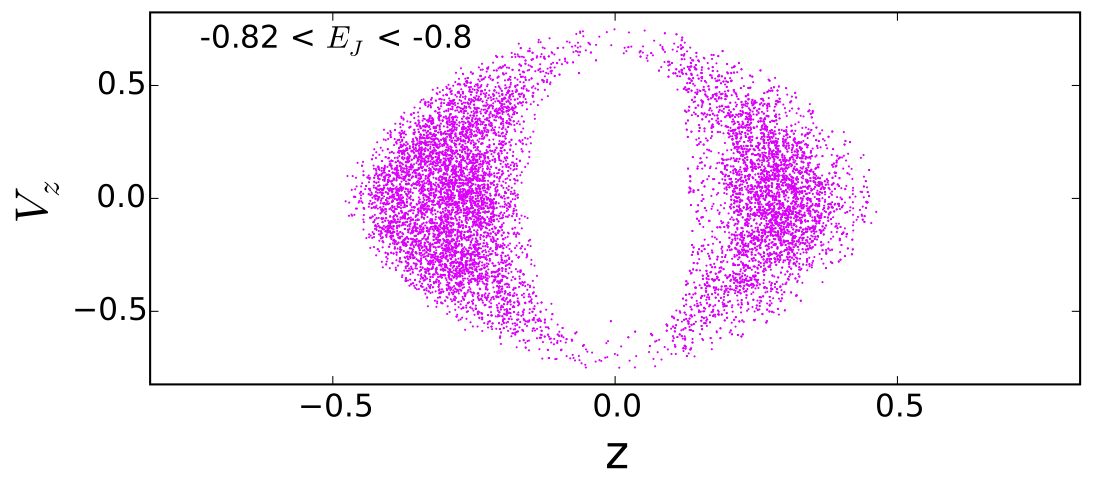} 
\includegraphics[trim=10.pt 0.pt 0.pt 0.pt,angle=0., clip,width=0.46\textwidth]{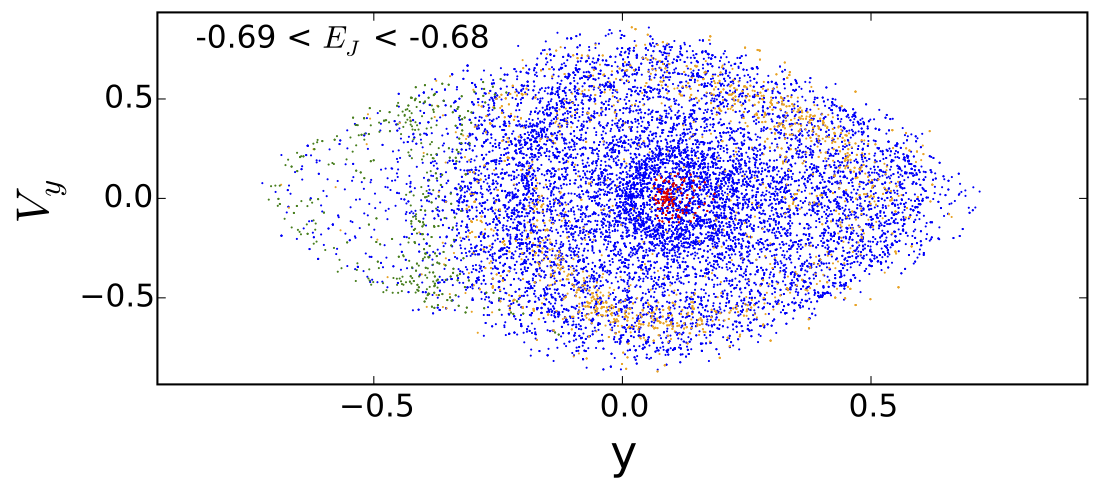} 
\includegraphics[trim=10.pt 0.pt 0.pt 0.pt,angle=0., clip,width=0.46\textwidth]{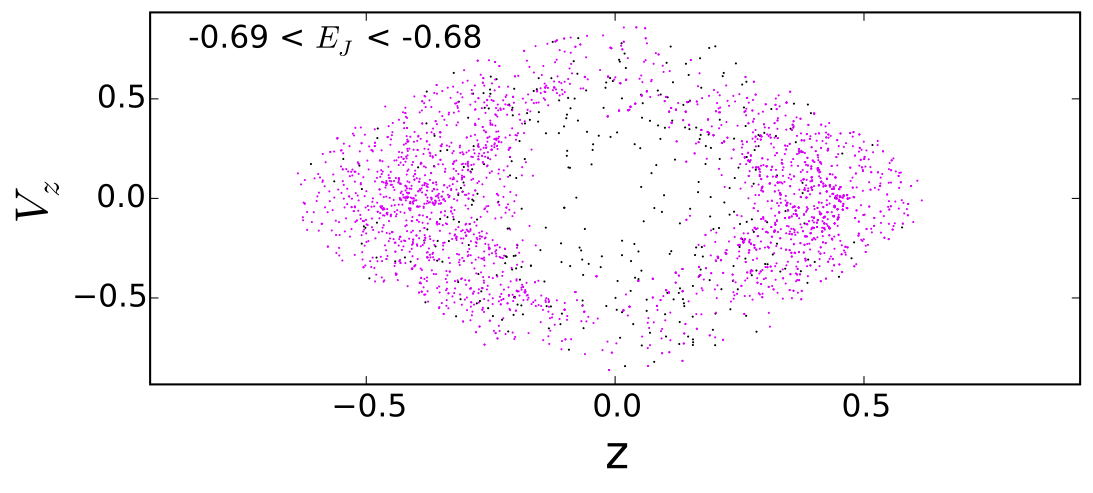} 
\includegraphics[trim=10.pt 0.pt 0.pt 0.pt,angle=0., clip,width=0.46\textwidth]{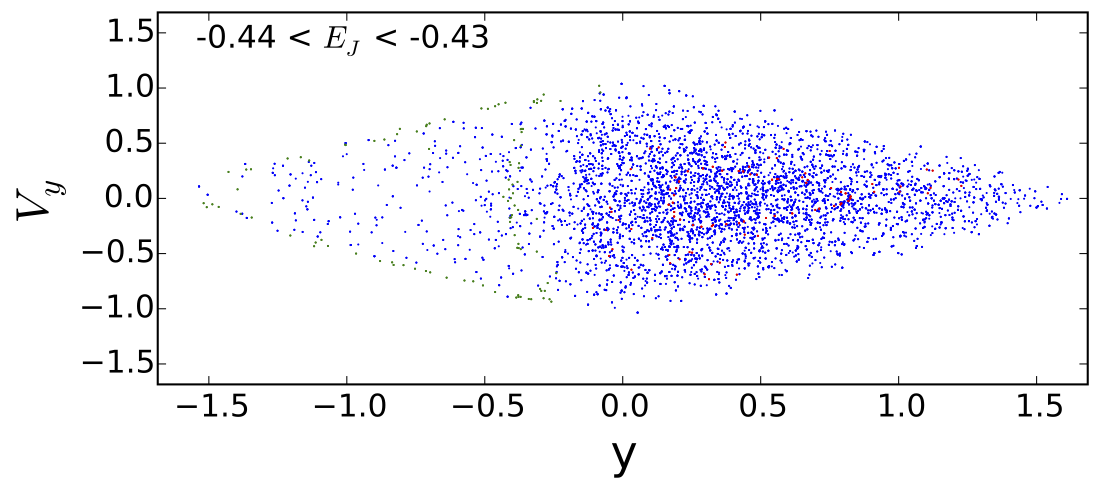} 
\includegraphics[trim=10.pt 0.pt 0.pt 0.pt,angle=0., clip,width=0.46\textwidth]{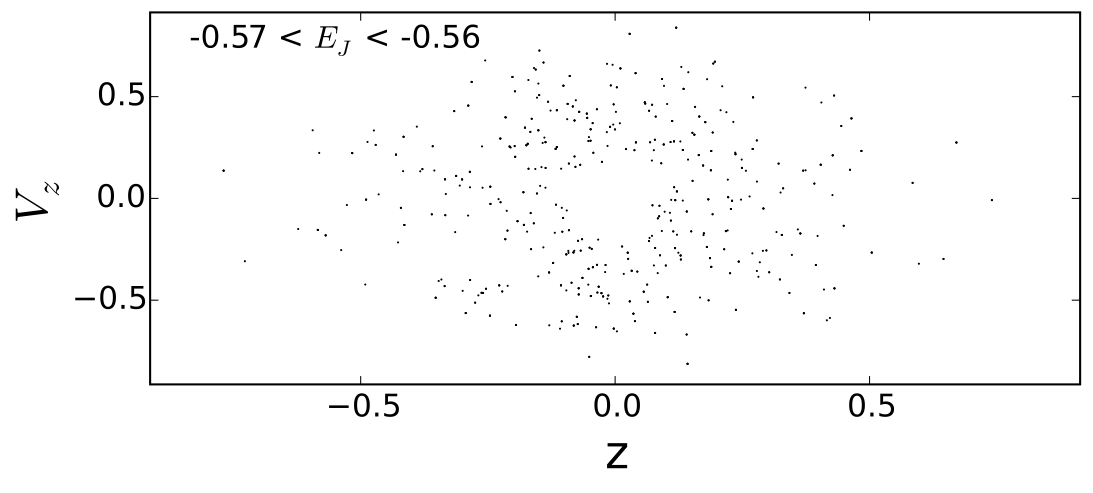} 
\includegraphics[trim=10.pt 0.pt 0.pt 0.pt,angle=0., clip,width=0.46\textwidth]{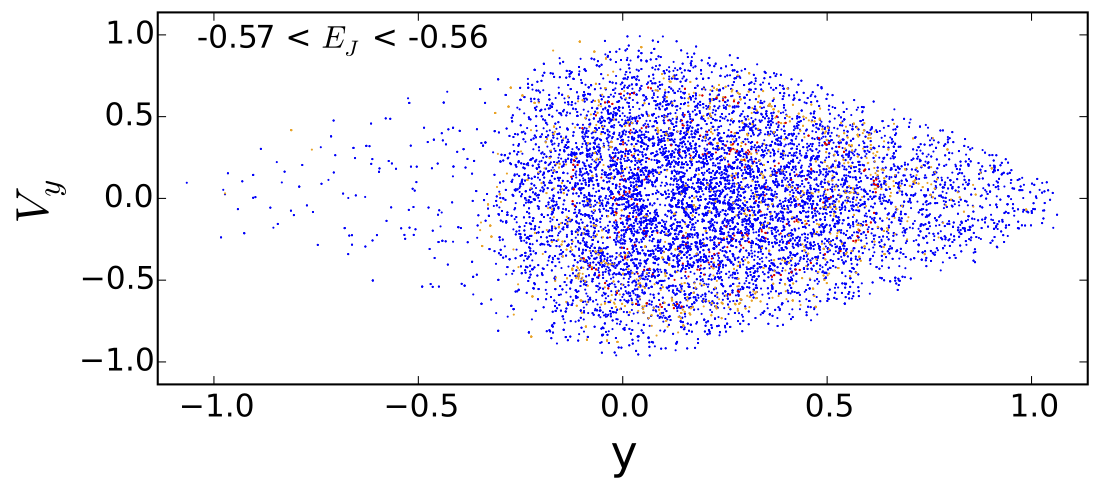} 
\includegraphics[trim=10.pt 0.pt 0.pt 0.pt,angle=0., clip,width=0.46\textwidth]{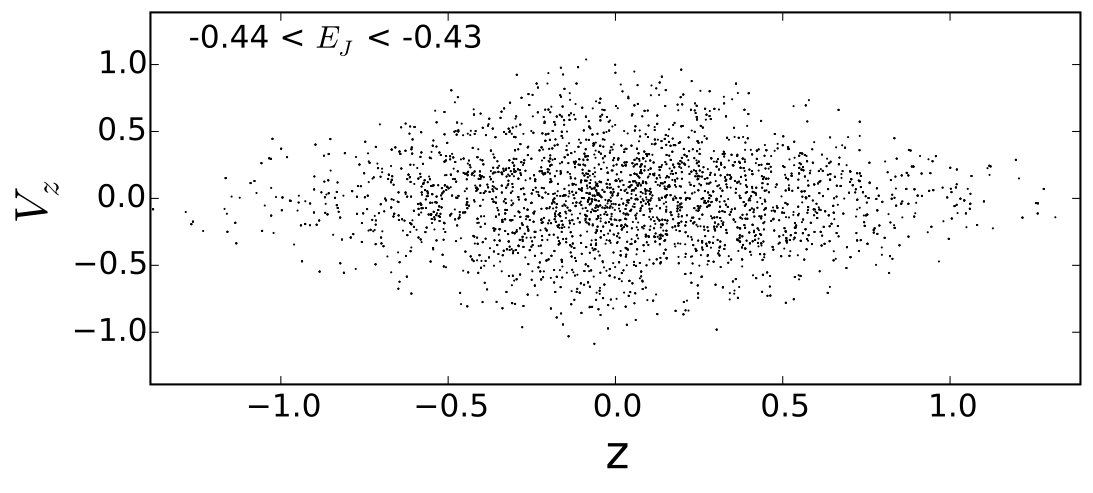} 
\caption{Surfaces-of-section (SoS) for orbits in Model A: left column shows $y$ versus $V_y$, while the right column  shows $z$ versus $V_z$.  Each row shows SoSs for 100 orbits in one of the 5 ranges in $E_J$  (shown as grey vertical bars in Fig.~\ref{fig:Ejhist}). The SoSs in the left hand column plot x1 orbits (red points), box orbits (blue points), 3:-2:0  orbits (orange points) and x4/x2/$z$-tube orbits (green points), while the SoSs in the right hand column plot only $x$-axis tubes (pink points) and chaotic orbits (black points).}
\label{fig:sosA}
\end{figure*}

A standard way to represent the phase space distribution of orbits in a bar is to plot a Poinc\'are surface-of-section (SoS) for a number of orbits at single value of the Jacobi Integral $E_J$. For two dimensional bars it is customary to construct a SoS by plotting the  velocity component $V_y$ as a function of  $y$  every time an orbit intersects the $x$ axis with a negative value of $V_x$ \citep[e.g.][]{sellwood_wilkinson_93}. In three dimensional bars one might choose to restrict orbits plotted on a SoS to those that are confined to the equatorial (i.e. $x-y$) plane  of the model \citep[e.g.][]{shen_sellwood_04}. When orbits of a single value of $E_J$ are plotted on a SoS, regular orbits follow thin closed or broken curves, resonant orbits form groups of islands, while chaotic orbits tend to fill larger areas.

Two problems arise in plotting an SoS for orbits from an $N$-body simulation: first all orbits occupy 3 spatial dimensions and are not well represented on a 2-dimensional SoS; second since orbits have a range of $E_J$  values (see Fig.~\ref{fig:Ejhist}) curves in the SoS appear fuzzy.  Figure~\ref{fig:sosA} show SoSs centered on five different $E_J$ values in  Model A. The left hand column shows  $y$ vs. $V_y$ when orbits cross the $y-z$ plane with positive $V_x$; the SoS appropriate for  x1 orbits (red dots), box orbits (blue dots), 3:-2:0 orbits (orange dots), x4 orbits (green dots). The SoSs in the right hand column show $V_z$ vs. $z$ when orbits  cross the $x-z$ plane with negative $V_y$; these coordinates are more meaningful for $x$-axis tubes (pink dots).  Chaotic orbits (black dots) were plotted in the right hand plots simply to avoid crowding. 

In Figure~\ref{fig:sosA} the spread in $E_J$ (as indicated in the legends) results in few clear curves in the SoSs.  Nonetheless one can see that each orbit family occupies a slightly different region of the SoS. x1 orbits (red dots) which are prograde about the $z$ axis  cluster around $V_y=0$ at positive $y$ values  (see dense red core at the bull's eye of the SoS in Figure~\ref{fig:sosA}). Box orbits (blue regions) lie on oval curves surrounding the x1 orbits, indicating that they belong to the same sequence. (This is much more clearly seen in  Figure 9 of SS04  which shows that the x1 orbits form the smallest ovals forming a ``bull's eye'', and this sequence of ovals  extends to larger radii.) Since the bigger circles (and the blue regions in our plots) can have negative $y$ values, it implies that their $J_z$ becomes negative as they acquire a box-like appearance. In an SoS resonant orbits  appear as multiple islands. Seen from the point of view of dynamical tools like the SoS, box orbits and x1 orbits do belong to the same family although they can have slightly different appearances.  SS04 show that a large number of resonances of various orders can (in principle) exist in an $N$-body bar potential. However our SoSs (and frequency maps in the next section) show only one prominent resonant family is populated in the distribution function. The 3:-2:0 family appears in the SoS in Figures~\ref{fig:sosA} as orange points which appear in multiple discontinuous islands in the region occupied by box orbits. In Figure~\ref{fig:sosA} retrograde $z$-tubes  (green dots) lie to the left of $y=0$ and there are no prograde (x2) orbits in this model. 

\begin{figure*}
\centering
\includegraphics[trim=10.pt 0.pt 0.pt 0.pt,angle=0., clip,width=0.46\textwidth]{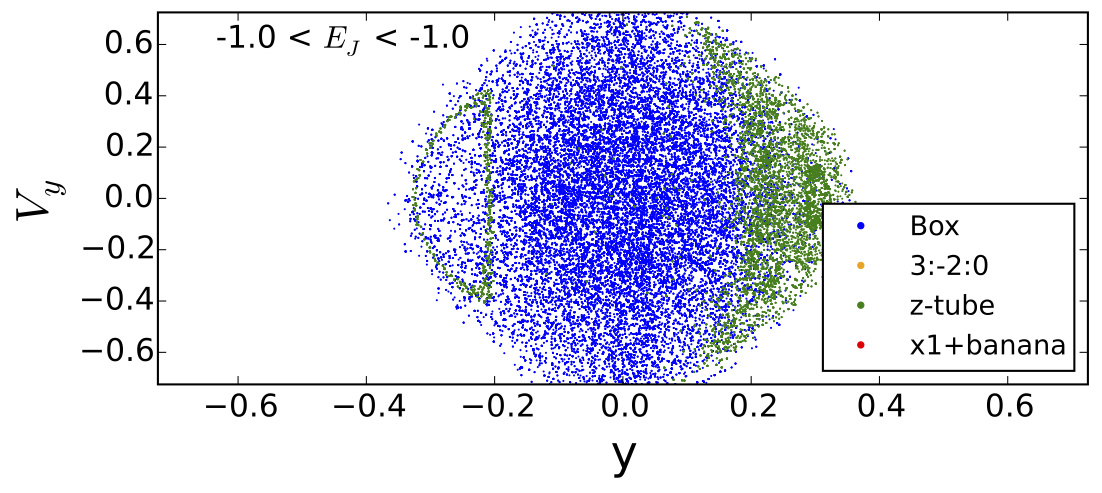} 
\includegraphics[trim=10.pt 0.pt 0.pt 0.pt,angle=0., clip,width=0.46\textwidth]{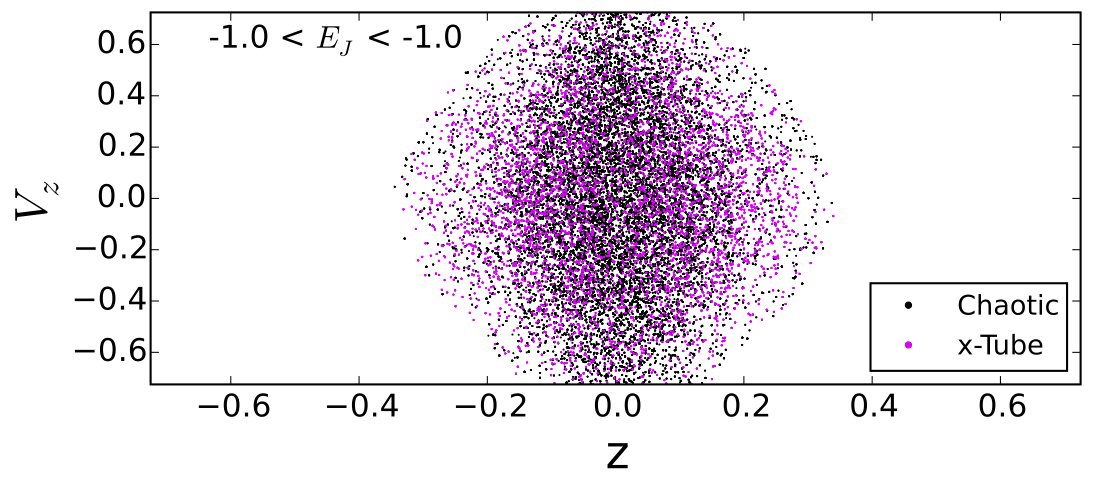} 
\includegraphics[trim=10.pt 0.pt 0.pt 0.pt,angle=0., clip,width=0.46\textwidth]{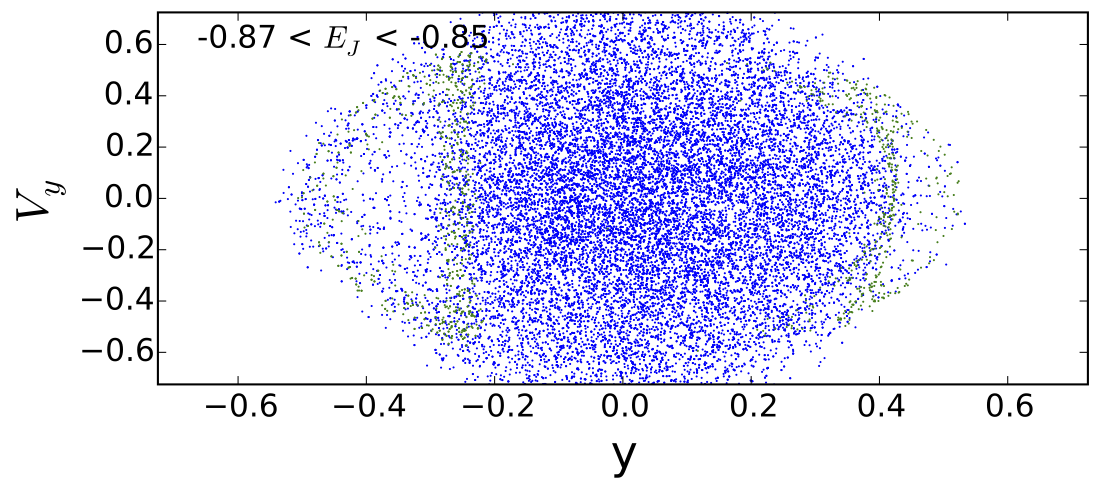} 
\includegraphics[trim=10.pt 0.pt 0.pt 0.pt,angle=0., clip,width=0.46\textwidth]{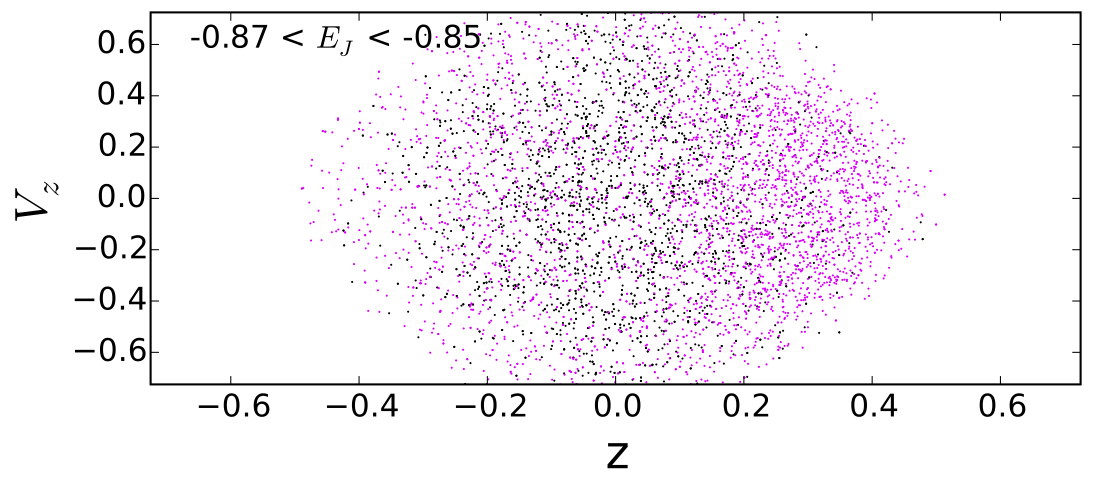} 
\includegraphics[trim=10.pt 0.pt 0.pt 0.pt,angle=0., clip,width=0.46\textwidth]{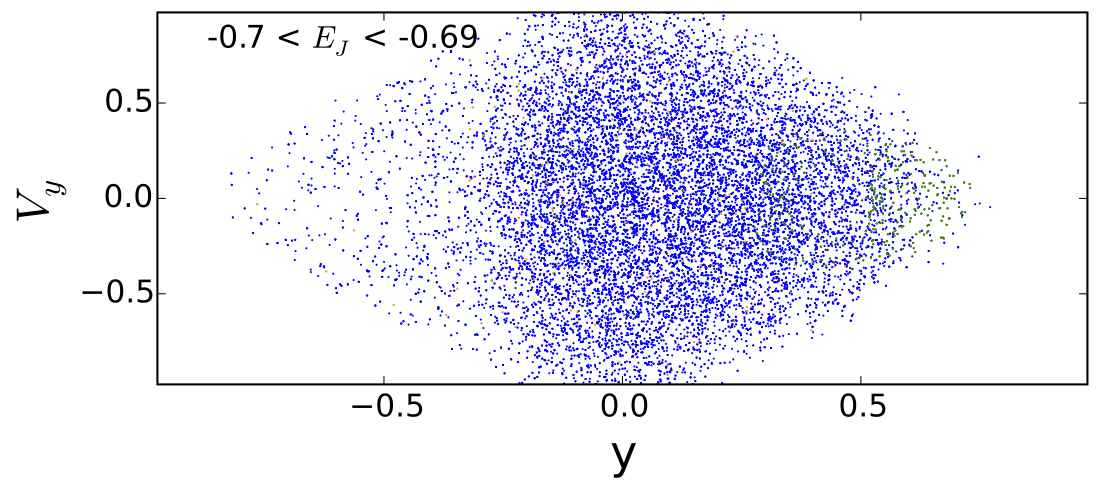} 
\includegraphics[trim=10.pt 0.pt 0.pt 0.pt,angle=0., clip,width=0.46\textwidth]{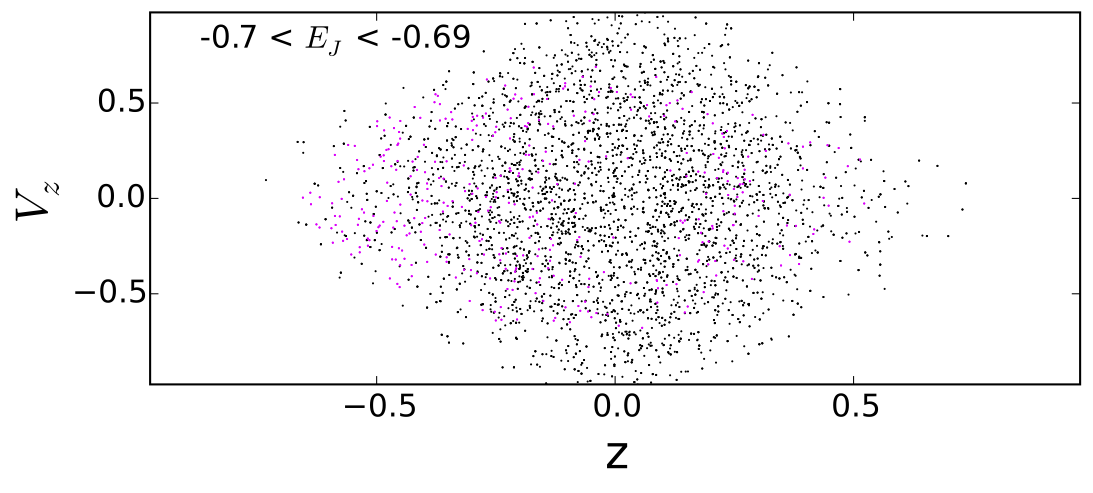} 
\includegraphics[trim=10.pt 0.pt 0.pt 0.pt,angle=0., clip,width=0.46\textwidth]{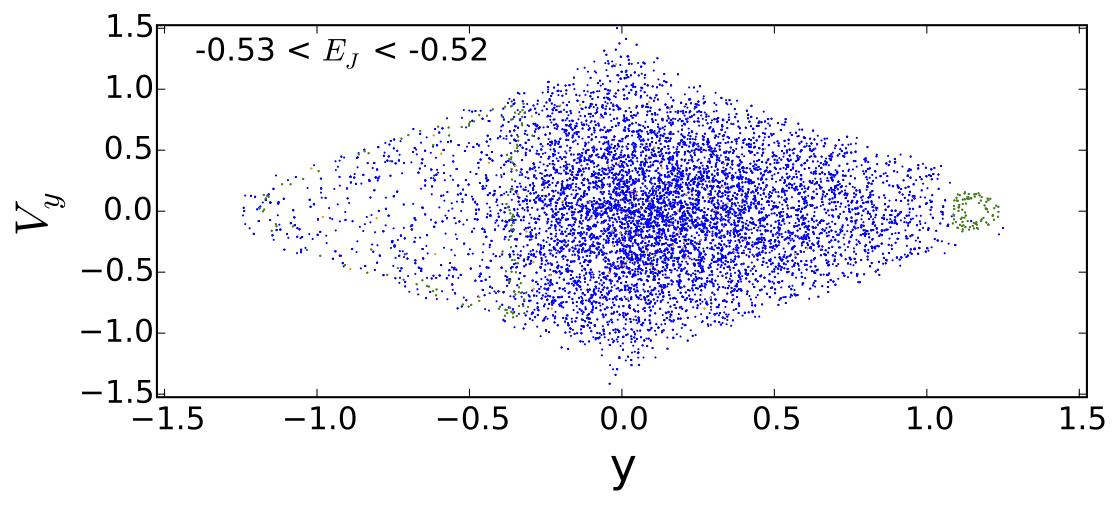} 
\includegraphics[trim=10.pt 0.pt 0.pt 0.pt,angle=0., clip,width=0.46\textwidth]{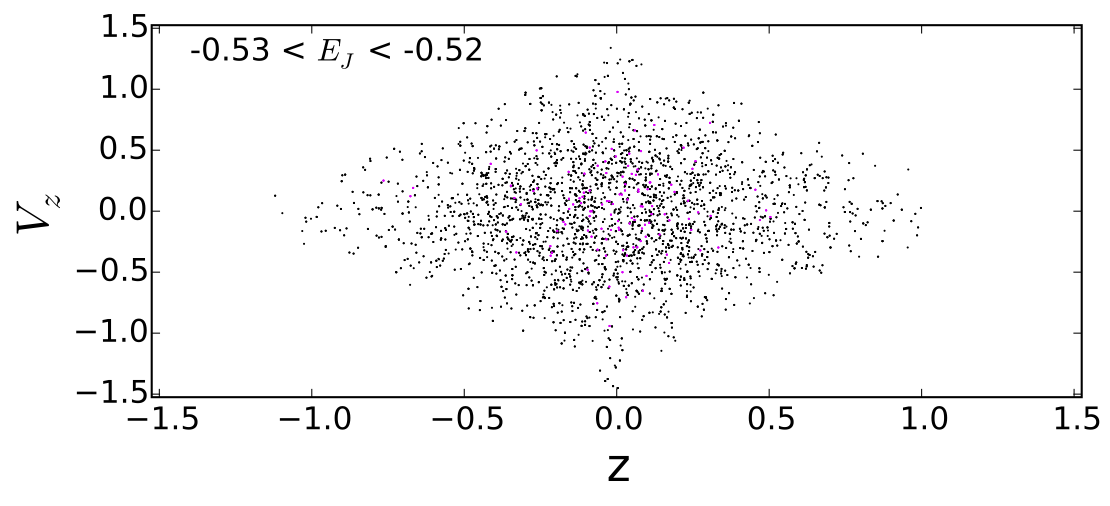} 
\includegraphics[trim=10.pt 0.pt 0.pt 0.pt,angle=0., clip,width=0.46\textwidth]{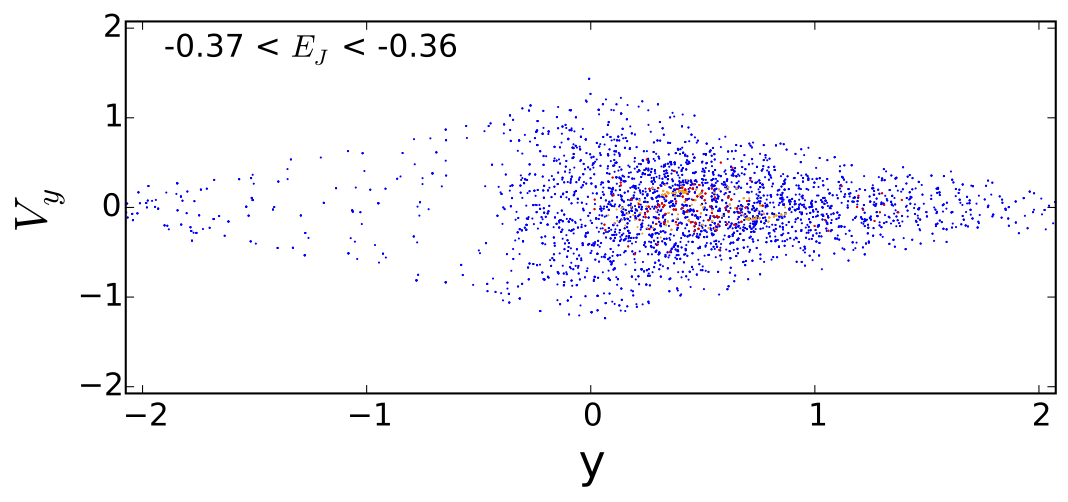} 
\includegraphics[trim=10.pt 0.pt 0.pt 0.pt,angle=0., clip,width=0.46\textwidth]{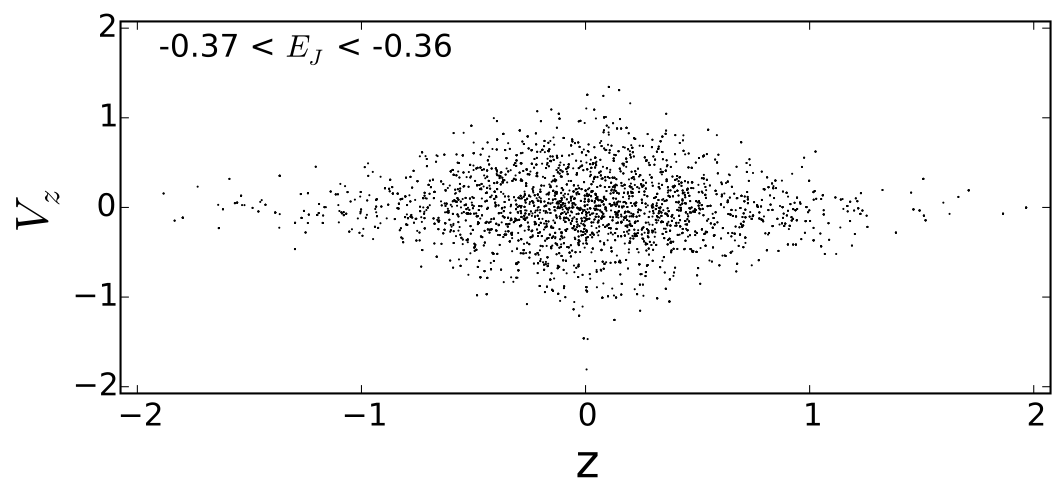} 
\caption{Same as Figure~\ref{fig:sosA} for orbits in Model B.}
\label{fig:sosB}
\end{figure*}

Figure~\ref{fig:sosB} shows surfaces-of-section after the growth of the SMBH (Model B).  
The most striking differences between Model B (after SMBH growth) and Model A  is the appearance of {\em prograde} $z$-tube orbits (parented by the x2 orbit). It is worthwhile reviewing briefly the results of \citet{brown_etal_13} which will shed light on the appearance of these orbits. \citet{brown_etal_13} carried out a detailed analysis of the  intrinsic velocity distributions (velocity dispersion profiles and velocity anisotropy) in the models  analyzed here as well as an axisymmetric disk in which a SMBH was grown in an identical  manner. In both a barred and axisymmetric galaxy, the growth of the SMBH increases the depth of the central potential (``adiabatic contraction''), but the density increase was significantly higher in the barred galaxy than in the axisymmetric one. Furthermore  as matter is dragged inwards in the axisymmetric galaxy the distribution function  becomes tangentially biased and inflow is limited by angular momentum conservation. However in a time-dependent barred galaxy the inward flow of mass is enhanced by outward transport of angular momentum. This results in a higher central stellar density in the bar than in the axisymmetric model. The differences between Model A and Model B arise due to two effects (a) elongated orbits e.g. x1 orbits and $x$-tubes are unable to survive the growth of the central point mass  which makes the potential more axisymmetric \citep{deibel_etal_11}; (b) inflow of matter from large radii with positive angular momentum results in the formation of prograde $z$-tube orbits seen in the SoS, which did not exist in the original bar. 


\subsubsection{Frequency Maps}     
\label{sec:freqmap}

An alternative way to represent the phase space distribution function is via a ``frequency map''.  Frequency mapping exploits the fact that most orbits in galaxies are approximately quasi-periodic (BT08), hence a Fourier transform of their space and/or velocity coordinates can be used to obtain the fundamental orbital frequencies that characterize each regular orbit (for a detailed description of the orbital spectral analysis method see Appendix~\ref{sec:frequencies}).   A frequency map is  useful for representing the phase space structure of a large sample of orbits even if they have a wide range of energies (or $E_J$). On such a map one can plot a large number of orbits that are representative of an entire distribution function (instead of just a subset at discrete values of $E_J$ as in the case of surfaces-of-section).  In an appropriate coordinate system, orbits belonging to different families will appear in different regions of the map providing an easy way to visually assess the importance of various orbit families  to the distribution function and the range of $E_J$  values for which each family is important. Frequency maps are also useful for identifying different resonances and their  importance to the distribution function \citep{valluri_etal_10, valluri_etal_12}.  

\begin{figure}
\includegraphics[trim=0.pt 10.pt 0.pt 20.pt, clip, width=0.5\textwidth]{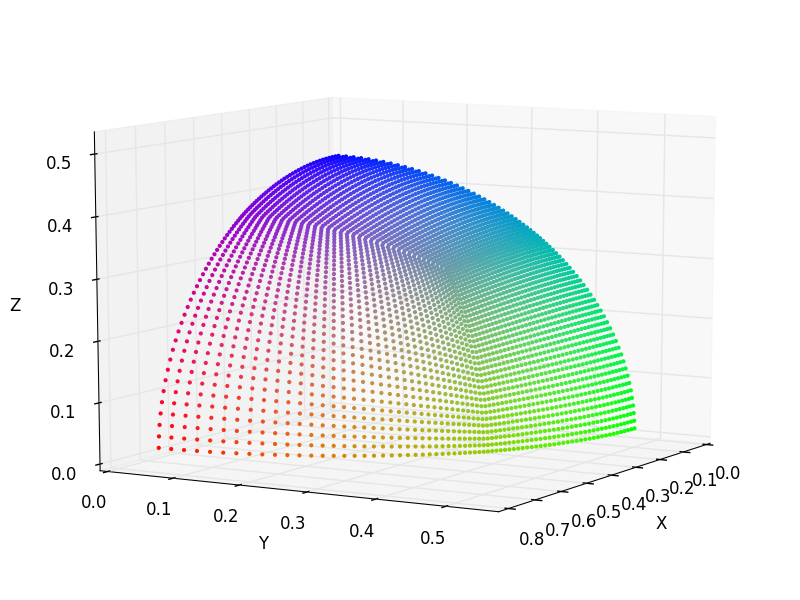}\\
\includegraphics[trim=0.pt 73.pt 0.pt 0.pt,clip, width=0.42\textwidth]{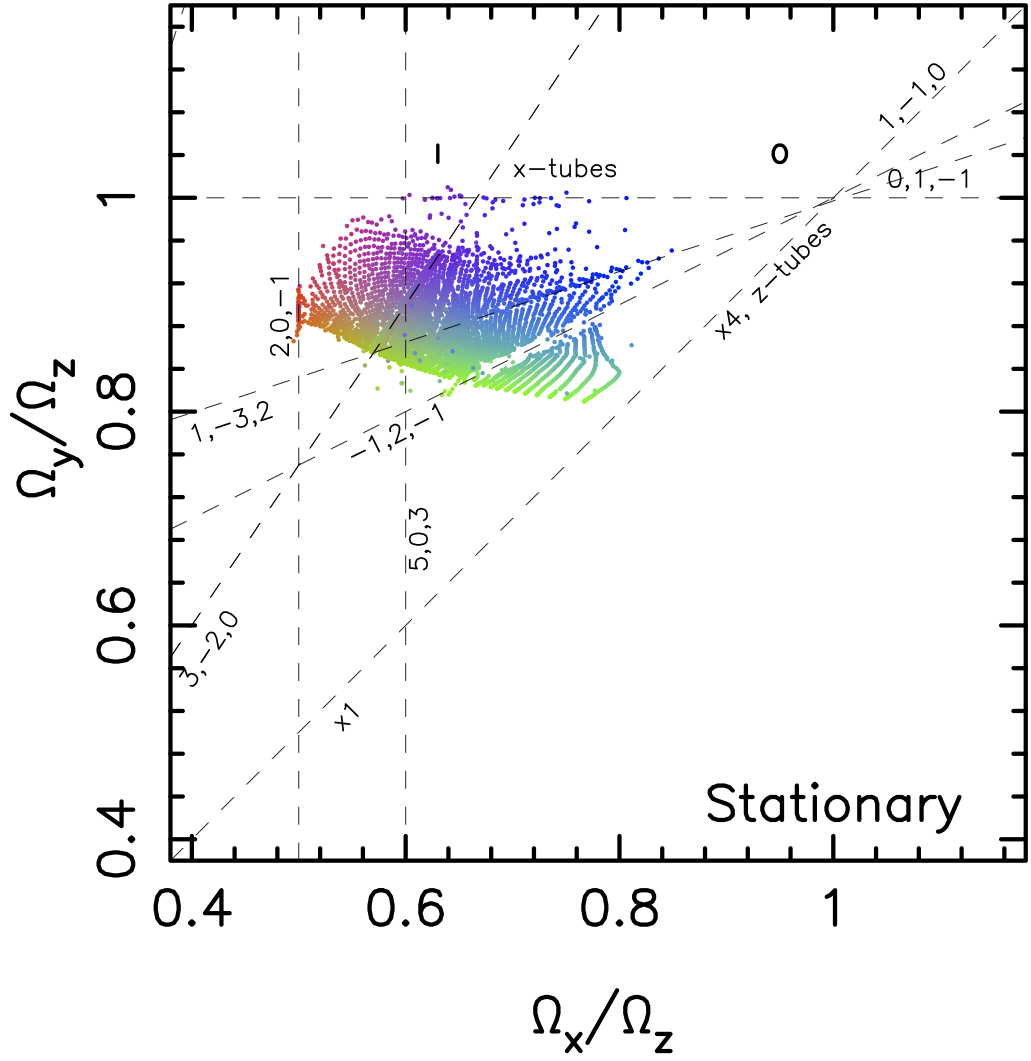}\\
\includegraphics[trim=0.pt 0.pt 0.pt 0.pt,width=0.42\textwidth]{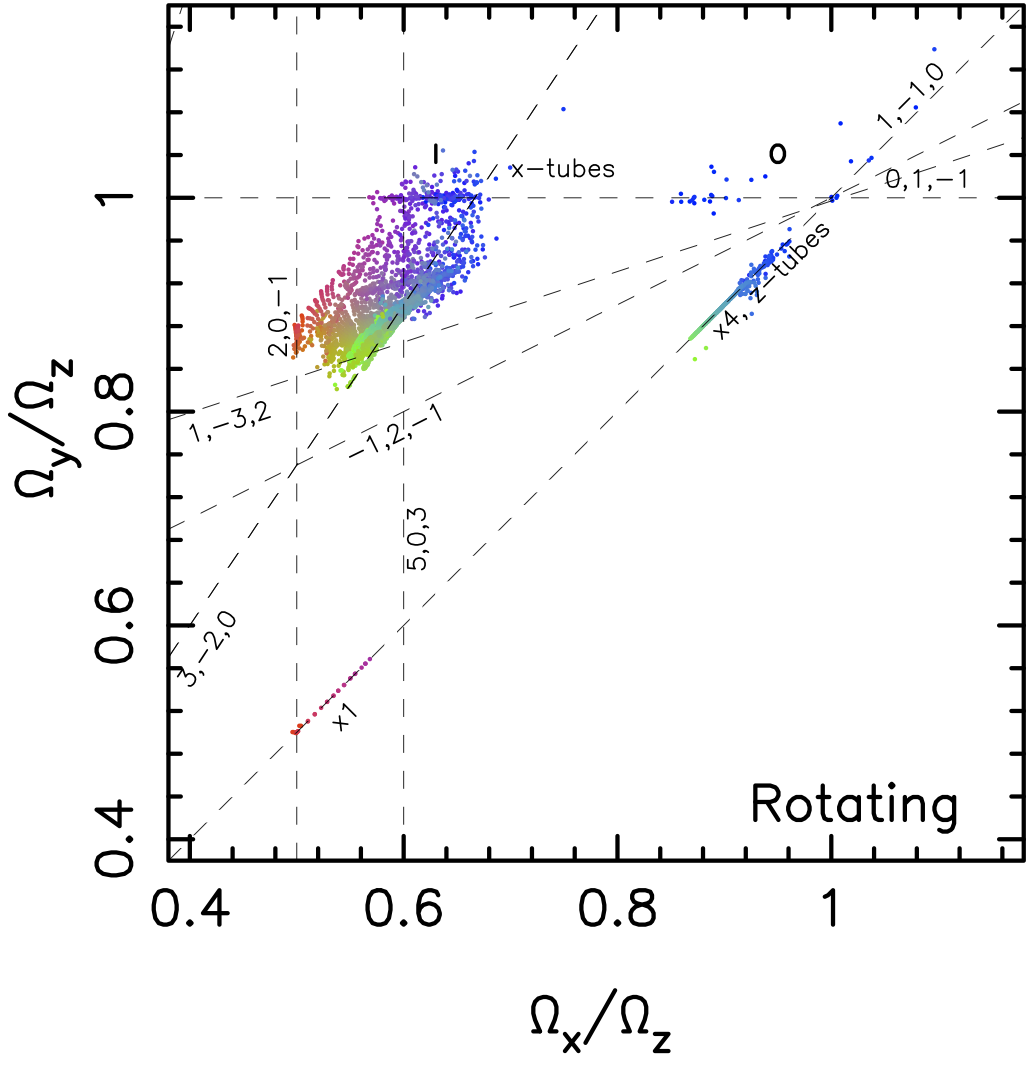}
\caption{Top: 3888 initial positions from which the box orbits are launched with zero initial velocity (i.e. box orbit initial conditions). Each point is given a unique color (determined by it's coordinate) to enable readers to map the start space to the frequency maps below. Middle: frequency map of the orbits launched from the initial conditions in the top panel in a stationary Dehnen model. Bottom: frequency map of the same initial conditions in the rapidly rotating Dehnen model. Several resonance lines are marked with dashed lines and resonance integers, others are clearly visible in the clustering of points but are not marked to avoid crowding.} 
\label{fig:freqmap_dehnen}
\end{figure}

    A frequency map is obtained by plotting pairs of fundamental orbital frequencies or ratios of such frequencies against each other.  {   We integrated a large number of orbits in the Dehnen model at a single energy level launched from ``stationary start space'' initial conditions  (i.e. zero initial velocity) to generate box orbits.  The same initial conditions were integrated with figure rotation. Figure~\ref{fig:freqmap_dehnen}~(top) shows the 3 dimensional distribution of these 3888 initial conditions with each point given a unique color determined by its coordinates ($x,y,z$). The RGB color indices are determined such that points that lie on the $x, y, z$ axes are red, green and blue respectively and the colors of other points reflect their distances from the principle axes.\footnote{The RGB color  index of a point with coordinate $x_i,y_i,z_i$ is given by  $x_i/x_{max}, y_I/y_{max}, z_i/z_{max}$, where $x_{max}, y_{max}, z_{max}$ mark intersections of the equipotential surface with the principle axes.} The middle panel show the frequency map in the stationary Dehnen model, while the lower panel shows the frequency map of the same orbits subjected to figure rotation. The colors of points in the frequency maps enable the reader to visually map points on the frequency maps to their initial launch positions in the top panel.

 In the stationary Dehnen model  (middle panel) most of the points lie in a fairly regular grid above the diagonal (1,-1,0) resonance line and below the horizontal (0,1,-1) resonance line. In the rotating Dehnen model (bottom panel) most of the points move to left, but a few  points appear along the (1, -1, 0) diagonal near the label ``x1''. Their red color  indicates that they were launched from near the $x$-axis and as expected they are converted to  ``x1'' orbits in the rotating frame.  There are box-like red points associated with the (2, 0, -1) ``banana resonance'' which is bifurcation of the x1 family (these are the box like orbits which also have the banana shape in $x-z$ projection e.g. 2nd row of Fig.~\ref{fig:Dehnenboxes} but are launched from some height above the $x-y$ plane. 
 
 As predicted by \citet{martinet_dezeeuw_88} $z$-axis orbits (and orbits launched close to the $z$-axis, shown in blue) with adequate angular momentum generate both retrograde $z$-tubes (parented by the x4 family) and the long axis tubes (parented by the stable anomalous orbits). These orbits  appear as blue dots along the horizontal (0, 1, -1) resonance and at the top of the diagonal (1,-1,0) resonance.  The label ``I" shows the location of the inner-long axis tubes and the label ``O''  shows the location of the outer long axis tubes. Finally, orbits launched close to the intermediate $y$-axis in the rotating Dehnen model with sufficient angular momentum  generate retrograde $z$-tube orbits (which appear as green dots along the  (1,-1,0) diagonal). }

\begin{figure*}
\centering
\includegraphics[trim=0.pt 45.pt 0.pt 0.pt,clip,width=0.45\textwidth]{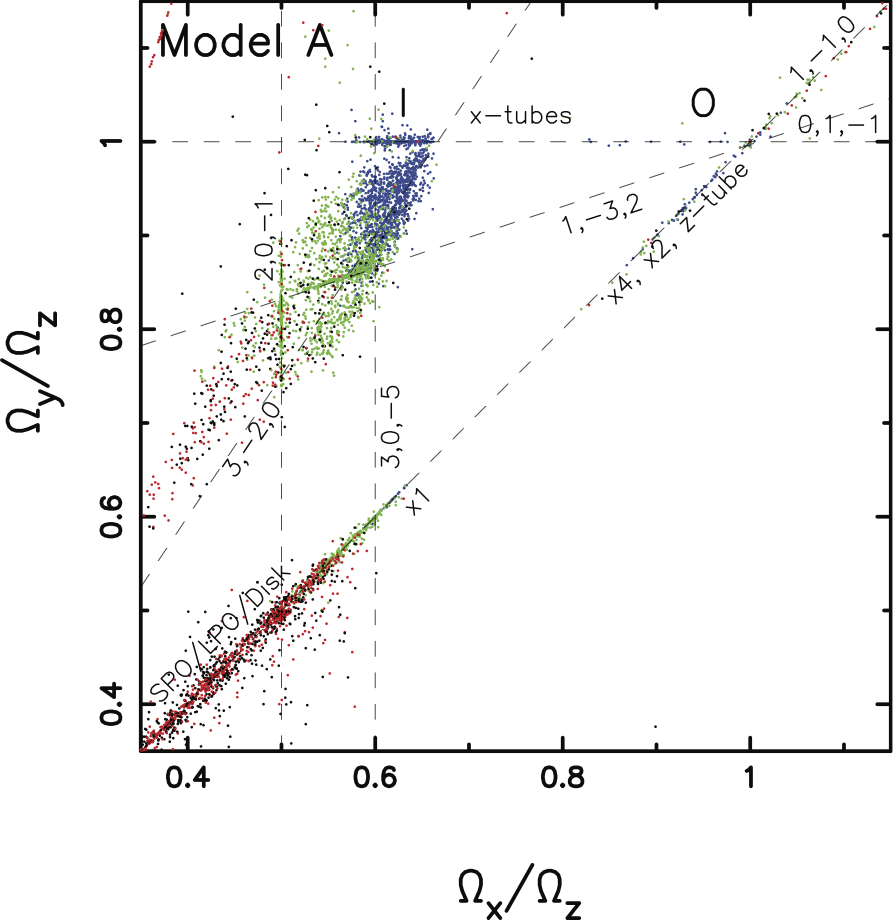}
\includegraphics[trim=0.pt 45.pt 0.pt 0.pt,clip,width=0.45\textwidth]{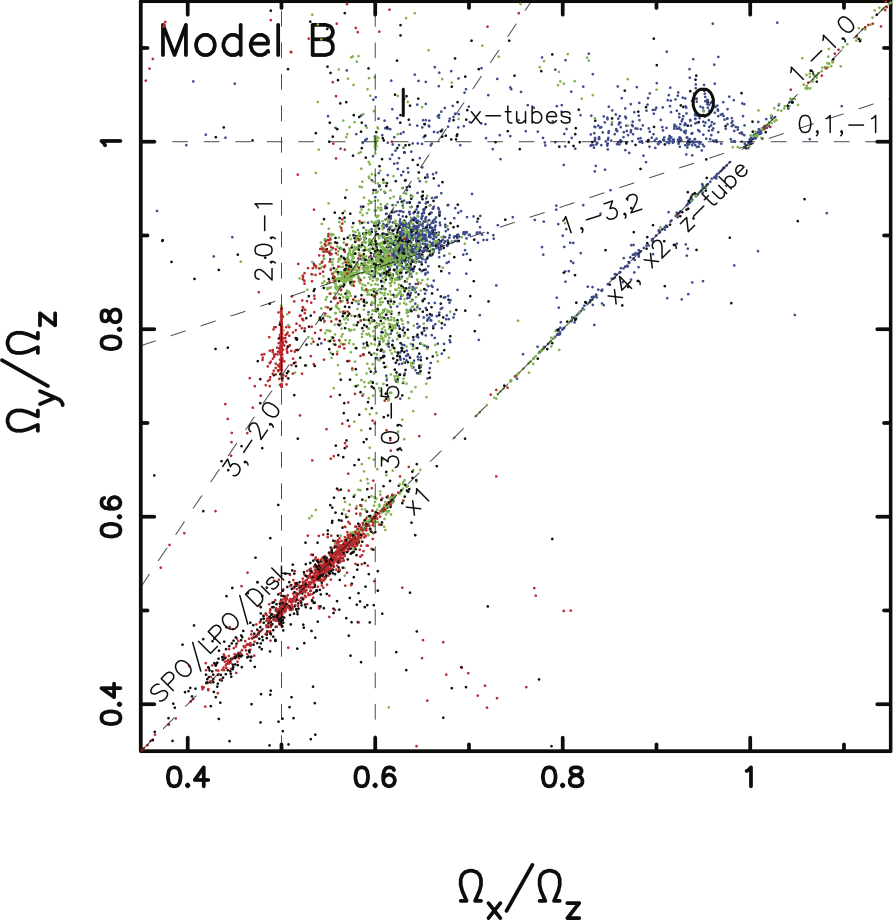}\\\includegraphics[trim=0.pt 0.pt 0.pt 0.pt,clip,width=0.45\textwidth]{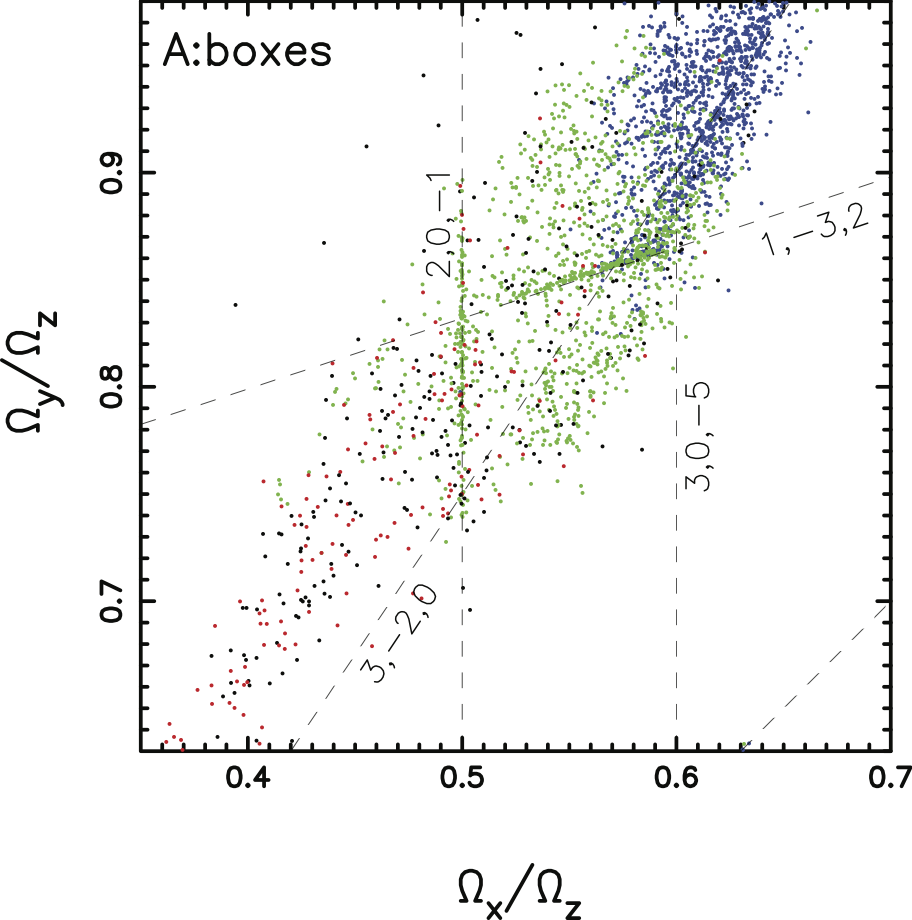}
\includegraphics[trim=0.pt 0.pt 0.pt 0.pt,clip,width=0.45\textwidth]{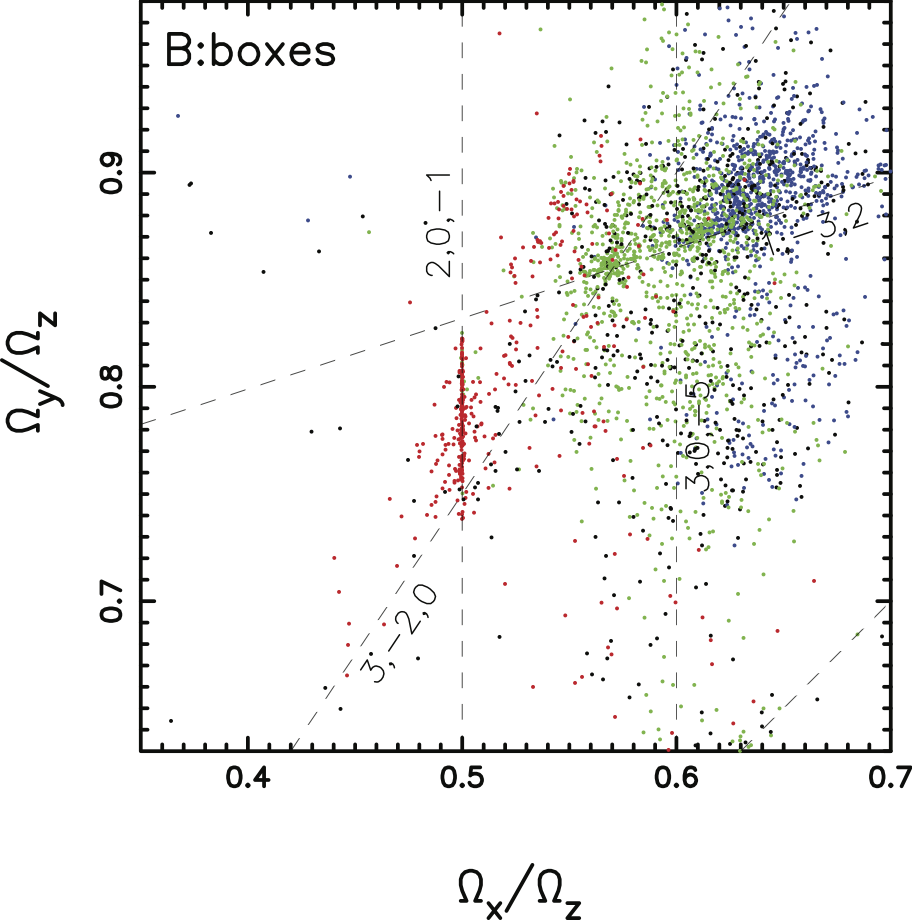}
\caption{Cartesian frequency map for orbits in Model A (top left) and Model B (top right) for orbits that were integrated for more than 20 orbital periods and with $R_{apo}<4$.  The colors signify the Jacobi integral $E_J$ of each orbit: blue particles have the most negative values, red have the least negative values and green lie in between, and grey points are chaotic with frequency drift parameter $\log(\Delta f) > -1.2$. The labeled dashed lines mark the main resonances (see text for details). The more scattered appearance of points in Model B is indicative of the larger fraction of chaotic orbits, especially close to the SMBH. The lower two panels show
a zoom in of the central region of the map with orbits colored by orbit family (grey points are chaotic orbits). } 
%
\vspace{1.cm}
\label{fig:freqmap}
\end{figure*}

We now compare these frequency maps with those generated for the orbits drawn from N-body bar distribution functions (Fig.~\ref{fig:freqmap}) for Model A (left) and Model B (right).  Accurate determination of fundamental orbital  frequencies require that orbits are integrated for at least 20 orbital periods.  Consequently we exclude all orbits which were integrated for less than 20 orbital periods from the frequency maps. After excluding orbits with short integration times, the frequency maps in the top row of this figure show 5704 orbits for Model A and 6168 orbits in Model B\footnote{The deeper potential resulting from the growth of the central point mass and associated angular momentum transport causes $\sim 450$ particles to be pulled inwards and hence they execute a large number of orbital periods in the same time interval.}. The two lower panels zoom into the region of each map occupied by boxes. In both models the excluded particles were visually classified as belonging to the disk, so the frequency maps are still excellent representations of the distribution functions of the bars. 

In Figure~\ref{fig:freqmap} the color of a point signifies the value of its Jacobi Integral $E_J$ (and not its initial position as in the previous maps). Particles in each map were divided into three equal groups in $E_J$ with the blue particles having the most negative values, the red particles have the least negative values and the green occupy the intermediate range in $E_J$.  Chaotic orbits are identified with black points and appear mixed in with the red points. In Model A most of the chaotic orbits are mixed in with red points in the transition region between the disk and the bar (this is expected e.g. from the work of \citet{voglis_etal_07}, \citet{contopoulos_harsoula_13}, \citet{Patsis_Katsanikas_14b}). In both maps points cluster along ``resonances''  marked by dashed lines labeled with frequency ratios. The x1 orbits and $z$-tube orbits parented by x2 and x4 orbits lie in the same part of the frequency map as they do in the rotating Dehnen model (Fig.\ref{fig:freqmap_dehnen}~(lower panel).   Along the diagonal we also find orbits not present in the Dehnen model that arise at the transition between the bar and disk: disk orbits, short period orbits (SPO) and long period orbits (LPO). SPO and LPO orbits are orbits that are temporarily trapped around the L4 and L5 Lagrange points. 
  
 A ``cloud'' of box orbits appears to the left of the (1,-1,0) line and below the (0, 1, -1) line  in the same part of the frequency map as box orbits in the rotating Dehnen model in Figure~\ref{fig:freqmap_dehnen}~(lower panel).  The lower two panels in Figure~\ref{fig:freqmap} zoom into this region to more clearly show the various resonant boxlets. The vertical line labeled ``2:0:-1'' marks the boxy banana orbits. The  x1 bananas lie at the intersection between ``2:0:-1'' and ``1:-1:0'' and are periodic orbits that satisfy the condition ``$\Omega_x:\Omega_y:\Omega_z=$2:-2:-1''  \citep[these are also referred to as ``x1v1'' orbits by][]{skokos1}. Also seen is the  3:-2:0 resonance. Although the  ``double pretzel''-like  orbits (Fig.~\ref{fig:boxlets}, top row) look slightly different from the ``double fish''-like orbits (Fig.~\ref{fig:boxlets}, 2nd \& 3rd rows), they both belong to this resonance family. 

The frequency map for Model B shows essentially the same features as Model A except that  a  number of orbits with low values of $E_J$ (blue)  are scattered around the map. These orbits were scatted by the growing central point mass but are regular in Model B and associated with inner $x$-tubes  (near the label ``I''),  outer $x$-tubes (near label ``O''), and prograde $z$-tubes. 

\section{Visual versus Automatic classification}
\label{sec:vis_auto}

\begin{table*}[t]
\center
\caption{Visual Classification versus Automatic Classification$^a$}
\begin{centering}
\begin{tabular}{llrrrrrrrr}\hline\hline
Model    &Classifier      &          $x$-tubes        & x4+x2+$z$-tubes  &x1+bananas     & Boxes               &  3:-2:0            & Chaotic           &  &Disk \\
\hline
\vspace{0.2cm}
              & Visual (CA) &         369  (7.0\%)    & 90 (1.6\%)            &	407 (7.2\%)   & 3831 (67.5\%) &  212 (3.7\%) &  740  (13.0\%)   &  & 4324 (N/A)\\
 {\it A}           & Auto    &         494  (8.5\%)     & 31 (0.5\%)            &  171 (3.0\%)    & 3707 (65.3\%) &  343 (5.9\%) & 1049 (18.5\%)  & & 4205  (N/A)\\
  \hline
 \vspace{0.2cm}
              & Visual (CA) &         305  (5.0\%)  &   81 (1.3\%)            &  360 (5.9\%)    &  3874 (64.0\%) & 45  (0.7\%) & 1392 (23.0\%)  & &  3706 (N/A)\\
{\it  B}    & Auto           &         250  (4.0\%)  & 147 (2.3\%)            &  196 (3.1\%)    &  4197 (66.7\%) & 119 (1.9\%  & 1395 (22.2\%) &  & 3696 (N/A)\\ 
\hline \hline
\multicolumn{10}{l}{$^a$In each column the integer indicates the number of orbits, while the quantity in parenthesis is percentage of that type in the {\em bar only}.}
\end{tabular}
\end{centering}
\label{tab:AVorbitclass}
\end{table*}

  We now compare the results of the automatic classification method for bar orbits (described in Appendix B) with our visual classification. In Table~\ref{tab:AVorbitclass} the integer in each column represent the total number of orbits  in that family (from a total of 10,000 orbits in each model), while the percentage of bar orbits (i.e. excluding disk orbits) contributed by this family is given in parenthesis. Note that for Model B we  exclude the 113 visually classified ``PKO'' orbits since the automated method is unable to distinguish these from boxes, long axis tubes, short axis tubes at large radii.

In Figures~\ref{fig:orbfracs},~\ref{fig:sosA},~\ref{fig:sosB} and Table~\ref{tab:AVorbitclass}  we present x1 orbits, 3:-2:0  orbits and boxes separately but remind readers that  they are all members of the box orbit family.  

The biggest difference between the automatic and visual classification is the fraction of orbits that are classified as ``chaotic'' or ``x1/x1+banana''. The majority of orbits that are visually classified as ``x1/x1+banana'' are automatically classified as boxes. As we have discussed this is expected since x1 orbits are the parents of boxes. The second difference is that while 90 orbits were visually classified as retrograde $z$-tubes orbits only 31 were automatically classified as such. Examination of the remaining 60 showed they were all identified as chaotic by the automatic classifier. The third major difference between the two classification methods is in fraction of chaotic orbits ({  as discussed before, visual classification of chaotic orbits is difficult}). Overall the differences between the automated classification and the visual classification are small (of order $\sim 3$\% at most) indicating that automated classification based on Cartesian fundamental frequencies (also used in triaxial ellipsoids) provides a robust way to classify bar orbits.

\begin{figure*}
\centering
\includegraphics[trim=0.pt 0.pt 0.pt 0.pt,angle=0., clip,width=0.45\textwidth]{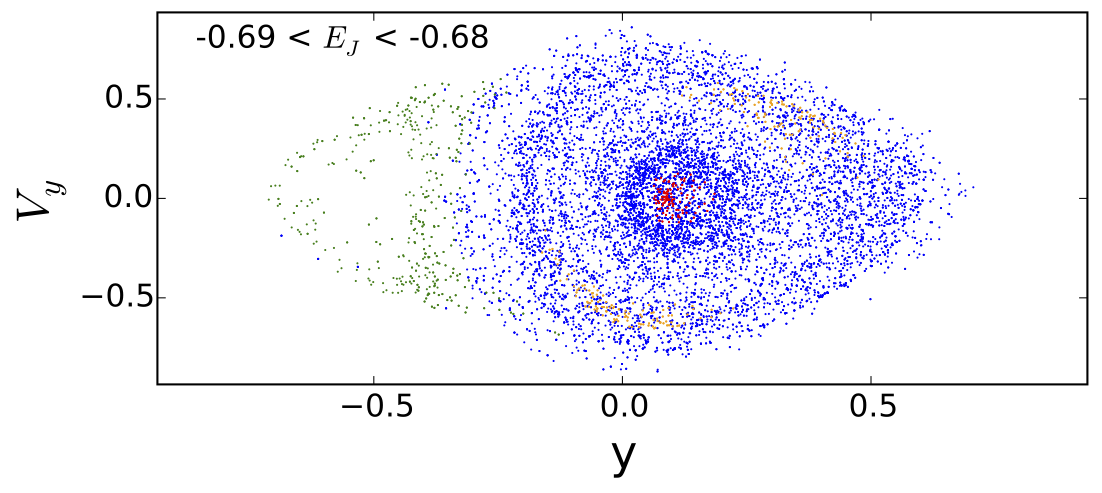} 
\includegraphics[trim=0.pt 0.pt 0.pt 0.pt,angle=0., clip,width=0.45\textwidth]{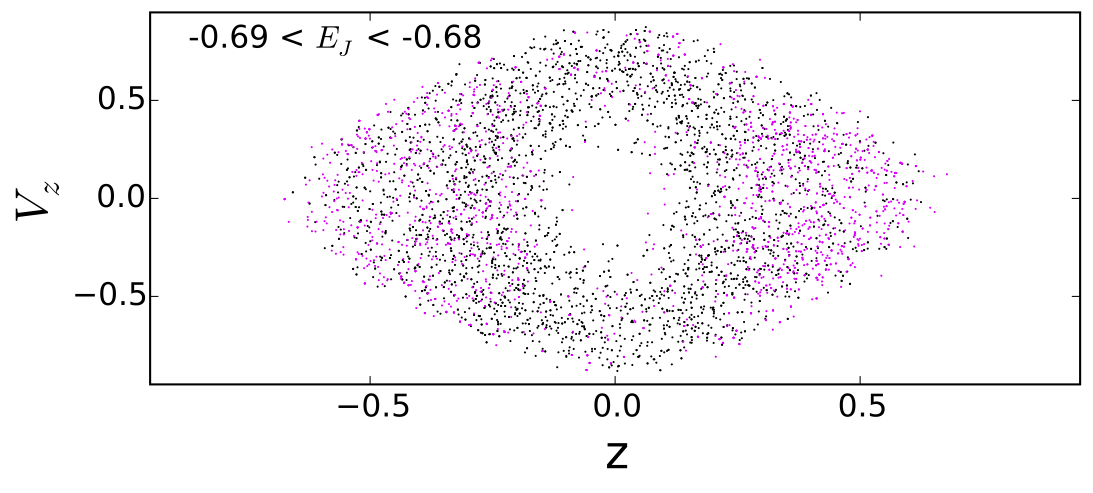}
\includegraphics[trim=0.pt 0.pt 0.pt 0.pt,angle=0., clip,width=0.45\textwidth]{Fig10e.png} 
\includegraphics[trim=0.pt 0.pt 0.pt 0.pt,angle=0., clip,width=0.45\textwidth]{Fig10f.png}
\caption{Comparison of visual classification (top row) and automatic classification (bottom row) using surfaces-of-section with points color coded by type: x1 orbits (red), box orbits (blue), 3:-2:0  orbits (orange) and $z$-tube orbits (green), $x$-tubes (pink) and chaotic orbits (black).}
\label{fig:sos_compare}
\end{figure*}

{  Figure~\ref{fig:sos_compare} shows SoSs for a small range of $E_J$ values (as before colors denote orbit type; red: x1/banana orbits; orange: 3:-2:0 orbits; blue: boxes; green: $z$-tubes). For the top row the color coding is based on visual classification while for the lower row it is based on automatic classification. The main difference is the size of the 3:-2:0 resonance which is large in the automatic classification (top left) but restricted more to the middle of the resonant islands in the visual classification (lower left). The automatic method identifies more orbits as associated with the 3:-2:0 resonance (orange points) since it is difficult to visually identify weakly resonant orbits. A  larger number of orbits are classified as chaotic by the automatic classifier in Model A since this method has more robust criteria for identifying chaotic  and resonant orbits. Apart from these differences the two methods give  similar SoSs. }

\section{Summary and Discussion}
\label{sec:conclude}

We have analyzed 10,000 orbits from each of two self-consistent $N$-body bar models. These 20,000 orbits were classified both visually and with a new automatic orbit classification method developed here. By  grouping individual bar orbits into major families we showed that each family has a counterpart in a rotating Dehnen model.  This is different from the textbook view that the main body of a bar is comprised primarily of prograde x1 orbits, prograde x2 orbits and their vertical bifurcations  (BT08). We examined the distributions of orbit families with radius and Jacobi integral ($E_J$). The phase space distributions were examined using surfaces-of-section and frequency maps.  The  main results of this analysis  listed below. 
\begin{itemize}
\item  It is well known that the x1 orbit is the same as the $x$-axis orbit in triaxial potential subjected to high pattern speeds \citep{schwarzschild_82,martinet_dezeeuw_88}, and that this orbit parents the box orbit family.  We demonstrate that although the x1 orbit is prograde, the box-like orbits parented by it have little or no net angular momentum about any axis. This family of non-rotating boxes dominates the distribution functions of the $N$-body bars and includes ``resonant'' boxlets  orbits  similar to those found in stationary  triaxial potentials (see Figs.~\ref{fig:x1-Box-bar}, \ref{fig:Dehnenboxes}, \ref{fig:boxlets}). 

\item $N$-body bars also contain a small fraction of long axis tube orbits which behave exactly like those in rotating triaxial ellipsoids. These orbits are parented by the stable anomalous orbits (periodic 1:1 loops rotating about the $x$ axis) which, as predicted by \citep{heisler_etal_82}, are tipped either clockwise or anticlockwise about the $y$-axis of the figure by the Coriolis force depending on the sign of their angular momentum $J_x$ (Fig.~\ref{fig:xtubes}). 

\item We find no prograde $z$-tubes parented by x2 orbits in the pure bar model (Model A) and only a small fraction of retrograde $z$-tubes (parented by x4 orbits) (Fig.~\ref{fig:x4}). This is consistent with the behavior of short axis tubes found in rotating triaxial potentials - retrograde members of the family dominate at high pattern speeds \citep{martinet_dezeeuw_88,deibel_etal_11}. However, in Model B which has a central point mass representing a 0.2\% SMBH, a significant fraction of the x1 and  $x$-tube orbits are destroyed and  replaced by prograde  $z$-tube orbits (parented by x2 orbits). This behavior in response to the growth of a central point mass is consistent with previous work on the self-consistent growth of central spherical components in triaxial potentials (e.g. SS04, \citet{valluri_etal_10}). 

\item Orbit families not seen in triaxial ellipsoids only appear at the interface between the bar and the disk. These include the short-period orbits (SPO) and long-period orbits (LPO) that circulate around the L4 and L5 Lagrange points of the bar as well as orbits that oscillate between L1 and L2 Lagrange points \citep{athanassoula_etal_09}.  There are also a large number of chaotic orbits  found in this region \citep{harsoula_kalapotharakos_09, contopoulos_harsoula_13}. 

\item The orbit families in bars with and without a SMBH are similar except in the central region. The growth of the SMBH causes  (a) a reduction in the fraction of x1 orbits and $x$-tubes; (b) an increase in the fraction of $z$-tubes, especially the prograde variety which were previously absent; (c) an increase in the fraction of chaotic orbits (Fig.~\ref{fig:orbfracs}); and (d) a new population of ``precessing Keplerian orbits" (PKOs). All these effects arise because SMBH scatters orbits with small pericenter radii and  enhances mass inflow due to adiabatic contraction, aided by angular momentum transport by the bar.

\item The three new families of PKOs are found near the SMBH: here the Keplerian potential is dominant but orbits experience precession orbits due to the rotating triaxial bar. Interestingly, when integrated for long times, the orbits in this region are found to belong to the same three major orbit families: boxes, long axis tubes and short axis tubes found in the main body of the bar. Only 1.8\% of bar orbits (113/6168 ) were PKOs, a mass comparable to that the SMBH itself. This family also has counterparts in stationary triaxial potentials with SMBH that are called ``saucers" ($z$-tubes) and ``pyramids'' (boxes) \citep{sridhar_touma_99, poon_merritt_01,merritt_vasiliev_10,Li_Bockelmann_Khan_15}.

\item A new automated orbit classification scheme for bar orbits based on Cartesian fundamental frequencies largely recovers the same classifications as those obtained by visually 20,000 examining individual orbits. A comparison of frequency maps of orbits in a rotating Dehnen model and orbits drawn from self-consistent bars provides further confirmation that the two systems are fundamentally similar.

\end{itemize}

Our effort to unify the orbital structure of bars and triaxial ellipsoids goes beyond mere theoretical curiosity. Recent developments make this new unified picture acquire important practical astrophysical applications. Recently we studied simulations of barred galaxies with SMBHs \citep{brown_etal_13,hartmann_etal_14} to show that a bar can cause an increase in the central line-of-sight velocity dispersion ($\sigma$) by about 7-25\% - an increase that is consistent with the average offset observed for barred galaxies relative to unbarred ones from the $\mbh -\sigma$ relation. In addition a more serious consequence of the presence of a bar is that its unique orbital structure (the combination of the radially biased bar orbits and the high bar pattern speed), has been shown to result in a high central velocity dispersion but negative 4th Gauss-Hermite parameters ($h_4$), even in the vicinity of the black hole. The  Schwarzschild orbit superposition method is currently  a popular way to measure super massive black hole masses from stellar kinematics.  \citet{brown_etal_13} showed that this unique combination of kinematical parameters (high central $\sigma$ and negative $h_4$) can result in a systematic over-estimate of the mass  of the SMBH  if the bar is modeled as if it is axisymmetric and the true nature of bar orbits is not taken into consideration. The axisymmetric  and stationary triaxial Schwarzschild modeling methods are currently considered the ``gold standards'' for dynamical black hole mass determination against which secondary methods (e.g. reverberation mapping) are calibrated. The over-estimate in $\mbh$ (which can be a factor of two or higher depending on the orientation of the bar) was dramatically illustrated in the case of the barred Seyfert 1 galaxy NGC 4151 \citep{Onken14}. In NGC 4151 although the central bulge appears very circular, there is clear kinematical evidence for a bar seen in the velocity fields e.g. clear isophotal twists in line-of-sight velocity and negative $h_4$ parameters along the length of the bar axis.  Although nearly 60\% of Spiral/S0 galaxies with existing stellar dynamical black hole mass measurements are in barred galaxies, they have been derived with axisymmetric models!  

Since the Schwarzschild method relies on superposition of orbits, the results are extremely sensitive to the completeness and accuracy of the library of orbits that is supplied to the superposition code \citep{valluri_etal_04, thomas_etal_04}. Currently there are  no orbit superposition codes designed to measure black hole masses from stellar kinematical data in bars. As the work in this paper clearly shows the textbook view that the orbits in a bar arise from perturbations to a series of prograde x1 and x2 orbits is incomplete since it ignores non-rotating boxes, retrograde $z$-tubes (parented by x4 orbits) and long axis tubes.  In fact, as shown in Table~\ref{fig:orbfracs}, the presence of a central point mass essentially destroys the majority of the  x1 orbits in the inner regions of the bar and prograde $z$-tubes (parented by x2 orbits) are only produced by the action of a SMBH or dissipative gas inflow \citep{debattista_etal_15_nucMW}.

There is a substantial body of literature on Schwarzschild modeling of triaxial ellipsoids with and without figure rotation \citep[e.g][]{schwarzschild_79, schwarzschild_82,vandenbosch_etal_08, vasiliev_13} which has recently been extended to modeling bars \citep{wang_mao_etal_13,vasiliev_athanassoula_15}.  A new unified framework for understanding the orbital structure of bars, especially bars with self-consistently grown SMBHs will make it possible to  construct more realistic Schwarzschild models for  barred disk  galaxies. This is important for ensuring that black hole masses at the low-end of the $\mbh-\sigma$ relation are accurately measured, since systematic overestimates of black hole masses in barred disks would erase morphological differences between the black hole scaling relations of disks and ellipticals, which could be crucial to understanding the co-evolution of black holes and their host galaxies. 

\section*{Acknowledgments}
The authors wish to thank  P.~T.~ de~Zeeuw and an anoynmous referee for constructive comments that significantly improved the presentation of this paper. MV and CA were supported in part by University of Michigan's Office of Research, HST-AR-13890.001, NSF award AST-0908346, NASA-ATP award NNX15AK79G.  The research presented here is partially supported by the 973 Program of China under grant no. 2014CB845700, by the National Natural Science Foundation of China under grant nos.11333003, 11322326, 11073037, and by the Strategic Priority Research Program  ``The Emergence of Cosmological Structures'' (no. XDB09000000) of the Chinese Academy of Sciences. This work made use of the facilities of the Center for High Performance Computing at Shanghai Astronomical Observatory. VPD is supported by STFC Consolidated grant \#~ST/M000877/1 and partially supported by the Chinese Academy of Sciences President's International Fellowship Initiative (PIFI, Grant No. 2015VMB004) .

\appendix

\section{A: Orbital Frequency analysis}
\label{sec:frequencies}

In this section we provide a brief description of orbital frequency analysis method first introduced by \citet{binney_spergel_82} and \citet{binney_spergel_84} and further developed by \citet{laskar_90, laskar_93}. In Hamiltonian dynamics the angle variables and their  canonically conjugate  actions $J_i$ uniquely define a regular orbit (BT08). The time derivatives of three angle variables are the fundamental frequencies $\dot{\theta_i}(t)=\Omega_i$.  Orbits in galaxies are approximately quasi-periodic (BT08), hence their space and velocity coordinates can be represented by time series of the form:
 \beq
 x(t) = \sum_{k=1}^{k_{\rm max}} A_{k} e^{i\omega_kt}
 \eeq
with similar expressions for $y(t), z(t)$ and velocity components, $V_x(t), V_y(t), V_z(t)$.  The  amplitudes $A_k$ and the frequencies $\omega_k$,  can be obtained by taking a Fourier transform (with a window function) of a time series $f(t)$ constructed from the spatial or/and velocity coordinates of an orbit. For a regular orbit in a three dimension potential, only three frequencies in the spectrum, $\Omega_i, i=1,..,3$ are linearly independent, and all other frequencies in the spectrum can be written as integer linear combinations of these three frequencies, therefore the  $\Omega_i, i=1,..,3$ are referred to as ``fundamental frequencies''.   
 \citet{laskar_90, laskar_93}   developed an accurate numerical technique ``Numerical Analysis of Fundamental Frequencies''  to recover frequencies in completely general potentials. In this paper we use the implementation of this algorithm\footnote{Made available publicly at\\ http://dept.astro.lsa.umich.edu/$\sim$mvalluri/resources.html} first presented in \citet{valluri_merritt_98} and subsequently modified to work with orbits in $N$-body potentials \citep{valluri_etal_10}.  With this code the frequency components in the spectrum can be recovered with high accuracy (1 part in $10^{-5}$ or better) in $\sim 20-30$ orbital periods. 
 Our code  uses  integer programming to recover orbital fundamental frequencies. We have applied  this frequency analysis method to orbits of  particles in simulations of triaxial halos \citep{valluri_etal_10,  valluri_etal_12}, to the orbits of halo stars and dark matter particles in fully cosmological-hydrodynamical simulations \citep{valluri_etal_13} and to orbits in rotating triaxial potentials \citep{deibel_etal_11}. 
 
 To determine the fraction of chaotic orbits in a potential, the orbital time series is divided into two equal segments and the orbital fundamental frequencies are computed in each time segment. Since regular orbits have fixed frequencies that do not change with time, the change in the frequency measured in the two time segments can be used to measure the drift in frequency space. The
``frequency drift'' for each frequency component $\Omega_i$ can be computed as:
\begin{eqnarray}
\log(\Delta f_i) = \log{\left|{\frac{\Omega_i(t_1)-\Omega_i(t_2)}{\Omega_i(t_1)}}\right|},\\
\end{eqnarray}
We define the frequency drift parameter $\log(\Delta f)$ (logarithm to base 10) for an orbit to be the
value associated with the fundamental frequency $\Omega_i$ with the largest amplitude in the Fourier spectrum. The larger the value of the frequency drift parameter, the more chaotic the orbit. Since the frequency difference is normalized by the absolute value of the frequency it is possible to compare diffusion rates of orbits with a wide range of orbital periods.  \citet{valluri_etal_10} showed that even for orbits in frozen $N$-body potentials (where most orbits are affected by discreetness noise), it is possible to distinguish between $N$-body jitter and true chaos. In their tests orbits with $\log(\Delta f) > -1.2$   were defined as chaotic. We use the same criterion in this paper.

\section{B: Automated bar orbit classification}
\label{sec:autoclassification}

Automated classification of orbits in triaxial $N$-body models is based on orbital fundamental frequencies. The method is well developed and has been utilized in the past \citep{carpintero_aguilar_98, valluri_etal_10}. We refer the reader to these papers for further details.  In this Appendix we describe a similar automatic classification scheme for orbits in  bars.   

\begin{figure*}
\centering
\includegraphics[trim=0.pt 0.pt 0.pt 0pt,width=0.45\textwidth]{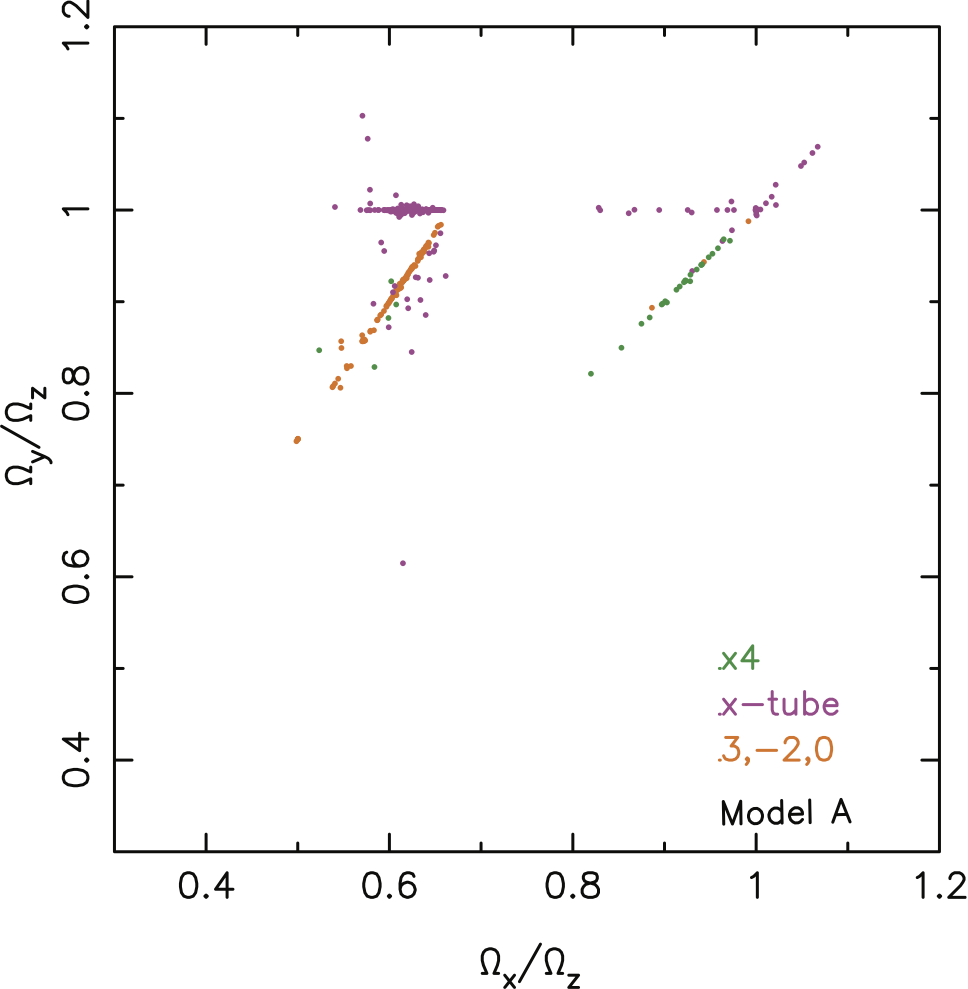}
\includegraphics[trim=0.pt 0.pt 0.pt 0pt,width=0.435\textwidth]{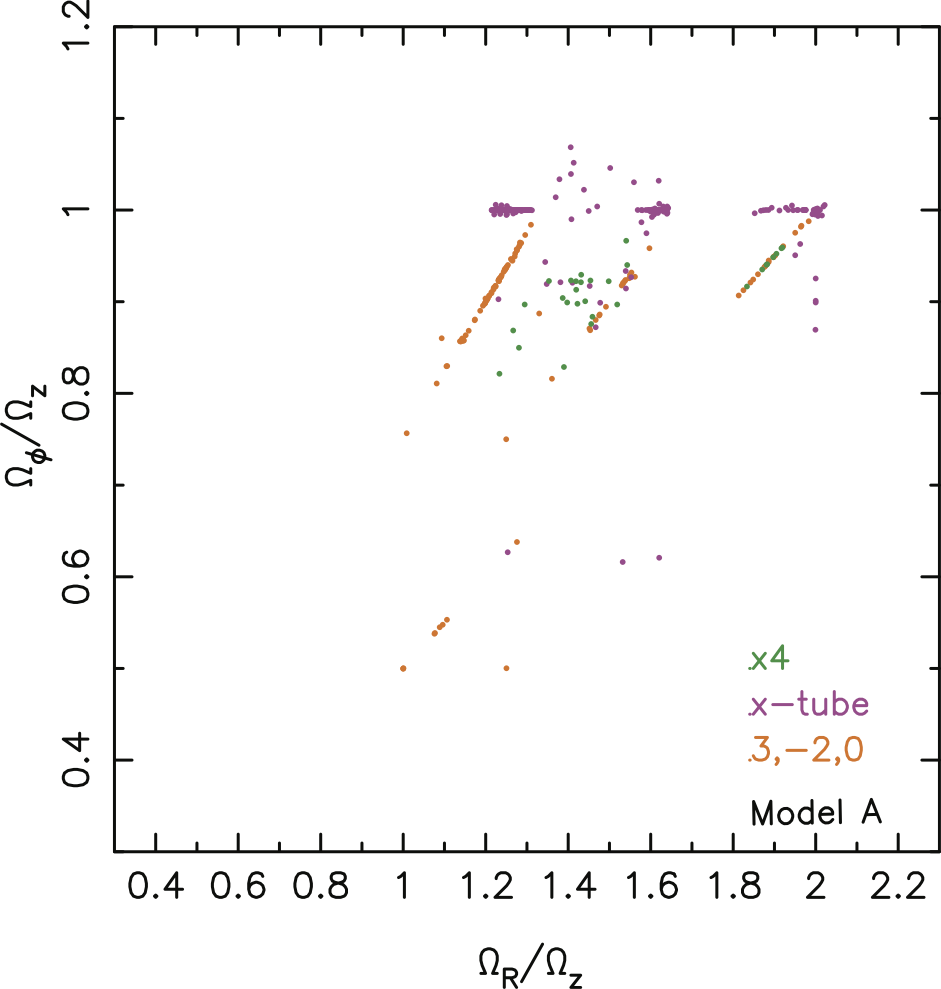}
\caption{Frequency maps constructed from  fundamental frequencies  in Cartesian coordinates (left) and cylindrical polar coordinates (right).    Only three orbit families from Model A,  color coded on the basis of their visual classification, are plotted: $x$-tubes (pink dots), retrograde $z$-tubes  parented by x4 orbits (green dots) and 3:-2:0 resonant orbits (orange dots). Resonance lines are not marked but orbits can be seen to clearly lie along 3 main resonances (each associated with one primary orbit family) in the Cartesian map (left). In the cylindrical frequency map the same families are split into multiple groups - many of which overlap - making it difficult to use these frequencies to classify orbits.}
\vspace{0.5cm}
\label{fig:cyl_cart}
\end{figure*}

 \citet{voglis_etal_07} used orbital fundamental frequencies in cylindrical polar coordinates in the frame co-rotating with the bar ($\Omega_R, \Omega_\phi, \Omega_z$) to make frequency maps and characterize regular and chaotic orbits in their $N$-body bars. However, as we now discuss, we find  that orbital frequencies in Cartesian coordinates provide a more robust means for bar orbit classification. 
 
Figure~\ref{fig:cyl_cart} shows frequency maps constructed from  fundamental frequencies  computed in Cartesian coordinates in the frame co-rotating with the bar (left) and in cylindrical polar coordinates (right).    Three orbit families, selected on the basis of visual classification of orbits in Model A  are plotted: $x$-tubes (pink dots),  retrograde $z$-tubes parented by x4 orbits (green dots) and 3:-2:0 resonant orbits (orange dots).  In the map on the left, the three families lie in reasonably (thought not perfectly) separate regions of the map. 
The separation of the different orbit families seen in a Cartesian frequency map (also in Fig.~\ref{fig:freqmap}) is quite similar to the separation found in self-consistent stationary triaxial potentials \citep{valluri_etal_12}.

In  the cylindrical frequency map (Fig.~\ref{fig:cyl_cart}~right) each orbit type (i.e. points of a single color) appears in multiple groups and there is significant overlap between the different types. We do not plot the box orbits in these plots  (the most numerous population)  to avoid overcrowding the maps. It can be seen in Figure~\ref{fig:freqmap} that the box orbits are concentrated in the cloud of points to the left of the diagonal. However in cylindrical coordinates this family is split into multiple groups which overlap with other families. \citet{voglis_etal_07} used frequency maps in cylindrical coordinates in which box-like orbits were separated into multiple groups. They used these grouping to classify orbits into x1, x4 orbits, several resonant families and two large groups that they called ``Group A'', and ``Group B''. Our maps in cylindrical coordinates also showed similar groups (e.g. Fig.~\ref{fig:cyl_cart}, right) but we found that the different groups were not characterized by unique qualitative or quantitative properties. 

Since our primary objective is to use orbital  frequencies to {\em automatically classify orbits}, the larger degree of separation of different orbit families  (boxes, $x$-tubes, $z$-tubes  and x1 orbits) in Cartesian frequency maps, though not perfect, in combination with orbital angular momentum, and orbital elongation results in the robust orbit classification scheme described below.
\begin{figure*}
\centering
\includegraphics[trim=0.pt 0.pt 0.pt 0pt,width=0.45\textwidth]{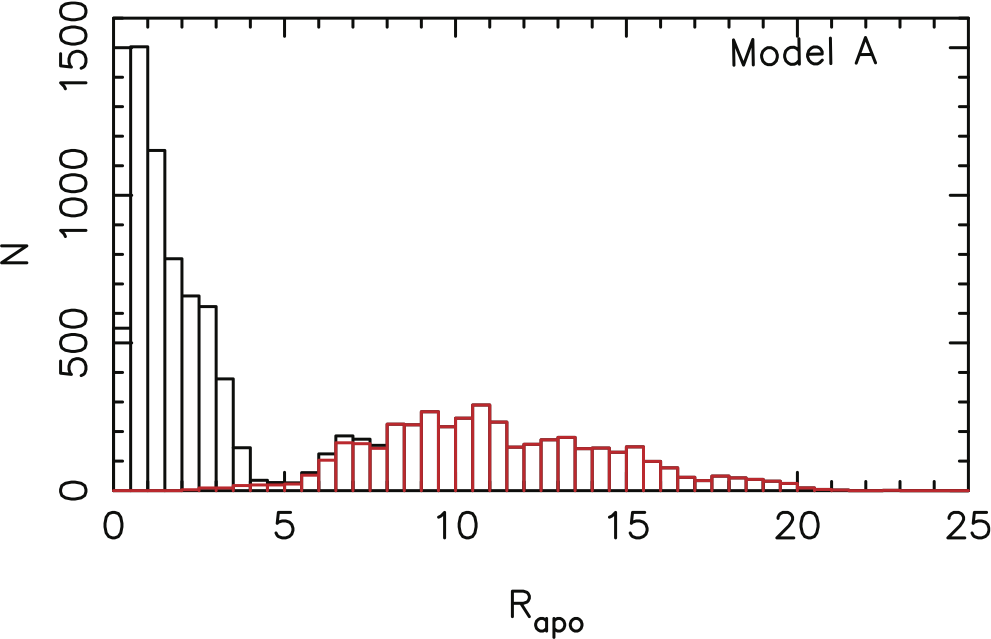}
\includegraphics[trim=0.pt 0.pt 0.pt 0pt,width=0.45\textwidth]{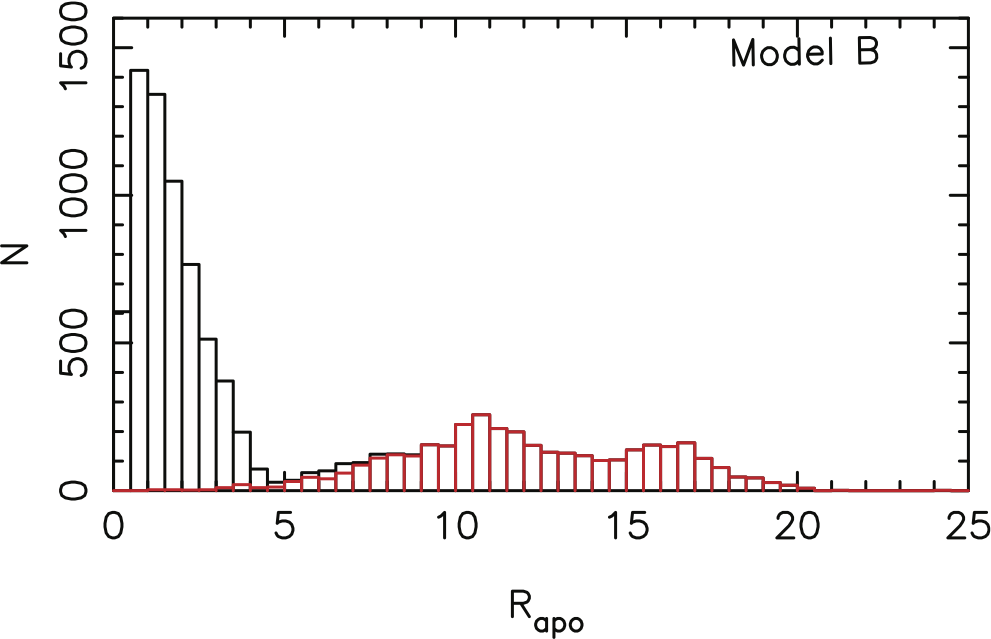}
\caption{Histograms of apocenter radii of 10000 orbits in Model A (left) and Model B (right). The black histograms shows apocenter radii of all 10,000 orbits in the model (determined by integrating each orbit for $t=1000$ in program units). The  red histograms shows the distribution of apocenter radii of all orbits that were visually classified as disk orbits.  The clear break at $\sim 4$  in both models coincides with the value of $r_{\rm apo}$ beyond which the majority of the disk orbits are found and is used to define the end of the bar. }
\label{fig:bar_apo}
\end{figure*}

Our orbit classification algorithm begins by identifying the location of the end of the bar. In Figure~\ref{fig:bar_apo} the black and red histograms shows the apocenter radii of all orbits in each of the two models. The red histograms show the distribution of apocenter radii of all orbits that were visually classified  as ``disk'' orbits (including the short/long period orbits).  Both histograms show a  clear break in $R_{\rm apo}$. This break coincides with the apocenter radius beyond which the majority of the visually classified disk orbits lie, consequently $R_{apo}=4$  defines the transition between the bar and the disk. The few disk orbits with $R_{\rm apo} < $ the break radius are classified as $z$-tubes by the automated classifier. All orbits that lie beyond the bar are automatically classified as ``disk'' orbits.  

For each orbit the 3 fundamental frequencies in Cartesian coordinates $\Omega_x, \Omega_y, \Omega_z$ are obtained using frequency analysis. We then determined which (if any) of the frequency ratios were rational i.e. if any pair of frequencies $\Omega_i/\Omega_j = n_i/n_j$ where $n_i, n_j,n_k < 10$ ($i, j,k$ corresponding to $x, y, z$  coordinates respectively). For determining whether the frequency ratios are rational we multiply both the denominator and numerator by successive integers ($<10$) and compute the difference between the new numerator (denominator) and the integer, until this difference is less than 0.01. This condition allows us to identify both strictly periodic or resonant orbits as well as those that are associated with a periodic or resonant family but are not strictly periodic (closed).

If all three frequency ratios are rational i.e. if  $\Omega_x/ \Omega_y= n_i/n_j$,  $\Omega_y/ \Omega_z=n_j/n_k$ and hence  $\Omega_z/ \Omega_x =n_k/n_i$ and $n_i\neq n_j \neq n_k$ the orbit is a closed (periodic) box orbit.  If $n_j= n_k \neq n_i$ the orbit is a  closed periodic long axis tube orbit.  However if  $n_i = n_j \neq n_k$ the orbit is a closed periodic  x1 orbit. (Note that if two pairs of frequency ratios are rational, then the third pair must be rational as well.)
  
Only one pair of frequency ratios is rational, e.g. if  $\Omega_y/ \Omega_z=n_j/n_k$  then if $n_j=n_k$ the orbit loops around the short axis ($z$ tube, x2, x4), but if  $n_j\neq n_k$ the orbit is an open (resonant) boxlet. Likewise if $\Omega_y/ \Omega_z=n_j/n_k$ and if $n_j =n_k$ then the orbit is an open long axis tube but if $n_j \neq n_k$ the orbit is also an open boxlet. 

These conditions are essentially identical to those used to classify orbits  in triaxial potentials into boxes, $z$-tubes and $x$-tubes \citep{carpintero_aguilar_98, valluri_etal_10}. In addition to these conditions on orbital frequencies, for some types of orbits we required that certain conditions on the time averaged normalized angular momenta to be satisfied. For retrograde $z$-tubes $\langle{J_z}\rangle <-0.95$ and for prograde $z$-tubes $\langle{J_z}\rangle >+0.95$. 

Classical x1 orbits  are those for which $\Omega_x/\Omega_y \sim 1$ and $\Omega_y/\Omega_z< 0.7$ (determined empirically from the frequency map). We also find that box orbits (as classified above) that have $\langle{J_z}\rangle >+0.6$ and $|y|_{\rm max}/|x|_{\rm max} < 0.35$ (see Fig.~\ref{fig:Jzvsybyx}) are classical x1 orbits. We therefore use these two criteria to reclassify orbits that may have been at first classified as boxes or x1 orbits.  Banana orbits are boxes or x1 orbits which also satisfy the condition $\Omega_z:\Omega_x=2:1$. Brezel orbits are boxes which also satisfy the condition 
$\Omega_z:\Omega_x=-5:3$.  The ``3:-2:0'' resonant orbits satisfy the condition $\Omega_x/\Omega_y = 3:-2$. 
      


\bibliography{barorbs}


\end{document}